\pdfoutput=1
\documentclass[11pt]{article}

\usepackage{etex}
\reserveinserts{28}

\usepackage[margin=1in]{geometry}
\usepackage{amsmath,amssymb,amsfonts}
\usepackage[multiple,stable]{footmisc}
\usepackage[amsmath,amsthm,thmmarks]{ntheorem}
\allowdisplaybreaks[4]
\usepackage[noconfig]{refstyle}
\usepackage[dvipsnames]{xcolor}
\usepackage[hyperfootnotes=false, linktocpage=true, colorlinks, citecolor=blue, linkcolor=blue, urlcolor=Maroon]{hyperref}
\usepackage{array,tabularx,booktabs,geometry,graphicx,longtable}
\usepackage{caption}
\usepackage{tikz}
\usetikzlibrary{shapes,arrows}
\tikzstyle{block} = [rectangle, draw, text width=7em, text centered, rounded corners, minimum height=3em]
\usepackage{graphicx,color}
\usetikzlibrary{positioning,arrows}
\usepackage{amssymb}
\usepackage{arydshln}
\usepackage{color}

\usepackage[subrefformat=parens,labelformat=parens]{subfig}

\usepackage{cite}
\usepackage{arydshln}
\usepackage{tikz}
\usetikzlibrary{positioning,automata,decorations.pathreplacing,shapes}
\usepackage[english]{babel}
\usepackage{microtype}

\usepackage[all,cmtip]{xy}
\usetikzlibrary{matrix, arrows}

\newcommand{\firstp}[0]{\mathbb{P}^2_{(1)}}
\newcommand{\secondp}[0]{\mathbb{P}^1_{(2)}}
\newcommand{\thirdp}[0]{\mathbb{P}^2_{(3)}}

\newcommand{\firstxone}[0]{x_{0}}
\newcommand{\firstxtwo}[0]{x_{1}}
\newcommand{\firstxthree}[0]{x_{2}}

\newcommand{\secondxone}[0]{y_{0}}
\newcommand{\secondxtwo}[0]{y_{1}}

\newcommand{\thirdxone}[0]{z_{0}}
\newcommand{\thirdxtwo}[0]{z_{1}}
\newcommand{\thirdxthree}[0]{z_{2}}
\newcommand{\divsec}[0]{\Sigma}

\let\eqref=\relax
\newref{eq}{name={eq.~},Name={Eq.~},names={eqs.~},Names={Eqs.~},rngtxt={-},refcmd=(\ref{#1})}
\newref{f}{name={footnote~},Name={Footnote~},names={footnotes~},Names={Footnotes~}}
\newref{app}{name={appendix~},Name={Appendix~},names={appendixes~},Names={Appendixes~}}
\newref{tab}{name={table~},Name={Table~},names={tables~},Names={Tables~}}
\newref{ch}{name={chapter~},Name={Chapter~},names={chapters~},Names={Chapters~}}
\newref{sec}{name={section~},Name={Section~},names={sections~},Names={Sections~}}
\newref{fig}{name={figure~},Name={Figure~},names={figures~},Names={Figures~}}
\numberwithin{equation}{section}

\newtheorem{theorem}{Theorem}[section]

\usepackage{hyperref}
\usepackage{extarrows}

\newcommand{\eref}[1]{(\ref{#1})}

\newcommand{\eeq}{\end{equation}}
\newcommand{\beq}{\begin{equation}}
\newcommand{\ba}{\begin{array}}
\newcommand{\ea}{\end{array}}

\newcommand{\rk}{{\rm rk}}

\def\a{\alpha}

\newcommand{\id}{\mathbf{1}}

\def\clap#1{\hbox to 0pt{\hss#1\hss}}

\newcommand{\be}{\begin{equation}}
\newcommand{\ee}{\end{equation}}
\newcommand{\bea}{\begin{equation}\begin{aligned}}	
\newcommand{\eea}{\end{aligned}\end{equation}}		

\newcommand{\field}[1]{\mathbb{#1}}

\newcommand{\C}{\field{C}}

\newcommand{\iddots}{\mathinner{\mkern2mu\raise1pt\hbox{.}\mkern2mu \raise4pt\hbox{.}\mkern2mu\raise7pt\hbox{.}\mkern1mu}}

\DeclareMathOperator{\SU}{SU}
\DeclareMathOperator{\SO}{SO}

\providecommand{\id}{\leavevmode\hbox{\small$\mathrm{1}$\kern-3.8pt\normalsize$\mathrm{1}$}}
\def\fnote#1#2{\begingroup\def\thefootnote{#1}\footnote{#2}
     \addtocounter{footnote}{-1}\endgroup}
     
\newcommand*{\conicbundle}[0]{\mathcal{B}_3}
\newcommand*{\conicbase}[0]{B_2}

\begin{document}

\title{ {\flushright \vspace{-1.5cm}
\normalsize UUITP--43/21\\[5mm]}  
\vspace{1cm}     
  {\Large \bf $\mathbb{P}^1$-fibrations in F-theory and String Dualities}}

\vspace{2cm}

\author{
Lara~B.~Anderson,${}^{1}$
James~Gray,${}^{1}$
Mohsen~Karkheiran,${}^{2}$ \\
Paul-Konstantin~Oehlmann,${}^{3}$ 
and
Nikhil~Raghuram${}^{1}$
}
\date{}
\maketitle
\begin{center} {\small ${}^1${\it Department of Physics, 
Robeson Hall, Virginia Tech \\ Blacksburg, VA 24061, U.S.A.} \\[0.2cm]
    ${}^2${\it Center for Theoretical Physics of the Universe, Institute for Basic Science,\\
       $~~~~~$ Daejeon 34051, South Korea.}\\[0.2cm]
           ${}^3${\it Department of Physics and Astronomy, Uppsala University,\\
       $~~~~~$ SE-751 20 Uppsala, Sweden.}}\\

\fnote{}{lara.anderson@vt.edu}
\fnote{}{jamesgray@vt.edu}
\fnote{}{mohsenkar@ibs.re.kr}
\fnote{}{paul-konstantin.oehlmann@physics.uu.se}
\fnote{}{raghuram.nikhil@gmail.com}

\end{center}

\begin{abstract}
\noindent In this work we study F-theory compactifications on elliptically fibered Calabi-Yau $n$-folds which have $\mathbb{P}^1$-fibered base manifolds. Such geometries, which we study in both 4- and 6-dimensions, are both ubiquitous within the set of Calabi-Yau manifolds and play a crucial role in heterotic/F-theory duality. We discuss the most general formulation of $\mathbb{P}^1$-bundles of this type, as well as fibrations which degenerate at higher codimension loci. In the course of this study, we find a number of new phenomena. For example, in both 4- and 6-dimensions we find transitions whereby the base of a $\mathbb{P}^1$-bundle can change nature, or ``jump", at certain loci in complex structure moduli space. We discuss the implications of this jumping for the associated heterotic duals. We argue that $\mathbb{P}^1$-bundles with only rational sections lead to heterotic duals where the Calabi-Yau manifold is elliptically fibered over the section of the $\mathbb{P}^1$ bundle, and not its base. As expected, we see that degenerations of the $\mathbb{P}^1$ fibration of the F-theory base correspond to 5-branes in the dual heterotic physics, with the exception of cases in which the fiber degenerations exhibit monodromy. Along the way, we discuss a set of useful formulae and tools for describing F-theory compactifications on this class of Calabi-Yau manifolds.
\end{abstract}

\thispagestyle{empty}
\setcounter{page}{0}
\newpage

\tableofcontents

\section{Introduction}\label{sec:Intro}

F-theory is a powerful framework for the ``geometrization" of effective field theories arising in string theory \cite{Vafa:1996xn}. As such, it is interesting to consider ways that the geometric backgrounds of F-theory -- namely, elliptically fibered Calabi-Yau $n$-folds -- can be systematically built utilizing the most general constructions possible.

In this work we consider the geometry of $K3$-fibered Calabi-Yau (CY) manifolds in compactifications of F-theory \cite{Morrison:1996na,Morrison:1996pp,Bershadsky:1996nh,Friedman:1997ih,Friedman:1997yq}. Within the context of F-theory, the geometric realization of the Type IIB axio-dilaton as a torus fibration constrains the $(n+1)$-dimensional CY compactification geometry, $Y_{n+1}$, to be elliptically fibered. That is, there exists a surjective map $\pi_f: Y_{n+1} \to \mathbb{B}_n$ with elliptic fiber, $\mathbb{E}$. Moreover, in the case that the geometry is also compatibly $K3$-fibered over a base $B_{n-1}$, the base to the elliptic fibration must be $\mathbb{P}^1$-fibered and the following relationships hold:

\begin{equation}
\begin{array}{lllll}
&~~~~Y_{n+1}&\xrightarrow{~~\mathbb{E}~~}&{\cal B}_n&\\
 &K3~\Big\downarrow&&~\Big\downarrow~\mathbb{P}^1& \\
&~~~~B_{n-1} &\xleftrightarrow{~~=~~} &B_{n-1}&
\end{array}
\label{nested_fib}
\end{equation}

Given a $K3$-fibered manifold as shown above, the effective physics of F-theory compactified on $Y_{n+1}$ is expected to be dual to that of the $E_8 \times E_8$ heterotic string compactified on an elliptically fibered CY $n$-fold, $\pi_h: X_n \to B_{n-1}$, where the base to the heterotic elliptic fibration is \emph{the same} as the base to the $K3$ fibration shown above. Heterotic/F-theory duality has long been a useful tool in the study of the resulting effective theories and has motivated much progress in F-theory physics over the past twenty years.

Independent from any interest in string dualities, the structure of the base geometry to the F-theory elliptic fibration shown in \eref{nested_fib} above -- \emph{namely that ${\cal B}_n$ is a $\mathbb{P}^1$-fibration} -- has proven to play a central role in characterizations of elliptic CY manifolds. More precisely, the seminal work of Grassi and Gross \cite{Gross:1993fd,grassi}, which led to the proof that all genus-one fibered CY threefolds are finite, characterized the possible base geometries of elliptic CY threefolds as living in the following set of complex surfaces: the Enriques surface, $\mathbb{P}^2$, Hirzebruch surfaces and blow-ups of Hirzebruch surfaces. Since Hirzebruch surfaces and their blow-ups are $\mathbb{P}^1$-fibered, it is clear that essentially all bases of genus-one fibered CY threefolds admit $\mathbb{P}^1$ fibrations. Moreover, in powerful recent work, Birkar, Di Cerbo and Svaldi \cite{Di_Cerbo_2021,birkar2020boundedness} have completed a similar characterization of the birational geoemtry of bases of elliptically fibered CY 4-folds. In that setting the birational characterization of the three complex dimensional base manifolds in terms of so-called ``Fano towers" of fibrations once more places $\mathbb{P}^1$-fibrations in a central role in the classification. Finally, this apparent ubiquity to $\mathbb{P}^1$-fibered bases is confirmed in a different way via recent scans of known CY datasets to identify $K3$ fibrations (see \cite{Anderson:2017aux,Gray:2014fla}). The results of that work indicate that for the CY datasets studied (the CY 3- and 4-folds formed as complete intersections in products of projective spaces) the vast majority of manifolds (more than $99\%$) are $K3$-fibered. In summary then, the results above indicate that a thorough understanding of $\mathbb{P}^1$-fibered base manifolds can shed light on the \emph{general} structure of genus one fibered CY $n$-folds. Thus far, they have also proved to be an important and tractable starting point in efforts to generate large datasets of elliptically fibered CY $4$-folds \cite{Friedman:1997yq,Anderson:2014gla,Halverson:2015jua}.

In view of these results then, the goal of the present work is to generalize the construction of elliptically fibered CY manifolds with a $\mathbb{P}^1$-fibered base considered in the literature to date and to study their consequences for heterotic/F-theory duality. To begin, it is important to note that $\mathbb{P}^1$-fibrations can be simply divided into two classes: 
\begin{itemize}
\item Those that are nowhere degenerate (i.e. $\mathbb{P}^1$-fibrations that are in fact \emph{$\mathbb{P}^1$-bundles}) and
\item Those that \emph{do} degenerate over a higher codimensional discriminant locus in $B_{n-1}$. 
\end{itemize}

In the present literature, only a simple class of the first type -- namely of $\mathbb{P}^1$-bundles formed via the projectivization of a sum of line bundles over $B_{n-1}$ -- has been systematically studied in the context of heterotic/F-theory duality \cite{Friedman:1997yq}. It is the goal of the present work to extend this geometry in two important ways in the context of 6- and 4-dimensional effective theories arising from heterotic string theory and F-theory:
\begin{itemize}
\item In 4-dimensional compatifcations of F-theory, we consider new classes of $\mathbb{P}^1$-bundles, defined as the projectivization of a \emph{general rank 2 vector bundle} over $B_2$. The Hartshorne/Serre construction (see e.g. \cite{Friedman:1998}) guarantees that any rank 2 vector bundle over a complex surface can be described via an extension sequence
\beq\label{serre_bundle}
0 \to L_1 \to V_2 \to L_2 \otimes {\mathcal I}_z \to 0
\eeq
where $L_1$ and $L_2$ are line bundles over $B_2$ and $\mathcal{I}_{z}$ is an ideal sheaf associated to a set of points $\{z\}$ on $B_2$.

We will build more general base manifolds ${\cal B}_3$ for F-theory as the projectivization, $\mathbb{P}(\pi: V_2 \to B_2)$, of the rank $2$ bundles shown in \eref{serre_bundle}. We will see in this case that rational sections can appear in the $\mathbb{P}^1$-bundle which have important consequences in heterotic/F-theory duality.
\item We will further analyze $\mathbb{P}^1$-fibrations which degenerate over some sublocus\footnote{Note we will refer to the discriminant of the $\mathbb{P}^1$-fibration as $\Delta_b \subset B_{n-1}$ to distinguish it from the discriminant of the CY elliptic fibration, $\Delta_f \subset {\cal B}_n$.} $\Delta_b \subset B_{n-1}$. In the mathematics literature, the Sarkisov program (see e.g. \cite{Sarkisov_1983}) has led to a systematic classification of such objects in terms of their birational geometry. The simplest example of a fiber which could degenerate consists of a $\mathbb{P}^1$ fiber which is described as a conic in $\mathbb{P}^2$: $\mathbb{P}^2[2]$. Over higher-codimensional loci in the base manifold, the defining equation of such a fiber can clearly factor into a product of two linear functions in $\mathbb{P}^2$, leading to a degeneration of the $\mathbb{P}^1$ fiber into \emph{two} distinct $\mathbb{P}^1$s over $\Delta_b$.
\end{itemize}
In each of these cases above we will review the geometry of $\mathbb{P}^1$-fibrations in as general a context as possible and comment on the effective physics of both F-theory and heterotic string theory defined over the relevant dual geometries (as in \eref{nested_fib}).

The geometry and topology of both more general $\mathbb{P}^1$-bundles as well as $\mathbb{P}^1$-fibrations is explored in the following sections. We analyze how to construct elliptic CY manifolds over such fibered bases and an overview of some of the important features in the F-theory effective physics.
We find that moving away from the standard set of $\mathbb{P}^1$-fibered bases ${\cal B}_n$ (and hence $K3$-fibered CY manifolds), the effective physics of heterotic/F-theory duality becomes different from that seen in the standard situation \cite{Friedman:1997yq}. In particular, we demonstrate that they can behave differently under weak-coupling heterotic limits (i.e. under generalized stable degenerations \cite{Donagi:2012ts,Heckman:2013sfa}) and can lead to previously unexplored structure in the dual heterotic geometry.

The structure of this paper is as follows. In Section \ref{sec:p1fibs} we provide the basic geometric ingredients for the present study -- namely the known properties and categorizations of $\mathbb{P}^1$-bundles and fibrations. In Section \ref{sec:6DFtheory} we apply these insights to the study of $6$-dimensional F-theory compactifications on elliptically fibered surfaces. We reframe the standard perturbative and non-perturbative heterotic/F-theory duality in terms of properties of the $\mathbb{P}^1$-fibration. Most of this section is review, however, even in this well-understood arena we find new physical phenomena are possible via the so-called ``jumping effect" of $\mathbb{P}^1$-bundles in Section \ref{sec:jumping}. By analyzing the structure of $\mathbb{P}^1$-bundles over $\mathbb{P}^1$ we consider the possibility that for special values of the complex structure of the CY threefold the base complex surface can ``jump" between distinct complex manifolds -- for example between the Hirzebruch surfaces $\mathbb{F}_1$ and $\mathbb{F}_3$ (two geometries which are diffeomorphic as real manifolds, but distinct as complex manifolds) -- and the consequences of this for F-theory and heterotic string theory. Next in Section \ref{Sec:4dftheory} we consider simple $4$-dimensional compactifications of F-theory arising from more general $\mathbb{P}^1$-bundle $3$-dimensional base manifolds, ${\cal B}_3$. We argue that in cases where a $\mathbb{P}^1$-bundle only admits rational sections, the heterotic dual to the associated F-theory compactification is an elliptic fibration over the section, and not the base $B_2$, of ${\cal B}_3$. Finally, in Sections \ref{sec:conicnomono} and \ref{sec:conicmonodromy} we consider degenerate $\mathbb{P}^1$-fibrations, with and without monodromy respectively, and their consequences for dual physical theories. We find that in the cases where the degenerate fibers exhibit no monodromy, the descriminant locus in $B_2$ over which the fibers degenerate is a curve upon which 5-branes are wrapped in the heterotic dual. In the case where monodromy is present, it seems that no such interpretation is available.

\vspace{5pt}

We begin by briefly reviewing some of the geometric properties of $\mathbb{P}^1$-fibrations.

\section{$\mathbb{P}^1$-fibrations in a nutshell}\label{sec:p1fibs}

\subsection{$\mathbb{P}^1$-bundles}\label{sec:p1bun}

Let $B_{n-1}$ be the base to a $K3$ fibration as in \eref{nested_fib}. One definition of the Brauer group $\textnormal{Br}(M)$ for some complex manifold $M$ concerns classifying those projective bundles over ${\cal B}_{n-1}$ that cannot be built as the projectivization of a vector bundle (see e.g. \cite{brauer}). In particular, if the Brauer group of $M$ is trivial then all projective bundles over it are projectivizations of rank $2$ vector bundles over $M$ \cite{HARTSHORNE,Projectivebundlesonacomplextorus}. We will consider in this work the case that the base, $B_{n-1}$, to the F-theory $K3$-fibration has a trivial Brauer group. Intuitively, projectivization of the fiber space of a rank $m$ vector bundle, $\pi: V \to B_{n-1}$, is simply the transition from a non-compact $\mathbb{C}^m$-dimensional fiber to its compactification\footnote{Note that since we consider complex manifolds only, throughout this work $\mathbb{P}^m$ refers to \emph{complex} projective space.}, $\mathbb{P}^{m-1}$, where the fiber coordinates are identified up to a scale that is chosen by $\wedge^{m} V$, point by point over the base \cite{HARTSHORNE}. The resulting $\mathbb{P}^{m-1}$-fibered manifold is denoted 
\beq
\mathbb{P}(\pi: V \to B_{n-1})
\eeq
or simply $\mathbb{P}(V)$. As an example relevant to the present work, the projectivization of any rank $2$ vector bundle, $\mathbb{P}(V_2)$ over a base manifold $B_{n-1}$ is a smooth, $n$-dimensional manifold with a nowhere degenerate $\mathbb{P}^1$-fibration. 

The simplest example of this is to consider a sum of two line bundles
\beq\label{line_sum}
V_2=L_1 \oplus L_2
\eeq
 over $B_{n-1}$. Since the projectivization of a bundle is invariant under twists by an Abelian line bundle -- i.e. $\mathbb{P}(\pi: V \to B_{n-1})=\mathbb{P}(\tilde{\pi}: V \otimes L \to B_{n-1})$ for any line bundle $L$ -- without loss of generality the sum of line bundles can be chosen to be 
 \beq\label{special_abelian}
 V_2={\cal O} \oplus {\cal O}(D)
 \eeq
for some line bundle ${\cal O}(D)$.  In the case that $B_{n-1}$ is a toric manifold and $V_2$ is abelian as in \eref{line_sum}, the $\mathbb{P}^1$-fibered manifold $\mathcal{B}_n$ obtained by projectivization is also manifestly toric by construction. It is this class of geometries that initiated the first systematic studies of heterotic/F-theory duality. Examples include:
 \begin{itemize}
 \item In 6-dimensional dual compactifications of heterotic/F-theory, the dimension of the base of the elliptic fibration is $n=2$ in \eref{nested_fib} and the ``shared" base to the heterotic elliptic fibration and F-theory K3 fibration is simply $B_1=\mathbb{P}^1$. In this case the projectivization of a sum of line bundles yields the well-known Hirzebruch surfaces \cite{Friedman:1998}
 \beq
 \mathbb{P}(\pi: \mathcal{O} \oplus \mathcal{O}(n) \to \mathbb{P}^1)=\mathbb{F}_n
 \eeq
 \end{itemize}
 which provided the context for the first $6$-dimensional studies of heterotic/F-theory duality (see e.g. \cite{Bershadsky:1996nh}).
 
 Likewise, over any complex surface, the projectivization of a sum of line bundles provides a simple $\mathbb{P}^1$ bundle threefold, $\mathcal{B}_3$. This geometry was first outlined in an F-theoretic context by Friedman, Morgan and Witten \cite{Friedman:1997yq} and crucially used in their explicit matching of degrees of freedom, anomaly cancellation, etc in heterotic/F-theory duality. The projectivization of a sum of line bundles over a \emph{toric} complex surface was later employed in the literature to systematically generate large classes of $3$-fold bases for elliptically fibered CY $4$-folds \cite{Anderson:2014gla,Halverson:2015jua}. A simple example is given below for a toric $\mathbb{P}^1$ bundle defined over $\mathbb{P}^2$ (i.e. a threefold analog of a Hirzebruch surface). The toric weight matrix is as follows.
\beq\label{hirzebruch_generalized}
\begin{array}{ccccc}
x_0 & x_1 & y_0 & y_1 & y_2 \\
\hline
1& 1 & 0& 0& 0  \\
0& n & 1 & 1& 1  
\end{array}
\eeq
This manifold is the projectiviation $\mathbb{P}({\cal O}_{\mathbb{P}^2} \oplus {\cal O}_{\mathbb{P}^2} (-n))$. Although the projectivization of a sum of line bundles has led to well-understood heterotic/F-theory dual pairs, it is far from the only possibility. At this point we move beyond the simple cases considered in the F-theory literature to date.

As described in Section \ref{sec:Intro}, vector bundles over complex surfaces can be simply classified thanks to the Serre Construction \cite{Friedman:1998}. A general rank 2 vector bundle over $B_2$ takes the form:
\beq\label{serre_bundle2}
0 \to L_1 \to V_2 \to L_2 \otimes {\mathcal I}_z \to 0
\eeq
where $L_1$ and $L_2$ are line bundles over $B_2$ and $\mathcal{I}_{z}$ is an ideal sheaf associated to a (possibly empty) codimension 2 subscheme -- i.e. a set of points $\{z\}$ on $B_2$. As mentioned above, since $\mathbb{P}(V) \simeq \mathbb{P}(L \otimes V)$ for any line bundle $L$ \cite{HARTSHORNE} -- we can without loss of generality write this sequence as
\beq\label{defV}
0\rightarrow \mathcal{O}\rightarrow V_2 \rightarrow \mathcal{O}(D)\otimes \mathcal {\mathcal I}_z \rightarrow 0,
\eeq
and it is this form that we shall use going forward\footnote{Note that due to the presence of the ideal sheaf in the last term in \eref{defV}, the extension sequence can be non-split even when $h^1(B_2,{\cal O}(-D))=0$.}.

In the case that the ideal sheaf is non-trivial, the resulting $\mathbb{P}^1$ bundle, ${\cal B}_3=\mathbb{P}(V_2)$, can have very different properties to those described for Abelian bundles in \eref{special_abelian} above (i.e. the case of trivial extension class and vanishing $\{z\}$). One important feature that we will explore further in subsequent sections is that \emph{sections} to the $\mathbb{P}^1$ fibration might be quite different in character. We will explore this in detail in Section \ref{Sec:4dftheory} and Appendix \ref{Section_appendix}, but briefly the projective bundle defined by \eref{special_abelian} will always admit two holomorphic sections (which play a key role in the stable degeneration limit of heterotic/F-theory duality). By contrast, the $\mathbb{P}^1$ bundle defined by \eref{serre_bundle2} need not have two holomorphic sections. Instead, it may admit only \emph{rational} sections (which are birational but not diffeomorphic to the base, $B_2$).

The projectivization $\mathbb{P}(V_2)={\cal B}_3$ always comes equipped with a canonical section to the $\mathbb{P}^1$ fibration (dual to the so-called ``tautological" line bundle \cite{HARTSHORNE}). That is, there exists a divisor, $S$, on $\pi: {\cal B}_3 \to B_2$ and associated line bundle $\mathcal{O}(S)$ such that 
\beq\label{sdef1}
\pi_*(\mathcal{O}(S))=V_2
\eeq
and the intersection of $S$ with the generic fiber of the $\mathbb{P}^1$ fibration is the hyperplane within that fiber (i.e. a single point). Thus, $S$ is a section to the $\mathbb{P}^1$-fibration (i.e. it induces a map $\sigma: B_2 \to {\cal B}_3$ such that $\pi \circ \sigma=id_{B_2}$). The intersection structure of the manifold guarantees that
\beq \label{elfredo}
S^2 = c_1(V_2) \cdot S -c_2(V_2)
\eeq
where $c_1(V_2)=D$ and $c_2(V_2)=[z]$ (see Appendix \ref{sometop} for more details).

The addition of this divisor completes the Picard group of $B_3$ and the K\"ahler structure of ${\cal B}_3$ is simply induced from that of the fiber and the base with 
\beq\label{simple_pic}
h^{1,1}({\cal B}_3)=1+ h^{1,1}(B_2)
\eeq
Moreover, the Chern classes are readily computed, as described in Appendix \ref{sometop}, to be
\begin{eqnarray}\nonumber
c_1({\cal B}_3) &=& c_1(B_2)-D+2 S, \\  \label{mrthechern}
c_2({\cal B}_3) &=& - c_1(B_2) \cdot D+2 c_1(B_2) \cdot S+ c_2(B_2)-D\cdot S+S^2 + [z], \\\nonumber
c_3({\cal B}_3) &=& - c_1(B_2) \cdot D\cdot S+ c_1(B_2) \cdot S^2+2 c_2(B_2) \cdot S.
\end{eqnarray}
It is worth noting that the expressions above can be re-written to put them in the same apparent form one would obtain from the projectivization of two line bundles, $\mathbb{P}({\cal O} \oplus {\cal O}(D))$. However this is misleading since the intersection relation in \eref{elfredo} has crucially changed (depending now on $c_2(V_2)$ and hence $\left[z\right]$). We will return to the physical significance of this term for anomaly cancellation in heterotic/F-theory duality in Section \ref{Sec:4dftheory}.

\subsubsection{Sections to $\mathbb{P}^1$-bundles}
Before concluding this section, we return now to the issue of whether or not there exist other sections to the $\mathbb{P}^1$-fibration, besides $S$ described in \eref{sdef1}. This geometric question is intrinsically linked to the form that heterotic/F-theory duality will take in this context \cite{Friedman:1997yq} (including so-called stable degeneration limits \cite{Friedman:1997yq,Aspinwall_1998} and their generalizations \cite{Donagi:2012ts,Heckman:2013sfa}). It will also be crucial to determine whether or not these sections are rational or holomorphic -- that is whether the zero-locus of the section is strictly birational to the base, $B_2$, or exactly diffeomorphic to it.  

As mentioned above, the projectivization of a bundle is invariant under twists by a line bundle. That is, the projectivization of the twist of a rank 2 vector bundle $V_{\cal L}=V_2\otimes {\cal L}$ by a line bundle produces the same projective bundle as that of $V_2$ itself.  For all choices of twist, the projectivization ${\cal B}_n=\mathbb{P}(V_2) =\mathbb{P}(V_{\cal L})$ comes equipped with a set of divisors that intersect a generic fibers $S_{\cal L}$ in a hyperplane, that is a point \cite{HARTSHORNE}. Rephrasing \eref{sdef1} this is
\beq\label{sdef}
\pi_*(\mathcal{O}(S_{\cal L}))=V_{\cal L} \;.
\eeq
The divisor classes $S=S_{\cal O}$ and $S_{\cal L}$ are related as $S_{\cal L}=S_{\cal O} + \pi^*c_1({\cal L})$.  The divisor class $S_{\cal L}$ defines a section iff it is effective. As shown in Appendix \ref{Section_appendix} this is the case iff $V_{\cal L}$ has global sections.

Let us briefly summarize the several important results that will play a key role in our discussion. If ${\cal B}_n=\mathbb{P}(V_2) =\mathbb{P}(V_{\cal L})$ then
\begin{itemize}
\item ${\cal B}_n$ admits two disjoint holomorphic sections iff the bundle $V_2$ splits as a sum of two line bundles. 
\item The base ${\cal B}_n$ admits a single holomorphic section iff the bundle $V_2$ can be written as a non-trivial extension of two line bundles.
\item The base ${\cal B}_n$ always admits at least one rational section (shown in \eref{sdef1} above). 
\end{itemize}

The proofs of the above statements can be derived in a straightforward way from results on projectivization (see e.g. \cite{Eisenbud_Harris}) and are summarized in Appendix \ref{Section_appendix}. We will explore each of these situations --and their consequences for heterotic/F-theory duality -- in later sections. 

For now we will simply comment that the results of Appendix \ref{Section_appendix} demonstrate that the existence of two distinct holomorphic sections reduces the $\mathbb{P}^1$ bundle to the projectivization of the familiar form $\mathbb{P}(L_1 \oplus L_2)$. In this case, we can without loss of generality twist $V_2$ to write $\mathbb{P}( {\cal O} \oplus {\cal O}(D))$ and the two defining sections of the $\mathbb{P}^1$ fibration are then the well known $S_0, S_{\infty}$ introduced in \cite{Friedman:1997yq}. Here $S_{0}$ is the canonical section defined in \eref{sdef} above and $S_{\infty}=S_0 + \pi^*D $ is defined so that $S_0 \cdot S_{\infty}=0$. These sections play a crucial role in defining standard stable degeneration limits \cite{Friedman:1997yq,Bershadsky:1996nh,Aspinwall_1998,Donagi:2012ts,Heckman:2013sfa} in heterotic/F-theory duality.

However, for the case that the additional sections are \emph{rational}, we must tread more carefully. Here, as we will see in examples in later sections, multiple sections can exist for the projectivization of non-trivial extensions of the form \eref{defV}. The form of heterotic/F-theory duality is far from clear in this context and we will explore it further in Section \ref{Sec:4dftheory}.

We will return to the geometry of sections in more detail, as well as their consequences for physical theories/dualities in later Sections.
 
\subsection{Conic bundles}\label{sec:conic_intro}
In this section we consider the geometry of more general $\mathbb{P}^1$-fibrations in the base manifolds, ${\cal B}_n$ of F-theory. As has been done elsewhere in the literature, we will simply refer to such geometries as ``conic bundles" after the simplest prototype of such a fibration in which the fiber is a degree two hypersurface inside $\mathbb{P}^2$. For example the following threefold.
\beq
{\cal B}_3=\left[
\begin{array}{c|c}
\mathbb{P}^2 & 2 \\
\mathbb{P}^2 & 2
\end{array}\right]
\eeq
This is a $\mathbb{P}^1$ fibration over $\mathbb{P}^2$ whose fiber is described as a conic in $\mathbb{P}^2$. As mentioned in Section \ref{sec:Intro}, this fiber can degenerate over higher-codimensional loci in the base $\mathbb{P}^2$. On this locus $\Delta_b \in B_2$, the conic in $\mathbb{P}^2$ can factor into \emph{two} linear equations and hence, lead to a fiber consisting of not one, but two $\mathbb{P}^1$s.

More formally, we define a conic bundle as \cite{Sarkisov_1983,2018RuMaS..73..375P}

\vspace{10pt}

\noindent {\bf Definition:} A \emph{conic bundle} is a proper flat morphism $\pi: {\cal B}_{n} \to B_{n-1}$ of smooth varieties such that it is of relative dimension one (i.e. the fiber is $1$-(complex) dimensional) and the anti-canonical divisor $-K_{{\cal B}_n}$ is relatively ample.

\vspace{10pt}

\noindent Roughly, the term ``relatively ample" above refers to the property that $-K_{{\cal B}_n}$ restricted to the fiber is ample.

In general, a conic bundle as defined above may have a variety of algebraic descriptions. However, the Sarkisov program \cite{Sarkisov_1983} has characterized these manifolds in terms of birational minimal models whose fibers are conics in $\mathbb{P}^2$. In general, a ``standard form" for a conic bundle is characterized by 
\begin{enumerate}
\item A discriminant locus, $\Delta_b \subset B_{n-1}$ (over which the fiber degenerates).
\item A generic twist to the $\mathbb{P}^1$ fibration (playing the role of ${\cal O}(D)$ in the usual case of the projectivization of two line bundles: $\mathbb{P}({\cal O} \oplus {\cal O}(D))$). 
\item A two-sheeted cover of $B_{n-1}$ defined as the multi-section defined by the hyperplane in the $\mathbb{P}^2[2]$ fiber in the standard form. 
\end{enumerate}
We will return to a number of these characterizing features as we study examples in later sections. 

It is worth noting here that since our primary motivation in this work is the study of heterotic/F-theory duality, we will be interested in conic bundles that \emph{admit sections}. This in turn is equivalent to the statement that the F-theory elliptically fibered CY manifold, $Y_{n+1}$, possesses a $K3$-fibration with section. It is in this context that we are sure that the $8$-dimensional heterotic/F-theory duality \cite{Vafa:1996xn} extends to a lower dimensional duality relating the effective theories. However, this choice causes us to diverge from the birational ``standard models" for conic bundles described above, since in general fibers of the form $\mathbb{P}^2[2]$ admit multi-sections only, rather than true sections. We will also see in Section \ref{sec:conicmonodromy} that for some corners of moduli space, rational sections can be tuned even for such ``standard" conic bundles.

It is clear that for general $\mathbb{P}^1$-fibrations we can characterize these fibrations in terms of those that can be simply related (i.e. via small resolutions) to $\mathbb{P}^1$-bundles and those that cannot. Morally, this is a question of whether one of the curves in the degenerate fibers can be shrunk to zero size, leading to a true $\mathbb{P}^1$-bundle\footnote{Note that the study of $\mathbb{P}^1$-fibrations and their degenerations in this context is intrinsically linked to the study of $K3$-fiber degenerations in F-theory \cite{Braun:2016sks}.}. 
A key feature that determines this property is whether or not the fibration exhibits monodromy over some higher-codimensional locus in $B_2$, that is, whether or not the multiple $\mathbb{P}^1$-components of singular fibers are homologically equivalent or not. We will explore this property in detail in Section \ref{sec:conicmonodromy}. For now however, it should be noted that this monodromy of the $\mathbb{P}^1$-fibration is a phenomenon that is intrinsic to 4-dimensional compactifications of F-theory. Although conic bundles can exist for 6-dimensional compactifications of F-theory (i.e. conic bundles defined over $\mathbb{P}^1$), the degenerate fibers occur at most at points in the $\mathbb{P}^1$ base and thus, do not exhibit monodromy. In that setting the total space of the $\mathbb{P}^1$ fibration is a complex surface that is \emph{always} birational to a $\mathbb{P}^1$ bundle over $\mathbb{P}^1$ (i.e. a blow up of a Hirzebruch surface). Examples of this type have been studied in the literature \cite{Candelas:1996ht} (see also Section \ref{sec:6DFtheory} below). 

Unlike in the case of $\mathbb{P}^1$-bundles, more general $\mathbb{P}^1$-fibrations can have topology that varies more widely from that of the simple case outlined in Section \ref{sec:p1bun}. Indeed, $h^{1,1}({\cal B}_3)$ can be much larger than the minimal case of $1+h^{1,1}(B_2)$ seen in \eref{simple_pic}. We will provide examples of such geometries in subsequent sections.

\section{$\mathbb{P}^1$-fibered bases in 6-dimensional F-theory compactifications}\label{sec:6DFtheory}
In this section we undertake a simple ``warm-up" and study $6$-dimensional compactifications before moving on to 4-dimensional compactifications of F-theory in Section \ref{Sec:4dftheory}. In particular, we consider the geometry and effective physics associated to $\mathbb{P}^1$-fibered base manifolds ${\cal B}_2$ for elliptically fibered CY threefolds, $\pi: Y_3 \to {\cal B}_2$. We will review several well known possibilities in this context and show that even in this simple setting, unexpected new phenomena are possible. 

\subsection{$\mathbb{P}^1$-bundles and $\mathbb{P}^1$-fibrations over $\mathbb{P}^1$}
Here we consider complex surfaces ${\cal B}_2$ that are $\mathbb{P}^1$-fibered. In this context $\pi: {\cal B}_2 \to \mathbb{P}^1$. As argued above, every non-degenerate $\mathbb{P}^1$ fibration can be written as projectivization of a rank 2 vector bundle defined over the base (in this case $\mathbb{P}^1$). Since every vector bundle splits as a sum of line bundles over $\mathbb{P}^1$, it is clear that the most general $\mathbb{P}^1$-bundle takes the form
\beq
\mathbb{P}({\cal O}_{\mathbb{P}^1} \oplus {\cal O}_{\mathbb{P}^1}(n))
\eeq
That is, the only $\mathbb{P}^1$-bundle surfaces are Hirzebruch surfaces.

In the case that we allow the $\mathbb{P}^1$-fibration to degenerate over points in the $\mathbb{P}^1$ base, this class can be extended to a wide range of surfaces, all birational to Hirzebruch surfaces \cite{Morrison:2012np}. For example, ${\cal B}_2=dP_2$ can be viewed as a blow up of $dP_1=\mathbb{F}_1$ as presented here as a codimension 2 complete intersection:
\beq\label{dp2}
{\cal B}_2=\left[
\begin{array}{c|cc}
\mathbb{P}^2 & 1 & 1 \\
\mathbb{P}^1 & 1 & 0 \\
\mathbb{P}^1 & 0 & 1
\end{array}\right] \;.
\eeq
Alternatively, many simple toric descriptions for ``conic bundles" over $\mathbb{P}^1$ exist in this context, for example
\beq\label{blow_up_hirzebruch}
\begin{array}{ccccc}
p&x_0 & x_1 & y_0 & y_1 \\
\hline
0 &1& 1 & 0& 0  \\
0 & 0& n & 1 & 1 \\
1  & 0& q & 1 & 0
\end{array}
\eeq
is a simple toric blow-up of $\mathbb{F}_n$ \cite{Candelas:1996ht}.

The chains of possible blow ups can be vast (see \cite{Morrison:2012np} for explicit enumerations of them in the toric context). Indeed the \emph{vast majority} of $2$-dimensional bases for elliptically fibered CY threefolds fall into this latter category \cite{Gross:1993fd}.

\subsection{Dual Heterotic/F-theory geometry}\label{sec:6d_hetF}
Since it will be useful in subsequent sections we briefly review here the well-known dictionary of heterotic/F-theory duality in 6-dimensions.

The perturbative $E_8 \times E_8$ heterotic theory compactified on a $K3$ surface is fully specified by two poly-stable vector bundles, $V_i$, $i=1,2$ satisifying $c_1(V_i)=0$ and $c_2(V_1) + c_2(V_2)=24$. The massless spectrum of the theory is fixed by the topology of $V_i$, specifically the second Chern class which we can parameterize as $c_2(V_{1/2})=12 \pm n$ where $0 \leq n \leq 12$.

In very early work on F-theory, a precise dictionary was established between CY threefold backgrounds of F-theory of the form $\pi: Y_3 \to \mathbb{F}_n$ and perturbative heterotic theories with bundles of the form described above.
In the so-called ``stable degeneration" limit of F-theory, the elliptically fibered $K3$ surfaces degenerates into a fiber product of two rational elliptically fibered surfaces (i.e. $dP_9$ surfaces), glued together along a shared $\mathbb{P}^1$-base: $K3 \to dP_9 \cup_{\mathbb{P}^1} dP_9$. This limit consists of a ``cylinderizing" of the $\mathbb{P}^1$-base in which the poles of the 2-sphere support $E_8$ gauge symmetries. The data of the heterotic gauge bundles is encoded in the complex structure of the two $dP_9$-fibered ``halves" of the degenerate CY threefold via the spectral cover construction \cite{Donagi:1998vw}.

The addition of NS5-branes to the heterotic theory increases the number of tensor multiplets \cite{Morrison:1996na,Morrison:1996pp,Candelas:1996ht,Aspinwall:1998bw}. In the dual F-theory geometry this process consists of blowing up the base manifold (here $\mathbb{F}_n$ initially) to likewise increase the number of tensors. For \emph{smooth} $K3$ surfaces the number of $5$-branes, $m$, is limited by
\beq
c_2(V_1) +c_2(V_2) + m=24 \;.
\eeq

Importantly, in 6-dimensional theories, small instanton transitions, in which 5-branes are emitted/absorbed by the $S^1/\mathbb{Z}_2$ fixed planes (from the perspective of heterotic M-theory) are always possible. That is, every theory that contains additional tensor multiplets contains a limit in which the corresponding cycles in $Y_3$ are sent to zero size (i.e. blown-down).

From the point of view of $\mathbb{P}^1$ fibrations in the base of the F-theory geometry we see that the structure is simple. Either a $\mathbb{P}^1$ fibration is of the simple form already studied in the literature (i.e. $\mathbb{P}({\cal O} \oplus {\cal O}(n))=\mathbb{F}_n$) or it is simply a blow up of this case. This latter possibility encompasses the conic bundles in this context -- that is those $\mathbb{P}^1$ fibrations which degenerate over points in the $\mathbb{P}^1$-base. No monodromy is possible for such degenerations and they can always be limited back to the standard case.

The basic physical summary then is as follows:

\vspace{5pt}

\noindent {\it In 6-dimensional heterotic F-theory duality, base surfaces, ${\cal B}_2$, that are $\mathbb{P}^1$-bundles are simply Hirzebruch surfaces and lead to perturbative heterotic string compactifications over $K3$. In the case that the $\mathbb{P}^1$-fiber degenerates at $m$ points over the base $\mathbb{P}^1$, this leads to an increase in $m$ tensor multiplets in the 6-dimensional theory and a non-perturbative heterotic theory compactified over a (possible singular) $K3$ surface with $m$ NS5-branes.}

\vspace{5pt}

The results above seems complete and it might be tempting to conclude from them that there is nothing new to observe about $\mathbb{P}^1$ fibered geometries in $6$-dimensional string compactifications. However, even in this relatively simple context intriguing phenomena are possible. 

As one example, first studied in \cite{Morrison:1996pp}, a single $\mathbb{P}^1$-fibered base manifold, ${\cal B}_2$, may admit \emph{more than one} distinct $\mathbb{P}^1$-fibration. In such cases the F-theory effective physics is invariant under the choice of $\mathbb{P}^1$ fibration. However, this simple observation on the F-theory side can lead to novel structure in the dual heterotic theory. The two possible interpretations of fiber and base in $\mathbb{F}_0 = \mathbb{P}^1 \times \mathbb{P}^1$ lead to a highly non-trivial strong/weak coupling heterotic duality studied by Duff, Minasian and Witten \cite{Duff:1996rs}. This effect was generalized in \cite{Anderson:2016cdu} to include more general $\mathbb{P}^1$ fibrations (i.e. conic bundles) and the corresponding heterotic duality includes a rich array of possible interchanges between the heterotic dilaton and the additional tensor multiplets arising from 5-branes.

We also find one other phenomena in the geometry of $\mathbb{P}^1$-bundles and F-theory that has been previously unexplored in the literature. It is possible for one $\mathbb{P}^1$-fibered manifold to ``jump" between two distinct complex surfaces, with interesting consequences in the dual heterotic geometry. We examine this effect in the next section.

\subsection{Jumping phenomena (base transitions)}\label{sec:jumping}
In this section we consider a mostly unexplored effect which is visible even in the well understood arena of $6$-dimensional dual compactifications of heterotic string theory/F-theory (see \cite{Morrison:1996na,Hubsch:1992nu,Berglund:2016yqo,Berglund:2016nvh,Braun:2018ovc} for related work.)\footnote{This effect also appears in $4$-dimensional compactifications. See Section \ref{Sec:4dftheory}.}. We will consider a base, ${\cal B}_2,$ to the F-theory elliptically fibered threefold, $Y_3$, that is itself defined as the projectivization, $\mathbb{P}(V)$ of a rank $2$ vector bundle $\pi: V \to \mathbb{P}^1$. 

To begin, let us consider a vector bundle defined via a non-trivial extension of two line bundles over $\mathbb{P}^1$. For example,
\beq\label{ext_p1}
0 \to {\cal O} \to V \to {\cal O}(2) \to 0
\eeq
since $H^1(\mathbb{P}^1, {\cal O}(-2))=\mathbb{C}$, this bundle admits a 1-dimensional family of non-trivial extensions. However, since it is well known that \emph{every} vector bundle over $\mathbb{P}^1$ splits as a sum of line bundles \cite{okonek}, $V$ in \eref{ext_p1} can clearly \emph{also} be written as an Abelian sum of two line bundles. In this case, it is easy to verify that 
\beq
V={\cal O}(1) \oplus {\cal O}(1)
\eeq
 has the same first Chern class (the only topological invariant in this case) as the extension bundle defined in \eref{ext_p1}. However, a non-trivial extension class is not the only possibility to be considered in \eref{ext_p1}. If the extension class is chosen to be trivial then the extension splits as the sum ${\cal O} \oplus {\cal O}(2)$. Thus in this case, the moduli space of the simple extension bundle in \eref{ext_p1} in fact consists of a point and a line. As the extension class is smoothly varied to zero the bundle ``jumps" from one holomorphic type to another. These two sums of line bundles are distinct as holomorphic objects, but isomorphic as real bundles. This interesting effect is a well-known phenomena in the moduli space of bundles over $\mathbb{P}^1$ (see e.g. \cite{donaldson1990geometry,loop_group}).
 
Now, to extend this observation to the geometry of $\mathbb{P}^1$-fibrations central to this work, consider the projectivization of the bundle given in \eref{ext_p1}. For non-zero values of the extension class, it can be verified that $\mathbb{P}(V)=\mathbb{F}_0$ and can be simply written as a codimension 2 complete intersection manifold
\beq\label{jumpF0_F2}
\left[
\begin{array}{c|cc}
\mathbb{P}^3 & 1 & 1 \\
\mathbb{P}^1 & 1 & 1
\end{array}\right]
\eeq
without loss of generality, the defining equations can be written as \cite{Hubsch:1992nu}
\begin{align}\label{jump_eqns}
&z_0 w_0 + z_1 w_1 = 0 \\
&z_2w_0 + \left[\sum_{i=0}^2 a_i z_i + \epsilon z_3\right]w_1=0
\end{align}
 where $\{z_0,z_1,z_2,z_3 \}$ are the homogeneous coordinates on the ambient $\mathbb{P}^3$ and $\{w_0,w_1\}$ those of the ambient $\mathbb{P}^1$. As pointed out in \cite{Hubsch:1992nu}, for generic values of the defining equations with $\epsilon \neq 0$ this surface is a smooth description of $\mathbb{F}_0$. However, for the special value of $\epsilon =0$, the surface ``jumps" to become $\mathbb{F}_2$. In each case the surface described by \eref{jumpF0_F2} is in fact rigid, if we restrict to deformations that maintain the nature of the Hirzebruch surface involved. Similarly to the observation above regarding bundles, these two surfaces are the same as real manifolds, but differ in their complex structure. More generally, such jumping can occur between any even Hirzebruch surface, $\mathbb{F}_{2m}$, and $\mathbb{F}_0$ and between any odd Hirzebruch surface, $\mathbb{F}_{2m+1}$ and $\mathbb{F}_1$. That is, as real manifolds all even Hirzebruch surfaces are diffeomorphic to one another and all odd Hirzebruch surfaces likewise are identified. 
  
 What happens then, when a CY manifold is defined as an elliptic fibration over such a ``jumping" base surface? It is straightforward to show that the CY threefold remains smooth and well behaved for all values of $\epsilon$ in \eref{jump_eqns}. In this case, the topology of the CY threefold cannot vary as the complex structure is changed and indeed, we find that it does not. The only thing that changes in the structure of the elliptic threefold is that its cone of effective divisors changes as this modulus is varied. As the $\epsilon$-parameter is varied to cause $\mathbb{F}_0$ to jump to $\mathbb{F}_2$, a curve of self intersection $-2$ becomes effective in \eref{jumpF0_F2} inducing a new effective divisor in the CY threefold $Y_3$ (which could be used as a locus on which to support gauge symmetry in the F-theory compactification).
 
 In summary, using the structure of $\mathbb{P}^1$ bundles we have demonstrated that it is possible for the base manifolds of CY elliptic fibrations to non-trivially ``jump" to distinct complex surfaces, while leaving the topological family of CY threefolds unchanged. From the point of view of F-theory, the changing effective cone of $Y_3$ allows for new singular limits/gauge enhancements in the effective theory. 

\subsubsection{Weierstrass models over jumping bases}
The jumping example above that transitions an $\mathbb{F}_0$ base to an $\mathbb{F}_2$ of an elliptic CY threefold happens at a \emph{smooth} locus in the complex structure moduli space of the CY threefold. We can inquire what happens to the Weierstrass CY threefolds (and their singularity structures) when we perform a general Hirzebruch base transition between even $\mathbb{F}_{2m} \to \mathbb{F}_0$ or odd $\mathbb{F}_{2m+1} \to \mathbb{F}_{1}$ cases?

In general, we can parameterize these transitions via the following base ${\cal B}_2$ described as a hypersurface in a toric variety with the following weight matrix:
\begin{eqnarray} \label{basematrix}
\begin{array}{ccccc|c}
z_0&z_1&z_2&x_0&x_1&p\\  \hline1&1&1&0&0 &1\\ 0&0&0&1&1&n
\end{array}
\end{eqnarray}
This is a degree $(1,n)$ hypersurface (i.e. $(p(z_\alpha,x_j)=0$) in $\mathbb{P}^2 \times \mathbb{P}^1$. A simple study of the geometry and effective curves of this surface reveals that, for generic defining relations it describes $\mathbb{F}_0$ if $n$ is even and $\mathbb{F}_1$ if $n$ is odd. However, if we choose to specialize the resulting defining relations by taking $p \neq p(z_2)$ (i.e. to be independent of one $\mathbb{P}^2$ coordinate, say $z_2$) then we find that (\ref{basematrix}) describes $\mathbb{F}_n$. In other words, starting at the tuned polynomial $p_0$ and varying the defining relation we can deform any even Hirzebruch surface $\mathbb{F}_{2m}$ to $\mathbb{F}_0$ and any odd $\mathbb{F}_{2m+1}$ to $\mathbb{F}_1$. 

With this description in hand, we must now consider (\ref{basematrix}) as the base of a CY Tate model. Consider the manifold described by the following charge matrix.
\begin{eqnarray} \label{cm}
\begin{array}{cccccccc|cc} X&Y&Z& z_0&z_1&z_2&x_0&x_1&p&w\\ \hline 2&3&1&0&0&0&0&0&0&6 \\ 0&0&-2&1&1&1&0&0&1&0 \\0&0&n-2&0&0&0&1&1&n&  0 \end{array}
\end{eqnarray}
This charge matrix describes a Tate model (given by the polynomial $w=0$) over  the  base (\ref{basematrix}). Within the context of this larger charge matrix the same specializations of $p(z_\alpha,x_i)$ that were considered above now cause a transition in the base of the elliptic fibration. Importantly, a direct calculation verifies that \emph{the base-transitioning deformations in the parameterization above are indeed complex structure moduli of the full $\pi:Y_3\to {\cal B}_2$}.  Thus, within the context of (\ref{cm}) there are complex structure deformations that take a Tate model over one Hirzebruch surface to one over a different such surface.

In the case of $\mathbb{F}_0$/$\mathbb{F}_2$, that is $n=2$ in (\ref{cm}), this structure is the same as that in \eref{jumpF0_F2} and has been observed previously in the context of F-theory \cite{Morrison:1996na}. For all other cases, however, the ability to transition from one Hirzebruch base to another naively seems problematic. For example if we consider an elliptic fibration over $\mathbb{F}_n$ with $n\geq3$ then the F-theory physics exhibits a non-Higgsable cluster \cite{Morrison:1996pp}. Since neither $\mathbb{F}_0$ or $\mathbb{F}_1$ exhibit any such structure however, it is not obvious how these transitions will appear or what effective physics should describe them. Thus, some further investigation is needed, which we now carry out.

The first observation to be made is that most of these transitions are not allowed within the context of well-controlled F-theory vacua. A direct calculation yields that a Weierstrass model over the parametric base above will become badly singular (with $(f,g,\Delta)$ vanishing to orders higher than $(4,6,12)$ over codimension $1$ in the base) at the transition point between most of the Hirzebruch surfaces for $n>3$. Thus we expect that most of these transitions are not physically realizable in F-theory.

However, there is a significant exception to the statement above. In addition to the smooth $\mathbb{F}_0 \to \mathbb{F}_2$ transition, \emph{there is one other} that appears to be realized at a well controlled regime in field-space (and geometry): namely a CY threefold tuning which realizes the base transition $\mathbb{F}_1 \to \mathbb{F}_3$. Unlike the jumping example of the previous sub-section, here the transition point between the base geometries is singular, but not so badly so that it cannot be crepantly resolved (as a transition between CY threefolds this is a topology changing transition).

Note that in this case, the generic Weierstrass model over $\mathbb{F}_1$ is smooth, while that over $\mathbb{F}_3$ is singular with an enforced $SU(3)$ gauge group realized over the $-3$ curve within the $\mathbb{F}_3$ base. This symmetry is a prototypical example of a ``non-Higgsable" cluster \cite{Morrison:2012js,Morrison:2012np} in that there exists no deformations of the Weierstrass model over $\mathbb{F}_3$ which can smooth the $SU(3)$ fiber singularity and physically there is no charged matter available to realize a Higgsing transition.

However, in the context of the jumping phenomena studied here, we find that the transition point between the two Weierstrass models is not realized at a generic point in the moduli space of the $\mathbb{F}_3$ theory, but a special one\footnote{Importantly, it should be noted that this transition is non-toric, which is perhaps why it is not been previously observed, despite Weierstrass models over toric realizations of $\mathbb{F}_1/\mathbb{F}_3$ being well-studied.}. More precisely, we find that at the locus in the moduli space of the $\mathbb{F}_1$ Weierstrass model at which the effective cone jumps and a $-3$ curve becomes effective, a gauge enhancement is \emph{required} over the $-3$ curve to an $SO(8)$ symmetry. Moreover, this $SO(8)$ divisor is intersected at a single point by another component of the discriminant, leading to vanishing orders of the Weierstrass coefficients of $(f,g,\Delta)=(4,6,12)$ -- i.e. to a single superconformal (i.e. E-string) locus.

\subsubsection{An intuitive first look at the $\mathbb{F}_1 \to \mathbb{F}_3$ transition}
 Before providing the full details of a CY Tate model over the base given in \eref{basematrix} and the transition between $\mathbb{F}_1 \to \mathbb{F}_3$, it is useful to take a brief schematic look at the geometry of the transition.  
 
 We will be considering a limit in the complex structure moduli space of the Tate model of an elliptic fibration over $\mathbb{F}_1$ where the compactification becomes a tuned example of the elliptic fibration over $\mathbb{F}_3$. For readers familiar with the toric description of the $\mathbb{F}_1/\mathbb{F}_3$ base geometries in F-theory, the existence of such a transition may come as somewhat of a surprise. Thus, to make things more transparent, we will first give an overview of the structure we will see without delving into algebraic details. This will be particularly important as the full calculation, as detailed in the next subsection, involves many effective divisors that do not descend simply from the ambient space (see e.g. \cite{Anderson:2015iia}).
  
 \vspace{0.2cm}
 
 Consider the toric description of $\mathbb{F}_1$.
 \begin{eqnarray}\label{toricf1}
 \begin{array}{cccc}
z_0&z_1&x_0&x_1\\  \hline1&1&0&0 \\ 0&1&1&1
\end{array}
 \end{eqnarray}
 This geometry admits a unique curve  of self intersection $-1$ which we denote by $s_{-1}=z_0$. It also admits a unique curve of self intersection $+1$ that can not be written as a product of $s_{-1}$ with a linear in the $x$'s, which we denote by $s_{+1}=z_1$. The anti-canonical divisor of this base is given by $K^{-1}_{\mathbb{F}_1}=2 D_1+3 D_2$ where $D_1$ and $D_2$ are the divisor classes associated to $z_0$ and $x_0$ respectively. 
 
In the limit that we will describe below, at a special locus in the complex structure moduli space of the CY threefold over $\mathbb{F}_1$, the cone of effective curves ``jumps" giving rise to a new effective curve of self-intersection $-3$. At this same locus the sections $s_{\pm1}$  to the $\mathbb{P}^1$-fibration of $\mathbb{F}_1$ both become reducible and each contain a common factor of that curve of self-intersection $-3$. See Figure \ref{curve_decomp} below.
 
\begin{figure}[t]
\centering
\includegraphics[width=0.5\textwidth]{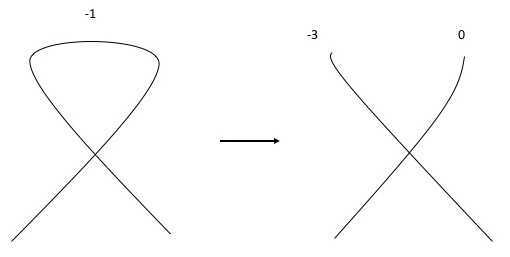}~
\includegraphics[width=0.5\textwidth]{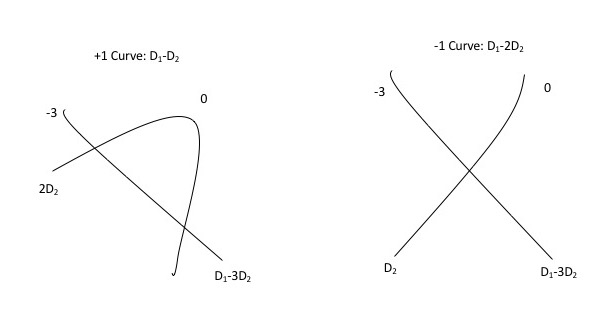}
\caption{ \emph{(Left): The decomposition of the irreducible curve of self intersection $-1$ is shown in the limit that $\mathbb{F}_1 \to \mathbb{F}_3$ in which it becomes a reducible curve. One component in this limit is the new effective curve of self intersection $-3$. (Right): The reducible limit of the sections of the $\mathbb{F}_1$ fibration (at the $\mathbb{F}_3$ locus). Both of the curves of self-intersection $\pm 1$ become reducible with a common factor of a curve self-intersection $-3$ and another factor of self-intersection $0$. Divisor classes of the curves are given in terms of the manifold shown in \eref{basematrix} with $n=3$.}}
\label{curve_decomp}
\end{figure}

 The Tate coefficients of the CY defining equation can be written in terms of sections of $H^0(\mathbb{F}_1, K^{-m}_{\mathbb{F}_1})$. For illustration, let us consider the first/simplest such coefficient: $a_1$ is an element of $H^0(\mathbb{F}_1,K^{-1}_{\mathbb{F}_1})$. This can be written, in generality, as follows.
 \begin{eqnarray} \label{a1toric}
 a_1 = s_{-1}^2 c(x) + s_{-1} s_{+1} q(x) + s_{+1}^2 l(x)
 \end{eqnarray}
 Here $l(x)$, $q(x)$ and $c(x)$ are a general linear, quadratic and cubic in the $x$'s respectively.
 
From this standard toric description, the first confusing aspect of a $\mathbb{F}_1 \to \mathbb{F}_3$ transition from this perspective is that there does not naively seem to be any tuning available in the Weierstrass model over the base (\ref{toricf1}) that would allow us to deform to a fibration over a base $\mathbb{F}_3$. This is simply because the associated deformation is not easily visible in this particular toric description. To add to the confusion however, even if one did know about the existence of such a transition, the structure of the Tate coefficients, such as (\ref{a1toric}) seem to imply that such a limit would be far to singular to provide a compactification of F-theory. As described above, in the transition the curves of self intersection $\pm 1$ in the $\mathbb{F}_1$ geometry both become reducible and each contain the unique curve of self intersection $-3$ (now effective in the $\mathbb{F}_3$ limit) as a component. Looking at the expression (\ref{a1toric}) it therefore seems as if $a_1$ would vanish to order $2$ on the $-3$ curve in the limit, with very high vanishing orders also appearing for the higher Tate coefficients. This would seem lead to a fibration far too singular to appear in an F-theory compactification.

The essential point is that the expansion of the Tate coefficients of $\mathbb{F}_1$ in terms of $\pm 1$ curves breaks down in the $\mathbb{F}_3$ limit that we will describe. That this might happen can be seen by considering a different parametrization of $a_1$. There is a curve of self intersection $+3$ on (\ref{toricf1}) which is given by $s_{+3} = z_0 \tilde{q}(x) + z_1 \tilde{l}(x)$. Here $\tilde{l}$ and $\tilde{q}$ are linear and quadratic functions in the $x$'s respectively. In terms of this, we can write the following parameterization.
\begin{eqnarray} \label{a1toric2}
a_1 = s_{-1}^2 c'(x) + s_{-1} s_{+1} q'(x) + s_{+1}^2 l'(x) + s_{+1}s_{+3} +\ldots
\end{eqnarray}
 The point of this rewriting is that there is a unique $+3$ curve that does not intersect the $-3$ curve in the $\mathbb{F}_3$ geometry. In the transitions we will describe below, the $+3$ curve in the $\mathbb{F}_1$ geometry transitions essentially unchanged to $\mathbb{F}_3$. Therefore we do not arrive at the same issue concerning vanishing orders if one parameterizes $a_1$ as in (\ref{a1toric2}). In particular, we will  find that the limit  of $a_1$ in (\ref{a1toric2}) vanishes to linear order in the $-3$ curve in $\mathbb{F}_3$ due to the $s_{+1} s_{+3}$ term.
 
 It is crucial to note that the above two parameterizations of $a_1$ are {\it equivalent} if we stay within the elliptic fibration over $\mathbb{F}_1$, and thus the (simpler) first one is most commonly used in the literature.  In particular all of the $+3$ curves can be written as combinations of $s_{+1}$, $s_{-1}$ and factors of $x$ in that geometry. The same is not true of the associated objects that these curves transition to in the $\mathbb{F}_3$ limit we will discuss however. Thus, while either parametrization (\ref{a1toric}) or (\ref{a1toric2}) can be used over $\mathbb{F}_1$, the first turns out to not give the most general limit that can be taken if used in approaching the $\mathbb{F}_3$ geometry.
 
 \vspace{0.2cm}
  
The above is of course very schematic and what we require is a concrete computation in which the transition between the Weierstrass models over $\mathbb{F}_1$ and $\mathbb{F}_3$ is explicit. It is to this that we turn in the next subsection. 

\subsubsection{The full $\mathbb{F}_1 \to \mathbb{F}_3$ transition}\label{f1f3sec}
To see this transition explicitly, let us begin by examining the general form of $w$ in (\ref{cm}) for the case $n=3$: 
\begin{eqnarray} \label{tateform}
Y^2+a_1 X Y Z +a_3 Y Z^3 - X^3 - a_2 X^2 Z^2 -a_4 X Z^4 - a_6 Z^6\;,
\end{eqnarray}
where the coefficients that appear are sections of the following bundles\footnote{Many of the cohomology calculations carried out in this paper made use of the ``CICY Package" \cite{CICYpackage}.}.
\begin{eqnarray} \label{tatetable}
\begin{array}{|c|c|c|c|} \hline  \textnormal{Tate coefficient} & \textnormal{Cohomology} & \textnormal{Dimension on } \mathbb{F}_1 & \textnormal{Dimension on } \mathbb{F}_3 \\ \hline
a_1 & H^0({\cal B}_2,{\cal O}(2,-1)) & 9& 9 \\a_2 & H^0({\cal B}_2,{\cal O}(4,-2))&25&26 \\a_3 & H^0({\cal B}_2,{\cal O}(6,-3)) & 49 & 51 \\ a_4 & H^0({\cal B}_2,{\cal O}(8,-4)) &81&84 \\ a_6 & H^0({\cal B}_2,{\cal O}(12,-6)) &169& 176\\ \hline
\end{array}
\end{eqnarray}

\vspace{0.1cm}

Using the techniques developed in \cite{Anderson:2015iia}, it can be shown that a basis of all of these cohomology groups can be constructed as sums of polynomials multiplied by combinations of three types of basic building block. These are,
\begin{eqnarray} \label{thisone}
A_i &\in& H^0({\cal B}_2,{\cal O}(1,-1)) \;\; \textnormal{where} \;\; i=1,\ldots, 3 \\ \nonumber
B_a &\in& H^0({\cal B}_2,{\cal O}(1,-2)) \;\; \textnormal{where} \;\; a=1,\ldots, n_B \\ \nonumber
C &\in& H^0({\cal B}_2,{\cal O}(1,-3) )\;.
\end{eqnarray}

Here $n_B=1$ for generic loci in complex structure moduli (i.e. the $\mathbb{F}_1$ geometry) and $2$ for the special locus (i.e. the $\mathbb{F}_3$ case) and $C$ consists of a single element which only exists for the $\mathbb{F}_3$ limit of the base. Note that for the generic loci in moduli space $B$ describes the section of the $\mathbb{P}^1$ fibration with self intersection $-1$ while the three $A$s correspond to the section with self-intersection $+1$ as well as two additional functions of the form $l(x)B$ where $l(x)$ is a linear function in $x$.   Finally, $C$ corresponds to the unique curve of self-intersection $-3$ which is only effective at the $\mathbb{F}_3$ locus in complex structure moduli space. 

We can find explicit representatives for all of these using the tools developed in \cite{Anderson:2015iia}. For example, writing the general defining relation of degree $(1,3)$ as,
\begin{eqnarray}\label{pdef}
p= l_1 x_0^3 + l_2 x_0^2 x_1 + l_3 x_0 x_1^2 + l_4 x_1^3\;,
\end{eqnarray}
where the $l_i$ are linears in the $z$'s, we can give an explicit expression for the basis $A's$ as follows\footnote{These fractional expressions are actually regular in the coordinate ring of (\ref{cm}). The reason for this is that when the denominators vanish, the numerators are also forced to vanish by the defining relation (\ref{pdef}). A complete discussion of such representations of cohomology can be found in \cite{Anderson:2015iia}.}.
\begin{eqnarray} \label{theAs}
A_1 =\frac{l_4}{x_0} \;\;,\;\; A_2 = \frac{l_1}{x_1} \;\;,\;\; A_3 = \frac{\sum_i l_i(z)}{x_0-x_1}\;.
\end{eqnarray}

These expressions hold both for the generic, $\mathbb{F}_1$, defining relation and for the $\mathbb{F}_3$ locus in moduli space. To go from the former to the latter we just set to zero the term in the $l(z)$'s containing $z_2$.

In contrast, the element $C$ only exists as an effective curve in the $\mathbb{F}_3$ limit with the associated cohomology $H^0({\cal B}_2,{\cal O}(1,-3))$ being zero dimensional for a generic $p$. In the $\mathbb{F}_3$ limit it can be written as follows. Writing the defining relation as,
\begin{eqnarray} \label{pbrewrite}
p_0(z) = C_0 z_0 + C_1 z_1\;,
\end{eqnarray}
we have,
\begin{eqnarray}\label{theC}
C= \frac{z_0}{C_1(x)} \;,
\end{eqnarray}
where $C_1(x)$ is a cubic in the $x$'s. 

The $B$'s are also easy to find explicitly at the $\mathbb{F}_3$ locus in parameter space. If we have,
\begin{eqnarray} 
C_0 = c_{01} x_0^3 +c_{02}x_0^2x_1 +c_{03} x_0 x_1^2 +c_{04} x_1^3 \;\;,\;\;C_1 = c_{11} x_0^3 +c_{12}x_0^2x_1 +c_{13} x_0 x_1^2 +c_{14} x_1^3\;,
\end{eqnarray}
then we can write the elements as follows.
\begin{eqnarray} \label{theBs}
B_1 &=&\frac{c_{01} z_0 + c_{11}z_1}{c_{11}(c_{02}x_0^2 +c_{03} x_0 x_1 +c_{04} x_1^2)-c_{01}(c_{12}x_0^2 +c_{13} x_0 x_1 +c_{14} x_1^2)} \\ \nonumber
B_2 &=&\frac{c_{04} z_0 + c_{14}z_1}{c_{14}(c_{01} x_0^2 + c_{02}x_0 x_1 +c_{03} x_1^2)-c_{04}(c_{11} x_0^2 +c_{12}x_0 x_1 +c_{13} x_1^2 )} 
\end{eqnarray}
where $c_{ij}$ are constants. The single $B$ that exists on the generic, $\mathbb{F}_1$ locus proves somewhat more elusive to write down as a function or complex structure, however. Nevertheless, it is simple, for any given choice of $p(x,z)$, to find it by brute force. One demands an element which has a numerator linear in the $z$'s and a denominator quadratic in the $x$'s, such that for any vanishing of the denominator on $p$ the numerator also vanishes. The result is unique up to an overall constant multiple if one demands that the denominator is not a perfect square. This is a necessary constraint in the gCICY method of construction \cite{Anderson:2015iia} if the vanishing orders of the numerator and denominator are to be matched correctly. Once this procedure has been carried out, one finds that the single $B$ that appears in the $\mathbb{F}_1$ case can be made to approach the $\mathbb{F}_3$ limit by the simple expedient of tuning any terms in the numerator proportional to $z_2$ to zero. When we do this we find that the result is exactly a linear combination of the elements in (\ref{theBs}) as one would expect.
Given these building blocks it is a lengthy but straightforward process to construct a complete basis for the cohomology groups that the Tate coefficients live in as claimed. This involves forming all possible combinations of the correct degree and performing a brute force linear dependence check to verify that the dimensions in (\ref{tatetable}) are correctly reproduced. 

\vspace{0.2cm}

Now that we have constructed the Tate model of interest explicitly, we can study its geometry. Let us start with the $\mathbb{F}_3$ locus where we expect an $SU(3)$ gauge theory on the locus\footnote{That this is the $-3$ curve can be seen in a straightforward manner. In particular, the self intersection of the associated divisor class, $D_1-3D_2$ in the obvious notation, can simply be directly calculated and gives the right number. In addition, this divisor class only becomes effective in the $\mathbb{F}_3$ limit as we would expect, as discussed above.} where $C=0$. This will give us another check that our construction of cohomologies is working correctly.

The general structure of the Tate coefficients in the $\mathbb{F}_3$ limit can schematically be written as follows.
 \begin{eqnarray} \nonumber
 a_1 &=& C \sum_{i=0}^1 C^i z_2^{1-i} m^{(1i)}_{(0,3i+2)} \\\nonumber
 a_2 &=& C \sum_{i=0}^3 C^i z_2^{3-i} m^{(2i)}_{(0,3i+1)}\\\label{mrtatethefirst}
 a_3 &=& C \sum_{i=0}^5 C^i z_2^{5-i} m^{(3i)}_{(0,3i)} \\ \nonumber
 a_4 &=& C^2 \sum_{i=0}^6 C^i z_2^{6-i} m^{(4i)}_{(0,3i+2)}\\\nonumber
 a_6 &=& C^2 \sum_{i=0}^{10} C^i z_2^{10-i} m^{(6i)}_{(0,3i)}\\\nonumber
 \end{eqnarray}

Some comments are in order to explain how this expression is obtained. For this geometry, all of the relevant cohomology elements associated to line bundles with negative degrees in their second entry can be written as powers of $C$ multiplied by a factor. Thus we must have enough powers of $C$ to account for the negative degrees in the line bundles of which the Tate coefficients are sections (\ref{tatetable}). We can not have more powers of $C$ than those indicated in the above expressions as this would require the coefficients to be associated to line bundles with a negative entry in the first degree and such objects have no global holomorphic sections. In any given term, the remaining required powers in the $\mathbb{P}^2$ coordinates of the ambient space of ${\cal B}_3$ are accounted for by powers of $z_2$. This is in fact general. If $C_1 \neq 0$ then we take $C= \frac{z_0}{C_1}$ and so the Tate coefficients above manifestly depend upon $z_0$ and $z_2$. Any $z_1$ dependence can be removed from the general expression simply by solving the equation $p_0=0$ for $z_1$ in terms of $z_0$. If $C_1=0$ then $C_0 \neq 0$ and we take $C=-\frac{z_1}{C_0}$ and so the Tate coefficients above manifestly depend upon $z_1$ and $z_2$. Any $z_0$ dependence can be removed from the general expression simply by solving the equation $p_0=0$ for $z_0$ in terms of $z_1$. 

Let us use (\ref{mrtatethefirst}) to examine the vanishing order of the Tate coefficients on the $-3$ curve, taking all of  the other polynomials involved in the equations to be sufficiently generic that they don't introduce any unexpected vanishing factors. We see, using the observation that the locus $C=0$ is the $-3$ curve, that the vanishing orders of $a_1,\ldots,a_6$ are $(1,1,1,2,2)$. Thus this is a type IV singularity. Which version of type IV is arising can be determined by looking at the leading non-vanishing coefficient in the $g$ of the associated Weierstrass model. Using the usual expression for $g$ in terms of Tate coefficients,
\begin{eqnarray}
g&=& -\frac{1}{864} \left( -\beta_2^3 + 36 \beta_2 \beta_4 - 216 \beta_6 \right) \;\; \textnormal{where} \\ \nonumber
\beta_2 &=& a_1^2+ 4 a_2 \\ \nonumber
\beta_4 &=& a_1 a_3+2 a_4 \\ \nonumber
\beta_6 &=& a_3^2 + 4 a_6 \;,
\end{eqnarray}
we see, using also (\ref{mrtatethefirst}), that the leading order term in $g$ near the -3 curve is simply $a_6$. This means, in particular, that the coefficient of the term giving rise to the second order vanishing is just that in the first term in the sum in the expression for $a_6$ in (\ref{mrtatethefirst}). This is a constant multiplied by $z_2^{10}$. Obviously this is a perfect square, and thus (following for example \cite{Katz:2011qp}) the gauge group on the $-3$ curve is $SU(3)$ as expected \cite{Morrison:2012np,Morrison:2014lca}.

\vspace{0.1cm}

Thus, we finally see that our description of $\mathbb{F}_3$, following from appropriately tuning $p$ in (\ref{cm}), correctly exhibits the known $SU(3)$ non-higgsable cluster. Now that we have reproduced these known results, we can proceed to examine the more interesting question: what happens if we try and deform $p$ in \eref{pdef} to pass between $\mathbb{F}_1$ and $\mathbb{F}_3$ bases.

\vspace{0.1cm}

We wish to examine the structure of the Tate model that can be obtained over the $\mathbb{F}_3$ locus of (\ref{cm}) as the limit of the general Tate model that exists in the $\mathbb{F}_1$ regime of moduli space. Let us first consider simply constructing a general Tate model in the $\mathbb{F}_1$ regime of moduli space. This analysis is somewhat different from that given above, because for a generic degree $(1,3)$ polynomial $p$, we do not have a $C$ building block and, in addition, we only have a single $B$ (which becomes a specific linear combination of the two that are available in the $\mathbb{F}_3$ limit). Therefore, we must restrict ourselves in building the Tate coefficients to only use this single $B$ and the $A$'s as building blocks. 
Given this, we can present the general $\mathbb{F}_1$ Tate coefficients (written in a manner which will make it easy to see what happens in the $\mathbb{F}_3$ limit). In these expressions we suppress the index running over $h^0({\cal B}_3,{\cal O}(1,-1))$ on the $A_i$ and the associated coefficients in order to keep the expressions from becoming too cluttered.
 \begin{eqnarray} \nonumber
 a_1 &=& \sum_{i=0}^2 A^i B^{2-i} n_{(0,3-i)}^{(1)i} + B n_{(1,1)}^{(1)} + A n_{(1,0)}^{(1)} \\ \nonumber
 a_2 &=& \sum_{j=2}^4 \sum_{i=0}^j A^i B^{j-i} n^{(2)i}_{(4-j,2j-i-2)} +B n^{(2)}_{(3,0)} \\ \label{mrtatethesecond}
 a_3 &=& \sum_{j=3}^6 \sum_{i=0}^j A^i B^{j-i} n^{(3)i}_{(6-j,2j-3-i)} + B^2 n^{(3)}_{(4,1)} + BA n^{(3)}_{(4,0)} \\\nonumber
 a_4 &=& \sum_{j=4}^8 \sum_{i=0}^j A^i B^{j-i} n^{(4)i}_{(8-j,2j-4-i)} + B \sum_{i=0}^2 A^i B^{2-i} n^{(4)i}_{(5,2-i)} + B^2 n^{(4)}_{(6,0)} \\ \nonumber
 a_6 &=& \sum_{j=6}^{12} \sum_{i=0}^j A^i B^{j-i} n^{(6)i}_{(12-j,2j-6-i)}+ B \sum_{j=3}^4 \sum_{i=0}^j A^i B^{j-i} n^{(6)i}_{(11-j,2j-4-i)} + B^3 n^{(6)}_{(9,0)}
 \end{eqnarray}
Now consider what Tate coefficients we can smoothly approach in the $\mathbb{F}_3$ limit, starting at a general $\mathbb{F}_1$ point. In this case, since there is no $C \in H^0({\cal B}_3,{\cal O}(1,-3))$ at a generic $\mathbb{F}_1$ point in moduli space all of the negative degree in the cohomologies in which the tate coefficients lie (\ref{tatetable}) must come from powers of $A$ and $B$ as above. These $A$ and $B$ become proportional  to $C$ in the limit, and so for some Tate coefficients this requirement of a smooth limiting process does not matter. However, for some it does. Take $a_3$ for example. In the generic $\mathbb{F}_3$ model we can saturate the negative 3 degree appearing in $a_3 \in H^0({\cal B}_3,{\cal O}(6,-3))$ by using a single factor of $C$ as can be seen in (\ref{mrtatethefirst}). The same is simply not true for a case which is obtained smoothly by taking a limit of an $\mathbb{F}_1$ geometry. Since we only have access to $A$'s and one $B$ we would need minimally either a term of the form $AB$ or $B^2$ to saturate the negative degree as in (\ref{mrtatethesecond}). Since both $A$ and $B$ are proportional to $C$ in the limit this leads to a minimal power of $C^2$ in the limiting $a_3$ coefficient. Similarly, for a general $\mathbb{F}_3$ model one can simply employ a minimal term  proportional to $C^2$ in order to saturate the negative degree in $a_6 \in H^0({\cal B}_3,{\cal O}(12,-6))$. For a model limiting from $\mathbb{F}_1$, however we would minimally require a term proportional to $B^3$ leading to a minimal term of $C^3$ in the limit. Thus, after these considerations, the most general $\mathbb{F}_3$ Tate coefficients which can be obtained as the smooth limit of an $\mathbb{F}_1$ geometry are as follows. 
  \begin{eqnarray} \nonumber
 a_1 &=& C \sum_{i=0}^1 C^i z_2^{1-i} n^{(1i)}_{(0,3i+2)} \\\nonumber
 a_2 &=& C \sum_{i=0}^3 C^i z_2^{3-i} n^{(2i)}_{(0,3i+1)}\\\label{mrtatethethird}
 a_3 &=& C^2 \sum_{i=0}^4 C^i z_2^{4-i} n^{(3i)}_{(0,3i+3)} \\\nonumber
 a_4 &=& C^2 \sum_{i=0}^6 C^i z_2^{6-i} n^{(4i)}_{(0,3i+2)}\\\nonumber
 a_6 &=& C^3 \sum_{i=0}^{9} C^i z_2^{9-i} n^{(6i)}_{(0,3i+3)}\\\nonumber
 \end{eqnarray}
Note that instead of viewing this as the result of continuously deforming a Tate model over $\mathbb{F}_1$ to one over $\mathbb{F}_3$, one could equally regard this procedure as performing a tuning in the $\mathbb{F}_3$ model, such that the resulting Tate model can be continuously deformed to one over $\mathbb{F}_1$. 

The Tate coefficients obtained in (\ref{mrtatethethird}) have vanishing orders of $a_1,\ldots,a_6$ along the $-3$ curve of $(1,1,2,2,3)$ and thus we have an $I_0^*$ singularity. Again we must ask what kind has manifested in this example. This time we find that the leading term in $f$ is proportional to $a_4$ and the leading term in $g$ is proportional to $a_6$. Following a similar line of reasoning to the previous case we then find that the monodromy triple cover \cite{Grassi:2011hq} factorizes as the coefficient of the leading terms in $f$ and $g$ are now $z_2^6$ and $z_2^9$ respectively.
\begin{eqnarray}
\psi^3 +  k_1 z_2^6 \psi + k_2 z_2^9 &=&0 
\end{eqnarray}
Here $k_1$ and $k_2$ are numbers and this polynomial factors  into three. Thus the gauge group has become $SO(8)$.

\vspace{0.2cm}

The next question that we must address is: what matter is present in the theory? Geometrically, this is the question of whether there are any loci on the -3 curve where the singularity structure enhances. Consider approaching the $-3$ curve along the locus $B=0$. The numerator of $B$ is simply a linear function in $z_0$ and $z_1$. If one substitutes this into the defining relation at the $\mathbb{F}_3$ locus, we actually find two branches of solutions. One is to just set $z_0=z_1=0$ which overlies the $-3$ curve (hence the above structure). The other just keeps the linear relation amongst $z_0$ and $z_1$ obtained from setting $B=0$ and gives a cubic constraint on the $x's$. Only one of the three solutions to the cubic keeps the denominator of $B$ non-vanishing and so, by the usual rules of the gCICY construction \cite{Anderson:2015iia}, this gives us our the second branch. If you approach the $-3$ curve along this second branch (which is allowed - the relation amongst the $x$'s does not make the numerator of $C$ vanish) then we come upon the $-3$ curve in a direction where $B$ is simply vanishing. The vanishing degree on the $-3$ curve, from (\ref{mrtatethesecond}), is then solely determined by the $A$ factors. Thus this should be a $(1,2,3,4,6)$ vanishing - more commonly known as an $(f,g,\Delta)=(4,6,12)$ point. Direct computation of this for an example choice of complex structure reveals exactly this $(4,6,12)$ point at the appropriate location\footnote{Note that because of the above composite structure of $B$ you can't see this single intersection point from intersection numbers alone. Indeed $B\cdot C=-2$ in the base.}.

In summary, we find that at the transition point between these base geometries, an $SO(8)$ symmetry on the $-3$ curve is realized and that the $I_1$ locus of the elliptic fiber discriminant intersects the $-3$ curve at a single point. On this intersection locus, the vanishing of $(f,g,\Delta)=(4,6,12)$ and hence we expect that this intersection does not represent ordinary, $SO(8)$-charged, bi-fundamental matter, but rather indicates a ``superconformal locus" in the theory where the physics is no longer described simply be an effective field theory.

Despite substantial progress, the effective physics of such ``superconformal loci" (and the transitions involving them) still remain somewhat mysterious. The SCFT associated to such a point can be properly studied in the decompactification limit of the CY geometry. There branches of the SCFT are well understood. From the starting point given above two branches are expected and well known. First, there is a Higgs branch. In this branch, the complex structure can be deformed to return the Weierstrass model to its generic form over $\mathbb{F}_3$, breaking the gauge group from $SO(8)$ to $SU(3)$. Next, there is also a clear tensor branch to the theory. This would consist of blowing up the base $\mathbb{F}_3$ surface at the $(4,6,12)$ point. This geometry then could be further blow-down to $\mathbb{F}_4$ with a generic $SO(8)$ symmetry on the $-4$ curve and no matter. This blowing-up and down process corresponds in the heterotic theory to a small instanton transition in which instantons are transitioned from one heterotic fixed plane to the other (see e.g. \cite{Morrison:1996pp}). However, the novelty in the present geometry is that there is an \emph{additional Higgs branch} that transitions the $SO(8)$ theory with an E-string locus to the $\mathbb{F}_1$ geometry, completely breaking the gauge group.

Much as in \cite{Anderson:2015cqy} (where SCFT loci facilitated transitions changing charged matter of the theory while leaving the gauge group unbroken), the exact field-theoretic nature of this transition is currently opaque. However, one can rigorously check the consistency of the EFTs before and after the transition as well as the geometry of the transition point itself.

As a final set of comments on this transition, it is also natural to ask what happens when gauge symmetry is tuned on the $-1$ curve of the $\mathbb{F}_1$ geometry before making the transition to $\mathbb{F}_3$? In the case described above we considered a \emph{generic} $\mathbb{F}_1$ configuration which was tuned to the $\mathbb{F}_3$ locus. It is of course also possible to first tune a non-trivial gauge symmetry into the $\mathbb{F}_1$ theory before attempting the transition. In this case then, the Weierstrass model over $\mathbb{F}_1$ is tuned to be singular and then made more so in order to perform the effective cone enhancement that leads to an effective $-3$ curve, and hence $\mathbb{F}_3$. It is clear that if a gauge symmetry $G$, of too high a rank is tuned on the $-1$ curve, the resulting singularities will be non-crepant (i.e. the vanishing order of $(f,g,\Delta)>(4,6,12)$ at the $\mathbb{F}_3$ limit. 

Direct calculation yields that the maximal symmetry that can be tuned without forcing the CY to be too singular is $SO(8)$ over the $-1$ curve. In this limit we find $SO(8) \times SO(8)$ symmetry, with one $SO(8)$ factor supported over the $-3$ curve and the other over the curve of self-intersection $0$ shown in Figure \ref{curve_decomp}. 

\subsubsection{Heterotic duals of base transitions}\label{het_jumping}
The base transitions presented in the previous sections are clearly a novel effect in F-theory. In the context of heterotic/F-theory duality, however, this ``jumping phenomena" is even more striking. For example, in jumping from an $\mathbb{F}_1$ to an $\mathbb{F}_3$ base we have jumped from an $E_8 \times E_8$ theory with gauge bundles whose second Chern classes are $c_2(V_1)=11$ and $c_2(V_2)=13$ to one in which $c_2(V_1)=9$ and $c_2(V_2)=15$. The matter spectrum of the theory depends on the topology of the $V_i$ and hence can also jump in this process! It is natural to ask -- what geometric/physical mechanism can cause such a dramatic shift in the heterotic theory? 

The first point to note is that the heterotic dual of the F-theoretic phenomenon we have been studying can not be seen in the weakly coupled limit of the theory, as given by stable degeneration. The stable degeneration limits associated to the $\mathbb{F}_1$ and $\mathbb{F}_3$ theories are distinct and incompatible, in the sense that they are dual to different limits in the weakly coupled heterotic moduli space. The $\mathbb{F}_1$ limit utilizes the $+1$ and $-1$ curves in an essential fashion, these being associated with the fixed planes in the heterotic M-theory description of the dual. These curves become degenerate in the $\mathbb{F}_3$ limit however and thus we do not obtain a simple perturbative $K3$ compactification. On the $\mathbb{F}_3$ locus in moduli space, the standard stable degeneration limit utilizes the $+3$ and $-3$ curves instead in order to obtain a conventional weakly coupled heterotic compactification. The weakly coupled duals of the $\mathbb{F}_1$ and the $\mathbb{F}_3$ F-theory models are only connected in weakly coupled heterotic theory by a tensor branch transition where, in the language of heterotic M-theory, an M5 brane passes across the interval between the two fixed planes.

However, this can not be the whole story. The heterotic and F-theory compactifications are supposed to be dual and thus a transition involving no tensor branch, of the form we have shown in the F-theory physics, must occur in the heterotic dual. Given the above discussion of stable degeneration limits we can see that this transition must be an inherently strongly coupled phenomenon. Indeed, there is some independent evidence for this. We see, in taking the limit of the $\mathbb{F}_1$ Tate model that the $\pm 1$ curves corresponding to the heterotic M-theory fixed planes are becoming reducible and sharing a component. This might indicate that the strongly coupled phenomenon involves a recombination of the fixed planes of the $S^1/\mathbb{Z}_2$ M-theory orbifold. In tuned cases one can already see some examples of non-standard heterotic physics, even in the stable degeneration limit. If we tune an $SO(8)$ on the $-1$ curve in the $\mathbb{F}_1$ model, for example, then the heterotic theory dual to the $\mathbb{F}_3$ limit has a gauge group of $SO(8) \times SO(8)$. One of these gauge factors arises from a singularity in the $K3$ of the heterotic compactification, lying on the component of the $-1$ curve arising in the limit which is not the $-3$ curve.

We believe that this novel heterotic phenomenon, whatever its detailed nature, does not just occur in this specific set up but is a more general feature of the theory. Evidence for this will be provided in the next section when we present 4-dimensional analogues of the above 6-dimensional F-theoretic phenomenon, illustrating that the physics we see here manifests in many different contexts.

  \section{Examples of $\mathbb{P}^1$-bundle bases for 4-dimensional F-theory compactifications}\label{Sec:4dftheory}
 
In the case of 3-dimensional $\mathbb{P}^1$-bundles (i.e. nowhere degenerate fibrations) which are bases for elliptic CY $4$-folds, there are three possible classes of models that can arise, each of which exhibits different structure in terms of the sections of the $\mathbb{P}^1$ bundle that are present. From the discussion in Section \ref{sec:p1fibs} and Appendix \ref{Section_appendix} these are
\begin{itemize}
\item A projectivization of a sum of two line bundles. Such a case admits two holomorphic sections which do not intersect.
\item A non-trivial extension of two line bundles. Such a case admits a holomorphic section, but not two that intersect to zero.
\item A non-trivial extension of the form (\ref{defV}), including a non-trivial ideal sheaf. Such a case admits at least one rational section, but no holomorphic sections.
\end{itemize}
In what follows we will consider each of these cases in turn and discuss their geometry and the form of Heterotic/F-theory duality that they exhibit.

\subsection{Projectivizations of sums of line bundles} \label{mrsum}

In the case that we have two holomorphic sections that do not intersect, the base ${\cal B}_3$ is the projectivization of a sum of two line bundles. This is the standard case which has been studied extensively in the literature (see e.g. \cite{Friedman:1997yq}). As such, we will not discuss this situation much here - but rather will content ourselves with a single example that demonstrates that some of the `base transition' behavior that we saw in 6-dimensions can also arises in 4-dimensions.

\subsubsection*{Example}

We can construct non-trivial 4-dimensional compactifications simply by considering elliptic fibrations over bases which are the direct product of those considered in Section \ref{sec:jumping} (there with base $\mathbb{P}^1$) with an additional $\mathbb{P}^1$. For example we can consider the following extension bundle over $\mathbb{P}^1 \times \mathbb{P}^1$.
\begin{eqnarray} \label{trivfac}
0 \to {\cal O} \to V_2 \to {\cal O}(n,0) \to0
\end{eqnarray}
The projectivization $\mathbb{P}(V_2)$ of (\ref{trivfac}) can be written as follows.
\begin{eqnarray} \label{cm2}
\begin{array}{cccccccccc|cc} X&Y&Z& z_0&z_1&z_2&x_0&x_1&y_0&y_1&p_b&p_t\\ \hline 2&3&1&0&0&0&0&0&0&0&0&6 \\ 0&0&-2&1&1&1&0&0&0&0&1&0 \\0&0&n-2&0&0&0&1&1&0&0&n&  0 \\0&0&-2&0&0&0&0&0&1&1&0&  0 \end{array}
\end{eqnarray}
For example, if we take $n=3$ this is the geometry for which detailed analysis was given, with an extra $\mathbb{P}^1$ factor added to the base.

In analyzing such a case we find that the discussion is only trivially changed from Section \ref{sec:jumping}. The conclusions for different $n$ still holds as before with the cases $n=2$ and $3$ having special features, with all higher $n$ leading to the generic base and its limit being at infinite distance in the F-theory moduli space. The relevant loci in the base where gauge symmetry is supported are the same as what they were in the six dimensional compactification times $\mathbb{P}^1$. Moreover, the locus where the orders $(f,g,\Delta)=(4,6,12)$ in the limit of the $n=3$ case appears now over a $\mathbb{P}^1$ not a point an so forth.

In short the physics is essentially unchanged in this case. Since the discussion of the heterotic dual would almost exactly mirror that of the 6-dimensional compactification we will forego it here.  Instead we will move on to study a case which has features which are more intrinsic to four (and less) dimensions.

\subsection{Projectivizations of extensions of line bundles}

As described in above, in the case that the projective bundle is a non-trivial extension of two line bundles, there will exist a holomorphic section to the $\mathbb{P}^1$-fibration. To illustrate this geometry, we turn to a particular example next.

\subsubsection*{Example}

Consider the projectivization of the bundle,
\begin{eqnarray} \label{mrext}
0\rightarrow \mathcal{O}\rightarrow V_2 \rightarrow \mathcal{O}(3,-1) \rightarrow 0 \;,
\end{eqnarray}
over $\mathbb{P}^1 \times \mathbb{P}^1$. It is straightforward, by comparing topology and cohomology, to show that this case cannot be written as a sum of two line bundles. The total space of this projectivization can be written in toric language by noting that we want a $\mathbb{P}^1$ bundle over $\mathbb{F}_0$ with the correct Chern characteristics and line bundle cohomology.
\begin{eqnarray} \label{cm3}
\begin{array}{cccccccccc|cc} X&Y&Z&y_0&y_1&x_0&x_1& z_0&z_1&z_2&p_b&p_w\\ \hline 2&3&1&0&0&0&0&0&0&0&0&6 \\ 0&0&-2&0&0&0&0&1&1&1&1&  0\\0&0&-5&0&0&1&1&2&1&3&3&0  \\0&0&-1&1&1&0&0&0&0&-1&0&  0 \end{array}
\end{eqnarray}
The split locus of the extension corresponds to taking a defining relation for the base $p_b$ where $p_b \neq p_b(z_2)$. At this locus both ${\cal O}(1,3,-1)$ and ${\cal O}(1,0,0)$ give rise to sections. Away from this locus, ${\cal O}(1,0,0)$ ceases to be effective.

\vspace{0.1cm}

Let us consider a generic point in the moduli space associated to (\ref{cm3}) and see what happens when we limit towards the split case. For generic enough maps in the Koszul sequence, direct computation shows that the cohomologies that describe the defining relation $p_b$ and all of the Tate coefficients appearing in $p_w$ are spanned by polynomial representatives. Thus, unlike in Section \ref{f1f3sec} the restriction of ambient polynomials to the coordinate ring of the manifold is sufficient to write down the Tate coefficients.

Given the Tate form (\ref{tateform}) what do the coefficients $a_i$ look like for this compactification? We, we have the following.
\begin{eqnarray}
H^0({\cal O}(2,5,1)) &\ni& a_1 = \ldots +z_0 z_2 m + z_1 z_2 m \\ 
H^0({\cal O}(4,10,2)) &\ni& a_2= \ldots +z_0z_1z_2^2m + z_0^2 z_2^2m+ z_1z_2^3m \\
H^0({\cal O}(6,15,3)) &\ni& a_3= \ldots +z_0^3 z_2^3 m+z_1^2 z_2^4 m +z_0 z_1 z_2^4 m\\
H^0({\cal O}(8,20,4)) &\ni& a_4= \ldots +z_0^4 z_2^4m+z_1^3 z_2^5m+z_1^2z_0 z_2^5m+z_1 z_0^2 z_2^5m+z_1^2 z_2^6m\\
H^0({\cal O}(12,30,6)) &\ni& a_6 = \ldots +z_0^6 z_2^6m+\ldots+z_1^4z_2^8m+z_1^3z_0 z_2^8m+z_1^2 z_0^2z_2^8m+z_1^3 z_2^9m
\end{eqnarray}
In the above, the $m$'s schematically denote a polynomial in the $x$ and $y$ variables that takes the associated term to the correct degree. Only some of the highest order terms in $z_2$ have been shown.

As in the transitions shown in Section \ref{sec:jumping} and the previous Subsection, this geometry allows for a transition realized by setting the extension class in \eref{mrext} to zero. Let us consider what happens in the limit, where all of the terms involving $z_2$ in the defining relation of the base are set to zero. In this limit the base defining relation becomes, schematically
\begin{eqnarray} \label{pbschem}
p_b = z_0 x + z_1 x^2
\end{eqnarray}
A new element of $H^0({\cal O}(1,0,0))$ that appears at this locus is then of the form
\begin{eqnarray} \label{100schem}
\frac{z_0}{x^2} \sim -\frac{z_1}{x}
\end{eqnarray}
The zero locus of the associated divisor is then simply given by $z_0=0$ which, given that the denominator in (\ref{100schem}) should not vanish, from the defining relation, also implies that $z_1=0$. What gauge symmetry appears on this divisor in the limit that it appears? To see this we can simply look at the leading terms in the Tate coefficients given above. We see that on this divisor we have a vanishing order of $(1,1,2,2,3)$ indicating that we have an $I_0^*$ singularity and we should look at the monodromy cover to get more information. The monodromy cover in this case can easily be computed to obtain the following.
\begin{eqnarray}
\psi^3 + m_{10} z_2^6 \psi + m_{15} z_2^9 
\end{eqnarray}
Here $m_{10}$ and $m_{15}$ are polynomials in the $y$'s of degree $10$ and $15$ respectively. They are not correlated, $m_{15}$ receiving a contribution from the leading term in $a_6$ whereas $m_{10}$ does not. By straightforward computation it can be shown that this can not be factorized. Therefore the gauge group is $G_2$ \cite{Grassi:2011hq}. 

What happens if we approach the gauge divisor along the divisor $z_1=0$ (associated to a global section of ${\cal O}(1,1,0)$)? Here, a quick examination of the forms of the Tate coefficients given above tell us that the result will give us a $(4,6,12)$ curve on the gauge divisor. Looking at (\ref{pbschem}) we see that if $z_1=0$ and we are not on the section of ${\cal O}(1,0,0)$ (i.e. we are approaching it along an intersecting locus) then the $x$'s are constrained by a single linear equation. Therefore this $(4,6,12)$ curve has multiplicity one. Note that approaching the gauge divisor along $z_0=0$ doesn't change the vanishing orders as compared to a generic point, and thus this is not a point of any particular interest.

In short, this example mirrors very closely the structure seen in the 6-dimensional case of Section \ref{sec:jumping}. The only major difference arises in the analysis of the monodromy cover. 

\vspace{.2cm}

It is worth commenting briefly on the possible heterotic duals of such a $\mathbb{P}^1$ fibration with only a \emph{single} holomorphic section. The standard notion of an $E_8 \times E_8$ stable degeneration limit is not generically available in this case as it relies crucially on \emph{two}, non-intersecting holomorphic sections. However, if we move to the tuned point in moduli space for which the base transitions and the extension class in \eref{mrext} splits, two holomorphic sections are available. Thus, all bases built as projectivizations of non-trivial rank 2 extension bundles have limits in their moduli space where the associated extension bundle splits as a sum of line bundles. The F-theory compactification over such a tuned base has a standard heterotic dual. If the Weierstrass model over the base at a generic point in its moduli space does not become too singular as we tune the base to this limit then the limit will be at a finite distance in moduli space. That is, if we can write down a one parameter family of valid F-theory compactifications which includes the split locus of the base as an end point, then the heterotic dual of the Weierstrass model over a generic ${\cal B}_3$ is some branch in the moduli space of a tuned version of the standard heterotic dual to the elliptic fibration over a projectivized line bundle sum. As noted in previous sections, this transition need not be realized in the weakly coupled limit of the heterotic theory (i.e. in a stable degeneration limit of the F-theory geometry). In summary, the possibility of such special loci and transitions is entirely similar to the discussion of the previous Section and comes with many questions/possibilities as noted in Section \ref{het_jumping}.

\subsection{Generic $\mathbb{P}^1$ bundles}

Generically, a projective bundle has no holomorphic sections whatsoever. It can only be written as the projectivization of a vector bundle of the form given in (\ref{defV}) (with a twist chosen to make the first line-bundle trivial and a non-trivial ideal sheaf ${\cal I}_z$). For such an example we cannot use the same trick that we employed in the proceeding subsection and look for a locus in moduli space where the extension splits. This would result not in a sum of two line bundles but rather the sum of a line bundle and a sheaf with fibers of varying rank (and hence, potentially, a singular base ${\cal B}_3$). One would not obtain two non-intersecting holomorphic sections in the limit as we did in the previous subsection.

\vspace{0.2cm}

In what follows, we will make a slight change of notation and write (\ref{defV}) as,
\beq\label{gen_2bun}
0 \to {\cal O} \to V_2\to L \otimes {\cal I}_{z} \to 0 \;.
\eeq
This will facilitate the comparison of our results with some of the existing literature in the next subsection. Recall that such a bundle corresponds to an effective section ${\cal S}$ as in \eref{sdef1} which is rational (given the presence of the non-trivial ideal sheaf).

As in previous sections, we can explore what features of F-theory (or its heterotic dual) we expect to be novel in this case. It is worth noting that since $c_1(V_2)=c_1(L)$ for $V_2$ as in (\ref{gen_2bun}), the first Chern class of ${\cal B}_3$ remains naively unchanged compared to the familiar case of a projectivization of two line bundles (i.e. $V_2={\cal O} \oplus L$) (as mentioned previously though, the section $S$ itself has changed, obeying a new intersection ring). However, the difference in $ch_2(V_2)$ above can lead to CY 4-folds constructed over this base can with a different structure in $ch_2(Y_4)$ and hence, potentially different $G$-flux compared to the familiar projectivization of two line bundles.

Since a detailed calculation of $ch_2(Y_4)$ is beyond the scope of the present work and most of the apparent physics visible from Weierstrass form will be familiar, we will not explore the F-theory physics of these bases ${\cal B}_3$ in detail. Instead, we will turn directly to a study of the heterotic dual and explore some apparently novel features of heterotic/F-theory duality.

 \subsection{Heterotic duality and anomaly cancellation in general $\mathbb{P}^1$ bundle bases}

Consider ${\cal B}_3=\mathbb{P}(\pi: V_2\to B_2)$ where $V_2$ is defined as in \eref{gen_2bun}. We can now ask how anomaly cancellation, via the $D3$-brane tadpole condition, differs in this case from that of the standard situation explored by Friedman, Morgan and Witten \cite{Friedman:1997yq} and what consequences does this have for heterotic/F-theory duality? In that paper, the authors considered bases ${\cal B}_3$ of the form ${\cal B}_3=\mathbb{P}(({\cal O} \oplus L) \to B_2)$, i.e. the case discussed in Section \ref{mrsum} where the projective bundle admits two disjoint holomorphic sections. 

The well-known D3-brane tadpole condition for a CY 4-fold
\beq
\frac{1}{24}\chi(Y_4)=\int_{Y_4} G \wedge G + \#D3s
\eeq
is directly related to the form of $\chi(Y_4)$ for a Weierstrass model over a $\mathbb{P}^1$-fibered base manifold ${\cal B}_3$. In the case of a projectivization of ${\cal O} \oplus L$, Friedman, Morgan and Witten showed that
\beq
\frac{1}{24}\chi(Y_4)=12 + 15 \int_{{\cal B}_3} c_1({\cal B}_3)^3\label{chi_start}
\eeq
With ${\cal B}_3$ the projectivization of a sum of line bundles, $c_1({\cal B}_3$) can be determined as in (\ref{mrthechern}) and the relations in (\ref{elfredo}) hold. In addition, as mentioned in \cite{Friedman:1997yq}, we have the relation $12=\int_{B_2} c_1(B_2)^2+c_2(B_2)$. Thus, the formula (\ref{chi_start}) reduces to a simple expression over the base of the K3-fibration, $B_2$, as follows \cite{Friedman:1997yq}.
\beq \label{mrthisone}
\frac{1}{24}\chi(Y_4)=\int_{B_2} c_2(B_2)+91c_1(B_2)^2+30c_1(L)^2
\eeq
In the absence of flux, this number is of course the minimal number of $D3$-branes required for the CY 4-fold geometry. Moreover, using independent arguments in a dual heterotic theory, they demonstrated that this same integer expression, (\ref{mrthisone}) over $B_2$, appears as the number of 5-branes wrapping the elliptic fiber in the dual heterotic theory. Thus, the geometry of $D3$-branes in F-theory and $5$-branes in the dual heterotic theory can be shown to match as expected.

\vspace{0.1cm}

In the following paragraphs we will repeat this calculation for the bundle defined as in \eref{gen_2bun} with the ideal sheaf present and ask how the anomaly cancellation condition changes compared to the case involving only line bundles.

\vspace{0.1cm}

We begin with \eref{chi_start} and note that $c_1({\cal B}_3)$ is unchanged compared to the standard case. However, with the ideal sheaf present in \eref{gen_2bun}, the relation (\ref{elfredo}) that is satisfied by the section picks up an extra term, which will then modify the above analysis that was used to obtain (\ref{mrthisone}). Consider
\begin{eqnarray}
c_1({\cal B}_3)^3&=&(2S-c_1(L)+c_1(B_2))^3 \\ \nonumber
&=&(4S^2-4Sc_1(L)+4Sc_1(B_2)- 2 c_1(L)c_1(B_2)+c_1(L)^2+c_1(B_2)^2)(2S-c_1(L)+c_1(B_2))
\end{eqnarray}
Now, by equation (\ref{elfredo}), we have $S^2= c_1(L)S- \left[z\right]$ and as a result the above expression can be simplified to
\beq 
c_1({\cal B}_3)^3=-8S\left[z\right]+6Sc_1(B_2)^2+2Sc_1(L)^2 \;.
\eeq
Plugging this into the expression for the Euler number we find
\beq
\frac{1}{24}\chi(Y_4)=12+15\int_{B_3} (-8S\left[z\right]+6Sc_1(B_2)^2+2Sc_1(L)^2)\label{euler_correct}
\eeq
As in \cite{Friedman:1997yq} it should be noted that $12=\int_{B_2} c_1(B_2)^2+c_2(B_2)$ for the complex base surface, $B_2$. Furthermore, since each term in the above expression for $c_1({\cal B}_3)^3$ contains a factor of the section, $S$, it is tempting to reduce this entire expression to an integral over $B_2$ as in \cite{Friedman:1997yq} and write\footnote{One might think that the integral over (\ref{needs_explanation}) would naturally occur over the rational section, not $B_2$. However, this expression is correct as can be seen in the following manner. The integral in (\ref{euler_correct}) can be thought of as computing some intersection numbers between 4-forms in the base and $S$. The 4-forms in the base are dual to points and, for a general element in the classes involved (even $[z]$) these points miss the exceptional locus over which the section wraps a $\mathbb{P}^1$. Thus in (\ref{euler_correct}) we can equally well write the integral over $B_2$ as over $S$.}
\beq
\frac{1}{24}\chi(Y_4)=\int_{B_2} c_2(B_2)+91c_1(B_2)^2+30c_1(L)^2-120\left[z \right]\label{needs_explanation} \;.
\eeq

At first pass this seems puzzling: by adding in the ideal sheaf of a set of points to the general description of the projectivization of $V_2$, we appear to have shifted the number of $D3$ branes by a factor of $120$ times the number of points! Since the naive dual heterotic calculation of anomaly cancellation carried out in \cite{Friedman:1997yq} would be unchanged, it is clear as well that this number can no longer match the number of fiber-wrapping 5-branes in a dual heterotic theory (which was shown in \cite{Friedman:1997yq} to match \eref{mrthisone}). Such a shift clearly requires an explanation.

The resolution can be found in the fact that we were slightly too quick in reducing the integral in \eref{euler_correct} to an integral over $B_2$. In \eref{mrthisone}, the reduction to an integral over $B_2$ is correct precisely due to the fact that the section, $S$ is \emph{a holomorphic} section to the $\mathbb{P}^1$ fibration. In that case, the zero-locus of $S$ is precisely diffeomorphic to the surface $B_2$ itself. 

In the case at hand however, the section to the $\mathbb{P}^1$ fibration is only a \emph{rational} section and thus, only birational to the base $B_2$. In view of this, it seems clear that the correct expression we should consider is not an integral over $B_2$, but one over the zero-locus of our rational section.

Let $\tilde{B}_2$ denote the complex surface that is defined by the rational section. Then $\tilde{B}_2$ is birational to $B_2$ and is the blow up of $B_2$ at exactly the number of points in the set $\{z\}$ defined by the ideal sheaf. For simplicity, let us begin by considering the case where $\tilde{B}_2$ is the blow up of $B_2$ at exactly one point. Then in order to write \eref{euler_correct} in terms of a surface integral, we will consider whether each term can be evaluated over $\tilde{B}_2$ (i.e. the zero-locus of the section, $S$). To begin, we are free to observe as was done above that since $\tilde{B}_2$ is a rational surface 
\beq
12=\int_{\tilde{B}_2} c_1(\tilde{B}_2)^2+c_2(\tilde{B}_2)
\eeq
 and thus \eref{euler_correct} can be written as
 \beq
 \frac{1}{24}\chi(Y_4)=\int_{\tilde{B}_2} c_1(\tilde{B}_2)^2+c_2(\tilde{B}_2)+15\int_{{\cal B}_3} (-8S\left[z\right]+6Sc_1(B_2)^2+2Sc_1(L)^2) \;.\label{euler_again}
 \eeq
 However, the terms in the second integral are clearly not written in a natural way for $\tilde{{\cal B}}_2$ and we must consider them more closely. First, let the birational morphism be denoted $g: S \to B_2$ and note that by standard results on the blow-up of a complex surface \cite{HARTSHORNE},
 \begin{align}
 &c_1(\tilde{B}_2)=g^*c_1(B_2)-e \\
 &c_2(\tilde{B}_2)=g^*c_2(B_2)+1\;,
 \end{align}
where $e$ is the exceptional divisor satisfying $e^2=-1$. Next we must consider $c_1(L)$. Fortunately, our construction has already made clear the relationship between the bundle $V_2$ on $B_2$ and on $\tilde{B}_2$. By \eref{pulledback_def} and \eref{pullback_v2} we have that the pulled-back bundle $\tilde{V_2}=g^*V_2$ on $\tilde{B}_2$ is defined in this case as
 \beq
 0 \to {\cal O} \to \tilde{V}_2 \to g^*(L) \otimes {\cal O}(e) \to 0
 \eeq
 Thus, we have that 
 \begin{align}\label{pullbackv}
& c_1(\tilde{V}_2)=g^*(c_1(L))+e \\
&c_2(\tilde{V}_2)=c_2(V_2)-[z]=0
 \end{align}
Finally, we can recall that the intersection numbers of $\tilde{B}_2$ satisfy $g^*(D)\cdot e=0$ for general divisors $D$ pulled back from $B_2$, the intersection numbers $g^*(D_1)\cdot g^*(D_2)$ remain unchanged and $e^2=-1$ as noted above. 

With these facts in hand, the key formula becomes
 \begin{align}
  \frac{1}{24}\chi(Y_4)&=\int_{\tilde{B}_2} c_1(\tilde{B}_2)^2+c_2(\tilde{B}_2) + 15(8\left[pt\right]+6(c_1(\tilde{B}_2)+e)^2+2(c_1(\tilde{V}_2)-e)^2) \\
  &=\int_{\tilde{B}_2} c_1(\tilde{B}_2)^2+c_2(\tilde{B}_2) +15(8\left[pt\right]+6c_1(\tilde{B}_2)^2-6\left[pt\right]+2c_1(\tilde{V}_2)^2-2\left[pt\right]) \\
  &=\int_{\tilde{B}_2} c_2(\tilde{B}_2)+91c_1(\tilde{B}_2)^2+30c_1(\tilde{V}_2)^2 \label{new_euler}
 \end{align}
 Thus we arrive at the usual formula for the Euler number of the CY 4-fold, but evaluated in terms of natural quantities over $\tilde{B}_2$ rather than $B_2$ and the factor of $120$ in \eref{needs_explanation} precisely accounts for this difference! We have illustrated this relationship for the blow-up of $B_2$ at a single point above, but the argument can be easily extended to the case of $n$ points blown-up.
 
 Thinking of heterotic/F-theory duality in this context it is natural to conjecture that the correct heterotic dual of this geometry must be formulated as an elliptically fibered CY threefold over $\tilde{B}_2$ rather than $B_2$ (so that the anomaly cancellation conditions will correspond as shown in \cite{Friedman:1997yq}). Indeed, the apparent `twist' in \eref{new_euler} defined by $c_1(\tilde{V}_2)$ in \eref{pullbackv} manifestly depends on the exceptional divisors in $\tilde{B}_2$ and as such will enter into the definition of the second Chern classes of the dual heterotic bundles. As further evidence of this, we will demonstrate in Section \ref{stable_degen} below that exactly this effect appears as expected in stable degeneration limits.

\subsection{Stable degeneration limits with rational sections}\label{stable_degen}

In this section we will apply standard techniques (see e.g. \cite{Donagi:2012ts}) to explore an explicit $E_8 \times E_8$ heterotic stable degeneration limit of a CY 4-fold defined as an elliptic fibration over the projectivization of a bundle of the form shown in \eref{gen_2bun}. The stable degeneration limits we consider are based on semi-stable degenerations of $K3$ surfaces (so-called Kulikov models \cite{Kulikov,Persson_Pinkham}) and the CY 4-folds inherit these degenerations through their $K3$ fibers. 

We will consider an elliptic CY 4-fold in Weierstrass form and fiber this manifold over a disc (a complex affine line) parameterized by a complex number $t$ to get a family of CY 4-folds, $\chi$ fibered over the disc, $ \Pi: \chi \to \Delta$. We will demand that $\chi$ be a semi-stable degeneration\footnote{See \cite{Aspinwall_1998} for discussions of the Clemens-Schmid sequence in this context.} -- i.e. that the fiber  $\Pi^{-1}(t)=Y_4(t)$ be smooth over generic $t$, but the central fiber can be reducible\footnote{In the case of the $E_8 \times E_8$ semi-stable degeneration limit of a $K3$ surface this degeneration leads to two rational elliptic surfaces at the central fiber.} and admit normal crossing singularities (but must be without infinitesimal automorphisms). We also demand that the anti-canonical bundle of the family is trivial. These conditions can be fulfilled after birational transformations.

To begin we will take $B_2=\mathbb{P}^2$ and consider a particular extension bundle of the form shown in \eref{gen_2bun}, which exhibits the ideal sheaf of a single point. That is the following bundle over $\mathbb{P}^2$
\begin{eqnarray}\label{sheaf_pt_eg}
0\rightarrow \mathcal{O} \rightarrow V_2 \rightarrow \mathcal{O}(H) \otimes \mathcal{I}_{p} \rightarrow 0.
\end{eqnarray}
As in previous Sections, it is straightforward to find a simple toric hypersurface description of this bundle projectivization\footnote{To see why this description holds, note that the bundle in \eref{sheaf_pt_eg} can be also written as $0\rightarrow \mathcal{O}(-1)\rightarrow \mathcal{O}^{\oplus 3} \rightarrow V_2 \rightarrow 0$.} of the full geometry $\mathbb{P}(V_2)$. Here this takes the form

\begin{eqnarray}
\mathbb{P}(\pi: V_2 \to \mathbb{P}^2)=\left[\begin{tabular}{c|c}
$\mathbb{P}^2$ & 1 \\
$\mathbb{P}^2$ & 1
\end{tabular}\right], \nonumber \\
F= u_1 x_0 + u_2 x_1 +\epsilon u_3 x_2 = 0,\label{defFrat}
\end{eqnarray}
where $u_i$ are the coordinates of the $\mathbb{P}^2$ base. Similarly to the discussion in Section \ref{sec:6DFtheory} we can identify the term $\epsilon u_3$ with a possible extension class defining the non-split sequence \eref{sheaf_pt_eg}. 

Over this $\mathbb{P}^1$-bundle base ${\cal B}_3=\mathbb{P}(\pi: V_2 \to \mathbb{P}^2)$ we can consider the form of a CY elliptic Weierstrass model
\begin{eqnarray}
W_F=y^2+x^3+f x z^4+g z^6=0,
\end{eqnarray}
where the coefficients take the form
\begin{eqnarray}
f = x_2^8 f_{0,8} + x_2^7 \sum_{j}^{1} x_0^j x_1^{1-j} f_{j,7} &+& \dots \nonumber \\
&+& x_2^4 \sum_j^4 x_0^j x_1^{4-j} f_{j,4}  \nonumber\\
&+& \dots\quad + x_2 \sum_{j}^7 x_0^j x_1^{7-j} f_{j,1}+ \sum_{j}^8 x_0^j x_1^{8-j} f_{j,0}, \label{frat} \\
g= x_2^{12} g_{0,12} + x_2^{11} \sum_{j}^{1} x_0^j x_1^{1-j} g_{j,11} &+& \dots\nonumber \\
&+& x_2^6 \sum_j^6 x_0^j x_1^{6-j} g_{j,6}  \nonumber\\
&+& \dots\quad +x_2 \sum_{j}^{11} x_0^j x_1^{11-j} g_{j,1}+ \sum_{j}^{12} x_0^j x_1^{12-j} g_{j,0}, \label{grat}
\end{eqnarray}
where $f_{i,j}$ and $g_{i,j}$ are degree 8 and 12 polynomials in terms of $(u_0,u_1,u_2)$. Therefore the F-theory geometry is an elliptically fibered CY fourfold which is also a $K_3$ fibration,
\begin{eqnarray}
\pi_{E} : Y_4 &\longrightarrow& {\cal B}_3 := \mathbb{P}(V_2) ,\label{Efibration} \\
\pi_{K_3} : Y_4 &\longrightarrow& \mathbb{P}_2.\label{K3fibration}
\end{eqnarray}

To find the Heterotic dual of this F-theory geometry, one first needs to choose a section for the $\mathbb{P}^1$-fibration. We choose $x_2$ as the (rational) section\footnote{All of the sections of this $\mathbb{P}^1$-fibration are birational to the base $\mathbb{P}^2$ as can be checked directly. This is related to the ideal sheaf $\mathcal{I}_{p}$ in defining $V_2$.}, and try to do the stable degeneration by constructing a family of Weierstrass CY four-folds parameterized by a complex variable $t$,
\begin{eqnarray}
f_{i,j} &\rightarrow& t^{4-j} f_{i,j},\quad \text{for $j\le 4$}, \\
g_{i,j} &\rightarrow& t^{6-j} g_{i,j}, \quad \text{for $j\le 6$}.
\end{eqnarray}
So there is a family 
\begin{eqnarray}
\Pi:\chi \longrightarrow \Delta,
\end{eqnarray}
where $\Delta$ is a disc parameterized by $t$. For generic $t\ne 0$ the fiber of $\Pi$, i.e. $\Pi^{-1}(t)$ is a generic smooth CY four-fold, but the central fiber on $t=0$ has a non-minimal $(f,g,\Delta)=(4,6,12)$ singularity on $x_2=0$. Then we make the following birational transformation to make the degeneration (semi)stable,
\begin{eqnarray}
t &\rightarrow& e t, \\
x_2 &\rightarrow& e x_2, \\
x &\rightarrow& e^2 x, \\
y &\rightarrow& e^3 y.
\end{eqnarray}
When these transformations are plugged into $W_F$, we find an overall factor of $e^6$ that must be removed. The resulting geometry thus has two branches (components) in the central fiber ($t=0$ and $e=0$). 

We begin by studying the geometry over $e=0$ first. The geometry on this locus can be expressed in the following toric form
\begin{eqnarray}
&\begin{tabular}{ccccccccc|cc} 
$x$ & $y$ & $x_0$ & $x_1$ & $x_2$ & $u_0$ & $u_1$ & $ u_2$ & $t$ & $F_r$ & $W_r$ \\ \hline
2 & 3 &0 & 0 & 1 & 0 & 0 & 0 & 1 & 0 & 6 \\
4 & 6 &1 & 1 & 1 & 0 & 0 & 0 & 0 & 1 & 12\\
4 & 6 &0 & 0 & 0 & 1 & 1 & 1 & 0 & 1 & 12 
\end{tabular},&\\
&F_r= u_1 x_0 + u_2 x_1 = 0,& \\
&W_r=y^2+x^3+f_r x +g_r =0.&
\end{eqnarray}
Note that the coordinate $z$ has disappeared, and the new Stanley-Reisner ideal of the ambient space is
\begin{eqnarray}
\langle x,y,x_2,t \rangle, \quad \langle x_0, x_1\rangle, \quad \langle u_0,u_1,u_2 \rangle.
\end{eqnarray}
This full 4-fold geometry is a fibration over $\tilde{B}_2$ (which will be explained bellow) with fiber given by
\begin{eqnarray}
\mathbb{P}^{2311}_{x,y,x_2,t}[6] \simeq dP_8. 
\end{eqnarray}
This $dP_8$ fiber is closely related to the usual $dP_9$ surfaces in the literature by blowing up the point $x_2=t=0$. The important point is that the complex structure moduli of this $dP_8$ is identified with the moduli of a flat $E_8$ bundle on the elliptic curve located in $t=0$ locus \cite{Friedman:1997yq,Friedman:1997ih,Donagi:2008ca}. More clearly note that the polynomials $f_r$ and $g_r$ in $W_r$ are given by
\begin{eqnarray}
f_r &=& x_2^4 \sum_j^4 x_0^j x_1^{4-j} f_{j,4}+ \dots + x_2 t^3 \sum_{j}^7 x_0^j x_1^{7-j} f_{j,1}+t^4 \sum_{j}^8 x_0^j x_1^{8-j} f_{j,0},\\
g_r&=& x_2^6 \sum_j^6 x_0^j x_1^{6-j} g_{j,6} + \dots + x_2 t^5 \sum_{j}^{11} x_0^j x_1^{11-j} g_{j,1}+ t^6 \sum_{j}^{12} x_0^j x_1^{12-j} g_{j,0}.
\end{eqnarray}

Now turning to the other component over $t=0$ we clearly have a Weierstrass fibration over $\tilde{B}_2$ 
\begin{eqnarray}
W_H=x^2+y^3+x x_2^4 \sum_j^4 x_0^j x_1^{4-j} f_{j,4} + x_2^6 \sum_j^6 x_0^j x_1^{6-j} g_{j,6}=0,
\end{eqnarray}
where $\langle x,y,x_2\rangle$ are now homogeneous coordinates of a weighted projective space $\mathbb{P}^{231}$. This is the dual Heterotic geometry, and the polynomial $W_r$ determines a flat $E_8$ bundles over the elliptic fibers of $W_H$. 

Finally, note that the base of the $dP_8$ fibration ($W_r=0$) and the Heterotic elliptic fibration $W_H=0$ is 
\begin{eqnarray}
&\tilde{B}_2=\begin{tabular}{ccccc|c}
    $x_1$ & $x_2$ & $u_0$ & $u_1$ & $u_2$ & $F_r$  \\ \hline
    1 & 1 & 0 & 0 & 0 & 1 \\
    0 & 0 & 1 & 1 & 1 & 1 
\end{tabular},& \label{HBase} \\ 
&F_r=u_1 x_0 + u_2 x_1 = 0. & \nonumber
\end{eqnarray}
Therefore $\tilde{B}_2=dP_1$ is a blow up of the base of $\pi_{K_3}$ in \eref{K3fibration}. It is also easy to check that the Heterotic geometry is indeed a CY threefold. 

This analysis leads to our first conclusion: the base of the dual Heterotic elliptically fibered CY is isomorphic to the section of the $\mathbb{P}^1$-fibration ${\cal B}_3$ \eref{Efibration}, and is birational to the base of the F-theory $K_3$-fibration \eref{K3fibration}. This agrees with the anomaly cancellation analysis of the previous Subsection and indicates that in the presence of rational sections, it is that zero-locus and not the base of the $K3$ fibration itself that determines the geometry of a heterotic/F-theory dual pair.

\vspace{.2cm}

Before concluding this discussion however, we must return to the $t=0$ locus. The Weierstrass four-fold is still non-minimally singular on this branch, and we need one further set of birational transformations to obtain a smooth (enough) model. In the standard case, where the $\mathbb{P}^1$-fibration has two sections, one blows up the other section and the coordinate $e$. However, in the cases we are studying here, there are only one section. So in this case we will attempt to parameterize the $t=0$ branch with another complex parameter $t'$. In other words, we consider another family $\chi'$ over a disc $\Delta'$ ($\Pi' : \chi'\longrightarrow \Delta'$) as 
\begin{eqnarray}
f &=& x_2^4 \left(x_2^4 e^4 f_{0,8} + x_2^3 e^3 t' \sum_{j}^{1} x_0^j x_1^{1-j} f_{j,7} + \dots + t'^4 \sum_j^4 x_0^j x_1^{4-j} f_{j,4} \right), \\
g&=& x_2^{6}\left(x_2^{6} e^6 g_{0,12} + x_2^{5}e^5 t' \sum_{j}^{1} x_0^j x_1^{1-j} g_{j,11} + \dots +t'^6 \sum_j^6 x_0^j x_1^{6-j} g_{j,6}\right).
\end{eqnarray}
Consider the following birational transformation,
\begin{eqnarray}
t' &\longrightarrow& e' t', \\
e &\longrightarrow& e' e, \\
y &\longrightarrow& e'^3 y, \\
x &\longrightarrow& e'^2 x.
\end{eqnarray}
After removing the overall $e'^6$ factor one gets a new family, and over $e'=0$ locus there will be another $dP_8$ fibration $\mathbb{P}^{2311}_{x,y,e,t'}$ and the interpretation is another flat $E_8$ bundle over the elliptic curve in $e=0$ locus (note that now $x_2\rightarrow 1$ on this branch).

We conclude this section by returning to an observation made previously that when $\epsilon =0$ in \eref{defFrat} $\mathbb{P}(V_2)$ becomes \emph{singular and non-flat}. This can be seen from the point of view of the extension bundle in \eref{sheaf_pt_eg} where
\begin{eqnarray}
\epsilon\rightarrow 0,\quad \Rightarrow \quad V_2 \rightarrow \mathcal{O} \oplus \mathcal{O}(H)\otimes \mathcal{I}_p.
\end{eqnarray}
Note that the second term above (i.e. the ideal sheaf contribution) corresponds to the locus in \eref{defFrat} where $u_1=u_2=0$. Therefore when $\epsilon=0$ and $u_1=u_2=0$ the rank of $V_2$ jumps to three and $B_3$ will be singular in this case.

Despite this singular nature of the split extension, in the limit $\epsilon\rightarrow 0$, there does exist another section to the fibration at $x_0=x_1=0$. It is possible to make this more explicit by blowing up ${\cal B}_3$
\begin{eqnarray}
x_0 &\rightarrow& w x_0, \\
x_1 &\rightarrow& w x_1.
\end{eqnarray}
After this birational transform we get a \emph{smooth, flat} $\mathbb{P}^1$-bundle over $dP_1$,
\begin{eqnarray}
&\begin{tabular}{ccccccc|c}
    $w$ & $x_0$ & $x_2$ & $x_2$ & $u_0$ & $u_1$ & $u_2$ & F  \\ \hline
    0 & 1 & 1 & 1 & 0 & 0 & 0 & 1 \\
    0 & 0 & 0 & 0 & 1 & 1 & 1 & 1 \\
    1 & 0 & 0 & 1 & 0 & 0 & 0 & 0
\end{tabular},& \\
&F= u_0 x_0+u_1 x_1.& \nonumber
\end{eqnarray}
There are two holomorphic sections $w=0$ and $x_2=0$ which don't intersect with each other. One can easily show that this birational transformation keeps the canonical bundle trivial, hence there is a standard heterotic dual over the same $dP_1$  which is defined in \eref{HBase}. The $f$ and $g$ polynomials of the Weierstrass fibration after the blow up in the base are
\begin{eqnarray}
f &=& x_2^8 f_{0,8} + x_2^7 w\sum_{j}^{1} x_0^j x_1^{1-j} f_{j,7} + \dots  + x_2 w^7\sum_{j}^7 x_0^j x_1^{7-j} f_{j,1}+ w^8 \sum_{j}^8 x_0^j x_1^{8-j} f_{j,0}, \\
g &=& x_2^{12} g_{0,12} + x_2^{11} w\sum_{j}^{1} x_0^j x_1^{1-j} g_{j,11} + \dots + x_2 w^{11} \sum_{j}^{11} x_0^j x_1^{11-j} g_{j,1}+ w^{12}\sum_{j}^{12} x_0^j x_1^{12-j} g_{j,0}.\nonumber \\
\quad                              
\end{eqnarray}
So another novelty is that in this particular example, one is able to do a conifold transition to reach to a geometry which can have a heterotic dual of the standard form.  This transition increases the $h^{1,1} ({\cal B}_3)$ by one. Also, in terms of the CY Weierstrass fibration, this birational transformation increases $h^{1,1}(X_4)$ by one, and reduces the $h^{3,1}(X_4)$ by one. The important point is that this heterotic dual is the same as the one derived earlier in this section.

\vspace{0.5cm}

The results of this section lead to a number of interesting open questions that could be explored in future work. One question in particular is how the detailed matching of massless degrees of freedom is realized across novel heterotic/F-theory dual pairs of the form considered here (with rational sections to the $\mathbb{P}^1$-fibration of ${\cal B}_3$)? Naively, it seems that the difference in $h^{1,1}$ between the base of the apparent dual heterotic elliptic fibration (here $dP_1$) and the base of the F-theory $K3$ fibration (here $\mathbb{P}^2$) might lead to a mismatch. However, this is far from clear due to the subtle structure of the stable degeneration limit constructed above. It might be possible that the F-theory degrees of freedom ``jump" to match those of the heterotic theory (i.e. the extra $h^{1,1}$ element) in the stable degeneration limit\footnote{It should be noted that a similar ``jump" in dual heterotic/F-theory degrees of freedom in the stable degeneration limit has been seen before in $6$-dimensional dual compactifications in the context of higher rank Mordell-Weil groups in the heterotic elliptic geometry \cite{Anderson:2015cqy,Cvetic:2015uwu}.}. To answer this question fully in the present context, an analysis of the degrees of freedom \emph{in the stable degeneration limit} must be completed (see e.g. \cite{Aspinwall_1998}) and we hope to return to this in future work. 

\vspace{.5cm}

For now, we conclude our study of $\mathbb{P}^1$-bundles in F-theory base geometries and turn to an exploration of the properties of more general $\mathbb{P}^1$-fibrations (also known as conic bundles).
 
\section{An example of a conic bundle threefold}
\label{sec:conicnomono}
We now turn to 4-dimensional F-theory models where the base ${\cal B}_3$ of the elliptic fibration is a conic bundle. Such a base is fibered, with the generic fiber being a $\mathbb{P}^1$. However, the fiber may degenerate over codimension one or higher loci in the base $B_2$ of the fibration ${\cal B}_3$. Section \ref{sec:6DFtheory} discussed the physics associated with conic bundles as bases for 6-dimensional F-theory models. While the 6-dimensional analysis should give significant insights into the 4-dimensional physics, there are new features in 4-dimensional situations that need to be considered. Chief among these is the possibility of a monodromy that exchanges the components in degenerate fibers, which we will discuss in Section \ref{sec:conicmonodromy}. However, there may be other differences between the physics in 6- and 4-dimensions, so it is worth analyzing 4-dimensional conic bundle models without the added complication of monodromy. Therefore, this section focuses on 4-dimensional F-theory models where the base of the elliptic fibration is a 3-dimensional conic bundle that lacks monodromy. The discussion of this section will largely be focussed upon examples, with a more general analysis being pursued in Section \ref{sec:conicmonodromy}.

\vspace{0.1cm}

We consider a specific example of a conic bundle $\conicbundle$ described as a hypersurface $P=0$ in a toric variety. The toric weight matrix and hypersurface constraint for this example is as follows.
\begin{equation}
    \begin{array}{cccccc|c}
    x_1 & x_2 & x_3 & y_1 & y_2 & y_3 & P\\\hline
     1  &  1  &  1  &  0  &  0  &  0  & 2\\
     0  &  1  &  0  &  1  &  1  &  1  & 1\\
    \end{array}
\end{equation}
The conic bundle has a $\mathbb{P}_{y}^2$ base given by the coordinates $[y_1:y_2:y_3]$. There is another $\mathbb{P}_{x}^2$ with coordinates $[x_1:x_2:x_3]$ fibered over this base, and the conic fiber is described as a quadratic curve in $\mathbb{P}_{x}^2$. The defining relation $P=0$ can be written explicitly as
\begin{equation}
    P = C_{11}x_{1}^2 + 2 C_{12}x_{1}x_{2}+2C_{13}x_{1}x_{3}+ 2C_{23}x_{2}x_{3} + C_{33} x_{3}^2,
\end{equation}
where the $C_{ij}$ are sections of line bundles on $\mathbb{P}^2_{y}$. Specifically, $C_{11}$, $C_{13}$, and $C_{33}$ are linear in the $y_{i}$, while $C_{23}$ and $C_{12}$ are constants that are generically nonzero. By redefinitions of $x_2$, we can therefore simplify the defining relation to 
\begin{equation}
    P =  2 C_{12}x_{1}x_{2}+2C_{13}x_{1}x_{3}+ 2C_{23}x_{2}x_{3}. \label{eq:exampleconicdefrel}
\end{equation}
For the remainder of this section, we work with the simplified form.

It will be helpful to describe some topological properties for this conic bundle. The important Hodge numbers for $\mathcal{B}_3$, which can be calculated using \emph{cohomCalg}  \cite{Blumenhagen:2010pv,cohomCalg:Implementation} for example, are
\begin{align}
    h^{1,1}(\mathcal{B}_3) =& 3 & h^{2,1}(\mathcal{B}_3) =& 0.
\end{align}
Two of the contributions to $H^{1,1}$ can be thought of as descending from the harmonic $(1,1)$ forms for the ambient space restricted to ${\cal B}_3$. We can call these forms $J_x$ and $J_y$, with
\begin{align}
    [x_1],[x_3]&\sim J_x & [y_1],[y_2],[y_3]&\sim J_y & [x_2]&\sim J_x+J_y.
\end{align}
However, because $h^{1,1}(\mathcal{B}_3)=3$, there is additional contribution beyond that from $J_{x}$ and $J_{y}$ (i.e. ${\cal B}_3$ is not K\"ahler favorable). This fact is closely tied to the presence of degenerate fibers consisting of two rational curves, which we discuss shortly. For an F-theory model that uses $\mathcal{B}_{3}$ as the base of the elliptic fibration, the resulting massless spectrum includes $h^{1,1}(\mathcal{B}_{3})$ K\"{a}hler moduli corresponding to the volumes of divisors in $\mathcal{B}_3$ \cite{Grimm:2012yq}. Meanwhile, if $h^{2,1}(\mathcal{B}_{3})$ is non-zero, the F-theory model built using $\mathcal{B}_{3}$ has $h^{2,1}(\mathcal{B}_{3})$ additional U(1) gauge bosons, which arise from the Type IIB $C_{4}$ form \cite{Grimm:2010ks, Grimm:2012yq}.

The base of the conic bundle is simply $\mathbb{P}^2_{y}$, and the Chern classes for ${\cal B}_{2}$ are therefore given by
\begin{align}
    c_{1}(B_2) =& 3 J_{y} & c_{2}(B_2)=& 3 J_{y}^2.
\end{align}
The Chern classes for $\mathcal{B}_3$, meanwhile, can be calculated from the formula
\begin{equation}
    c(\mathcal{B}_{3}) = \frac{\left(1+J_{x}\right)^2\left(1+J_{x}+J_{y}\right)\left(1+J_{y}\right)^3}{1+2J_{x}+J_{y}}.
\end{equation}
Expanding this expression gives us
\begin{align}
    c_{1}(\mathcal{B}_3) =& J_x + 3 J_y & c_{2}(\mathcal{B}_3) =& J_x^2 + 4 J_x J_y +3 J_y^2
\end{align}

We will utilize these formulae below in analyzing the physics associated to this geometry. Next we turn to the presence of sections to the $\mathbb{P}^1$ fibration.

\subsection{Sections} \label{bob}
We are interested in exploring the heterotic duals of F-theory models constructed using the conic bundle $\conicbundle$. In the standard heterotic/F-theory duality, the base of the F-theory elliptic fibration is a $\mathbb{P}^1$ bundle with two sections. Gauge groups supported on these two sections in the F-theory geometry are dual to the gauge groups coming from the two $E_8$ factors in the heterotic model. Sections play a similarly important role when the F-theory elliptic fibration is constructed over a conic bundle. We should therefore determine the possible sections of the conic bundle described above.

Any section must satisfy certain topological criteria. Let $\divsec$ be the divisor class of a section, $\hat{D}^b_{\alpha}$ with $\alpha=1,\ldots h^{1,1}(\conicbase)$ be a basis of divisor classes for the base $B_2$, and ${D}^b_{\alpha}$ be the pullback of $\hat{D}^b_{\alpha}$ to the full conic bundle $\conicbundle$. $\divsec$ must satisfy the Oguiso condition
\begin{equation}
    \divsec \cdot \prod_{k=1}^{2}D^b_{\alpha_k} = \prod_{k=1}^{2}\hat{D}^b_{\alpha_k}
\end{equation}
for all $2$ tuples $(\alpha_1,\alpha_{2})$, where each $\alpha_k$ is an integer from 1 to $h^{1,1}({\cal B}_{2})$. For the case at hand, $h^{1,1}(B_2)$ is 1, and the corresponding divisor $\hat{D}^b_{1}$ is the hyperplane class on $\mathbb{P}^2_{y}$. Therefore, the right-hand side of this equation is $1$ for our example. Additionally, $\divsec$ must satisfy a modified version of a condition given in \cite{Morrison:2012ei} that accounts for the non-CY nature of $\conicbundle$:
\begin{equation}
    \divsec\cdot\left(\divsec - [c_1({\cal B}_{3})]\right)\cdot D^b_{\alpha_k} = - [c_1({\cal B}_{2})]\cdot\divsec\cdot D^b_{\alpha_k}. 
\end{equation}

In the context of heterotic/F-theory duality, we want at least two sections to the conic bundle whose divisor classes, $\divsec_1$ and $\divsec_2$, satisfy these equations. The divisor classes $\divsec_1$ and $\divsec_2$ may not be the same. Based on the analysis of \cite{Friedman:1997yq}, we want these classes to satisfy
\begin{equation}
    \divsec_1\cdot\divsec_2 = 0
\end{equation}
The condition can be thought of as asking that the divisors corresponding to the $E_8$ factors do not intersect; this ensures that, as in usual, non-singular heterotic models, there is no matter jointly charged under gauge factors contained in distinct $E_8$ factors.

There are three sections that can be easily read off from \eref{eq:exampleconicdefrel}:
\begin{align}
    \hat{s}_{13}:&\{x_1=x_3=0\} & \hat{s}_{12}:&\{x_1=x_2=0\} & \hat{s}_{23}:&\{x_2=x_3=0\}.
\end{align}
The sections $\hat{s}_{12}$ and $\hat{s}_{23}$ have the same divisor class, which we call $D_{12}$. We refer to the divisor class of the section $\hat{s}_{13}$ as $D_{13}$. These classes satisfy the conditions above. The Oguiso condition essentially states that a section should hit each fiber once, which is clearly true for the sections above. We have that
\begin{equation}
    D_{12} \cdot \prod_{k=1}^{2}D^b_{\alpha_k} = D_{13} \cdot \prod_{k=1}^{2}D^b_{\alpha_k} = 1.
\end{equation}
The sections cannot intersect either, as $[x_{1}:x_{2}:x_{3}]$, the coordinates of $\mathbb{P}^2_{x}$, cannot simultaneously vanish. This implies that $D_{12}\cdot D_{13}=0$. For the remaining condition, let us denote the divisors associated to $J_x$ and $J_y$ as $[J_x]$ and $[J_y]$. Then, $D^b_{\alpha_{k}}$ is $[J_y]$ for all values of $k$, and the condition becomes
\begin{equation}
    \divsec\cdot\left(\divsec-[J_x]-3[J_y]\right)\cdot [J_y] = -3[J_y]\cdot\divsec\cdot[J_y].
\end{equation}
This equation can be simplified to
\begin{equation}
    \divsec\cdot\left(\divsec-[J_x]\right)\cdot [J_y]= 0.
\end{equation}
To evaluate the left-hand side of this equation, it is helpful to rewrite $[J_x]$ in terms of $D_{12}$ and $D_{13}$. The divisor $\{x_1=0\}$ has divisor class $[J_x]$, and from the structure of the defining relation \eref{eq:exampleconicdefrel}, we see that
\begin{equation}
    \{x_1=0\} = \{x_1=x_3=0\}\cup\{x_1=x_2=0\}.
\end{equation}
Therefore,
\begin{equation}
    [J_x] = D_{13}+D_{12}.
\end{equation}
With this result, we see that for either $\divsec=D_{13}$ or $\divsec=D_{12}$, the condition reduces to
\begin{equation}
    -D_{13}\cdot D_{12}\cdot [J_y]=0,
\end{equation}
which is satisfied due to the previously discussed result that $D_{13}\cdot D_{12} =0$. 

To find F-theory constructions with dual heterotic models, we want to choose a section of class $D_{13}$ and a section of class $D_{12}$. For $D_{13}$, the natural choice of section is $\hat{s}_{13}$. There are many possible choices for $D_{12}$, but here we simply choose $\hat{s}_{12}$ for convenience.

\begin{figure}[t!]
\begin{center}
\begin{picture}(0,120)
\put(-150,0){\includegraphics[scale=0.7]{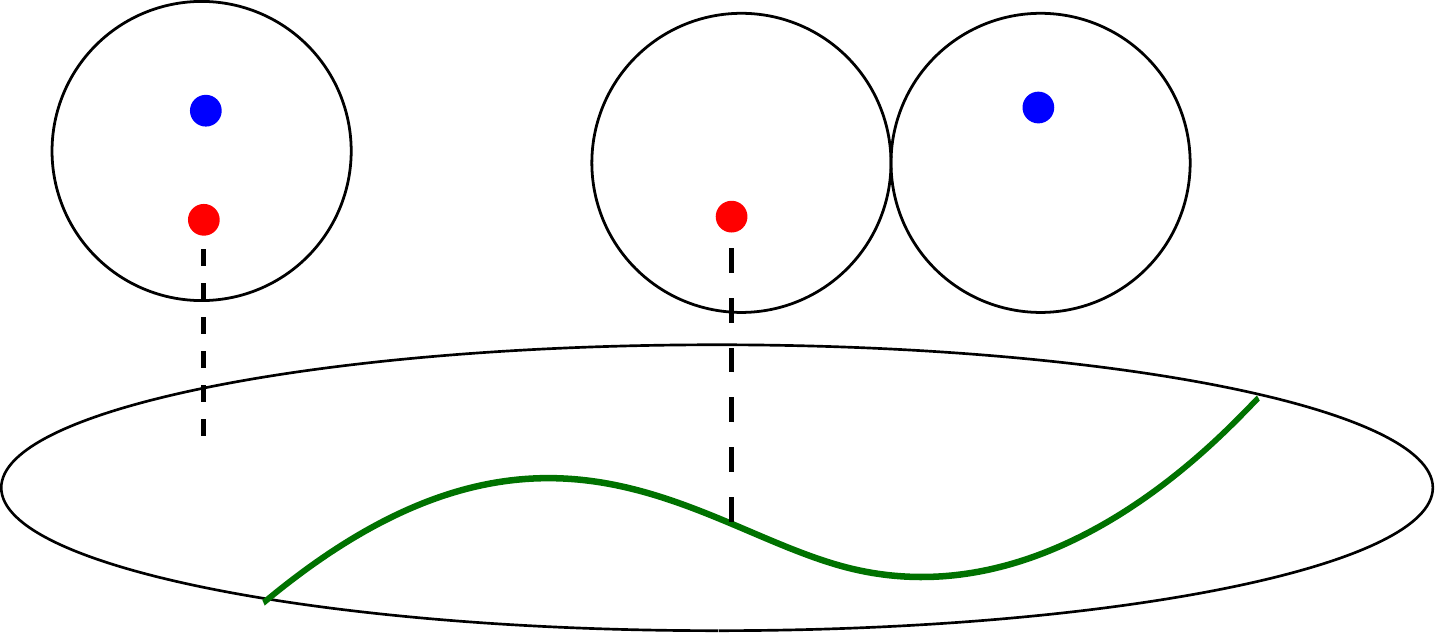}}
\put(-130,80){$\hat{s}_{13}$}
\put(-130,105){$\hat{s}_{12}$}

\put(-22,80){$\hat{s}_{13}$}

\put(40,105){$\hat{s}_{12}$}
\put(90,25){$C_{13}=0$}
\end{picture}
\caption{\label{fig:ConicBundle}\emph{Illustration of a generic conic bundle with two generic sections. Over a codimension one locus, the $\mathbb{P}^1$ degenerates into two components, each intersected by one of the sections.}}
\end{center}
\end{figure}
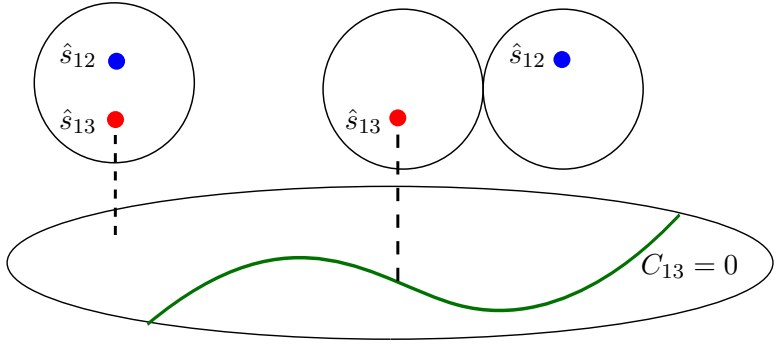

\subsection{Degenerations}
At generic points in the base, the fiber is given by a non-factorizable quadratic in $\mathbb{P}_{x}^2$, which is a conic. But at special points\footnote{Since $C_{12}$ and $C_{23}$ are constants, there are no points on the base where these parameters vanish. Hence, there are no degenerations associated with $\{C_{12}=0\}$ or $\{C_{23}=0\}$.} in the base, namely those along $\{C_{13}=0\}$, the defining relation factors as 
\begin{equation}
    2x_{2}\left(C_{12}x_{1}+C_{23}x_{3}\right)=0.
\end{equation}
The fiber at such points degenerates into two rational curves, $\{x_{2}=0\}$ and $\{C_{12}x_{1}+C_{23}x_{3}=0\}$, which intersect at the single point
\begin{equation}
    [x_{1}:x_{2}:x_{3}] = [-C_{23}:0:C_{12}].
\end{equation}
The section $\hat{s}_{13}$ hits the $\{C_{12}x_{1}+C_{23}x_{3}=0\}$ components, while the sections $\hat{s}_{12}$ and $\hat{s}_{23}$ (and any other section of class $D_{12}$) hit the $\{x_{2}=0\}$ component as shown in Figure~\ref{fig:ConicBundle}. 

Importantly, $\{C_{13}=x_{2}=0\}$ and $\{C_{13}=C_{12}x_{1}+C_{23}x_{3}=0\}$ are both genuine divisors of $\mathcal{B}_3$. The classes of these two divisors are related. The set $\{C_{13}=0\}$, which is associated to a divisor of class $[J_{y}]$, can be decomposed as
\begin{equation}
    \{C_{13}=x_{2}=0\}\cup\{C_{13}=C_{12}x_{1}+C_{23}x_{3}=0\},
\end{equation}
and thus, the classes of these two divisors sum to $[J_y]$. However, there is still one independent divisor class associated with the degenerations. We refer to the class associated with $\{C_{13}=C_{12}x_{1}+C_{23}x_{3}=0\}$ as $\tilde{D}$. This divisor class represents the extra contribution to $h^{1,1}(\mathcal{B}_{3})$ seen earlier. In turn, the additional modulus in the F-theory model is associated with the ability to change the relative volume of the two components in the degeneration. 

In 6-dimensions, the degenerations of the form described above were interpreted in terms of bulk M5 branes in a dual heterotic/M-theory picture. In particular, the extra modulus corresponds to the position of the M5 brane along the $S^1/\mathbb{Z}_2$ interval, and shrinking one of the components represents moving the M5 branes to one of the $E_8$ walls. One would expect a similar interpretation to hold for the types of degenerations in $\mathcal{B}_3$ discussed here. In particular, shrinking the $\{C_{13}=x_{2}=0\}$ component represents moving the M5 brane to the $E_8$ wall corresponding to $\hat{s}_{12}$, while shrinking the other components represents moving the M5 brane to the $E_8$ wall for $\hat{s}_{13}$. We will see further evidence for this assertion when we examine elliptic fibrations built over $\mathcal{B}_3$.

\subsection{Elliptic fibrations over $\mathcal{B}_3$}
We now discuss how to build elliptic fibrations over $\mathcal{B}_{3}$, focusing on elliptic fibrations in the Weierstrass form
\begin{equation}
    Y^2 = X^3 + f X + g.
\end{equation}
which we require to be a CY manifold. Therefore, $f$ and $g$ are sections of the line bundles $\mathcal{O}(-4K_{\mathcal{B}_3})$ and $\mathcal{O}(-6K_{\mathcal{B}_3})$, respectively. Since $-K_{\mathcal{B}_{3}}=J_{x}+3J_{y}$, we can expand $f$ and $g$ in series of degrees $4$ and $6$ in the $x_{i}$, respectively:
\begin{align}
    f =& \sum_{i=0}^{4}\sum_{j=0}^{4-i} f_{i,j}(y)x_{1}^{4-i-j}x_{2}^{i}x_{3}^{j} & g =& \sum_{i=0}^{6}\sum_{j=0}^{6-i} g_{i,j}(y)x_{1}^{6-i-j}x_{2}^{i}x_{3}^{j}
\end{align}
The $f_{i,j}(y)$ and $g_{i,j}(y)$ are homogeneous polynomials of degree $12-i$ and $18-i$ in the $y_i$ coordinates. This expansion can be simplified. Because $C_{23}$ is a non-zero constant, we can use the conic bundle defining relation  \eref{eq:exampleconicdefrel} to replace all occurrences of $x_{2}x_{3}$ with $x_{1}x_{3}$ and $x_{1}x_{2}$. We can therefore rewrite the $f$ of the Weierstrass model as
\begin{align}
    f=&\sum_{i=1}^{4}f^{\prime}_{i}(y)x_{2}^{i}x_{1}^{4-i} + f_{0}(y)x_{1}^4 + \sum_{j=1}^{4}f^{\prime\prime}_{j}(y)x_{3}^{j}x_{1}^{4-j}. \label{eq:fexconic}
\end{align}
The $f^{\prime}_{i}(y)$ are homogeneous polynomials of degree $12-i$ in the $y_i$, while $f_{0}(y)$ and the $f^{\prime\prime}_{j}(y)$ are homogeneous polynomials of degree $12$ in the $y_{i}$. Similarly, $g$ can be rewritten as
\begin{align}
    g=&\sum_{i=1}^{6}g^{\prime}_{i}(y)x_{2}^{i}x_{1}^{6-i} + g_{0}(y)x_{1}^6 + \sum_{j=1}^{6}g^{\prime\prime}_{j}(y)x_{3}^{j}x_{1}^{6-j}.\label{eq:gexconic}
\end{align}
The $g^{\prime}_{i}(y)$ are homogeneous polynomials of degree $18-i$, and $g_{0}(y)$ and the $g^{\prime\prime}_{j}(y)$ are homogeneous polynomials of degree $18$.

It is useful to compare this construction to a Weierstrass model where the base is a $\mathbb{P}^1$-{\it bundle} over $\mathbb{P}^2_{y}$ with twist $J_{y}$. If we let the coordinates of the $\mathbb{P}^1$ fiber be $[u:v]$, the $f$ and $g$ of this Weierstrass model would take the forms
\begin{align}
    f=& \sum_{l=0}^{8}\tilde{f}_{l}(y)u^{l}v^{8-l} & g=& \sum_{m=0}^{12}\tilde{g}_{m}(y)u^{m}v^{12-m},
\end{align}
where the $\tilde{f}_{l}(y)$ are homogeneous polynomials of degree $16-l$ in the $y$'s and the $\tilde{g}_{m}(y)$ are homogeneous polynomials of degree $24-m$ in the $y$'s. 

In the context of heterotic/F-theory duality, we now wish to perform a small instanton transition in this model. We first tune
\begin{align}
    \tilde{f}_{l}\to&\alpha^{4-l}\tilde{f}^\prime_{l}\text{ for }l<4 & \tilde{g}_{m}\to&\alpha^{6-m}\tilde{g}^\prime_{m}\text{ for }m<6,
\end{align}
where $\alpha$ is a linear expression in the $y$'s. The $\tilde{f}^\prime_{l}$ all have degree $12$, while the $\tilde{g}^\prime_{m}$ all have degree $18$. We then blow up the base at $\{\alpha=u=0\}$ by letting $\alpha\to \beta \alpha$, $u\to \beta u$ and dividing $f$ and $g$ by $\beta^4$ and $\beta^6$. This leads to a Weierstrass model described by
\begin{align}
    f=& \sum_{l=0}^3 \tilde{f}_{l}^\prime \alpha^{4-l}u^{l}v^{8-l} + \tilde{f}_{4}u^4 v^4 + \sum_{l=5}^8\tilde{f}_{l}\beta^{l-4}u^{l}v^{8-l},\\
    g=& \sum_{m=0}^5 \tilde{g}_{m}^\prime \alpha^{6-m}u^{m}v^{12-m} + \tilde{g}_{6}u^6 v^6 + \sum_{m=7}^{12}\tilde{g}_{m}\beta^{m-6}u^{m}v^{12-m}.
\end{align}
The structure of this Weierstrass model is remarkably similar to that described by \eref{eq:fexconic} and \eref{eq:gexconic}. In particular, the degrees of the $\tilde{f}^\prime_{l}$ and $\tilde{g}^\prime_{m}$ match with those of the $f^{\prime\prime}_{j}$ and $g^{\prime\prime}_{j}$, while the degrees of the $\tilde{f}_{l}$ and $\tilde{g}_{m}$ match with those of the $f_{i}^\prime$ and the $g^{\prime}_{i}$. These similarities suggest that the two Weierstrass models describe similar physical situations. Since the small instanton transition physically involves moving an M5 brane into the bulk in a dual heterotic/M-theory picture, the Weierstrass model for the conic bundle should also involve bulk M5 branes. This provides some additional evidence for the asserted relation between the degenerate $\mathbb{P}^1$ fibers and bulk M5 branes. 

\subsubsection{Degenerations and $(4,6)$ loci}
To further explore the relationship between the degenerate conic fibers and the presence of dual bulk M5 branes, we would like to show that blowing down one of the components in the degenerations produces codimension-two loci in the base of the elliptic fibration where $f$ and $g$ vanish to orders 4 and 6. We can approach this problem by trying to describe the conic fibers as $\mathbb{P}^1$'s. At generic points, the fiber is simply a rational curve, and we should be able to describe points on this curve in terms of $\mathbb{P}^1$ coordinates $[U:V]$. Therefore, we want to find a map between the $x_{i}$ coordinates and $[U:V]$, which can be done by means of a Veronese map. We first expand the $x_{i}$ as quadratic expressions in $U,V$:
\begin{equation}
    x_{i} = a_{i}U^2 + b_{i} U V + c_{i} V^2.
\end{equation}
We then plug these expansions into the conic bundle defining relation \eref{eq:exampleconicdefrel} and adjust the $a_{i}$, $b_{i}$, and $c_{i}$ to satsify the equation. One such map between the $x_{i}$ coordinates and $[U:V]$ takes the form
\begin{align}
    x_{1}=&C_{23} U V & x_{2}=& C_{13}\left(-U V + V^2\right) & x_{3}=&C_{12} \left(U^2-U V\right) .\label{eq:veronesemap}
\end{align}
Note that $[U:V]=[0:1]$ corresponds to the section $\hat{s}_{13}$, while $[U:V]=[1:0]$ corresponds to the section  $\hat{s}_{12}$.

While this map should pose no issues at most points in $B_2=\mathbb{P}^2_{y}$, special behavior should occur at the degeneration locus $\{C_{13}=0\}$. Indeed, when $C_{13}$ is 0, all values of $[U:V]$ map to points on the conic fiber where $x_{2}$ vanishes. All of these points belong to the $\{C_{13}=x_{2}=0\}$ component; the only point on the $\{C_{13}=C_{12}x_{1}+C_{23}x_{3}=0\}$ component with a corresponding value of $[U:V]$ is the point $[x_1:x_2:x_3]=[-C_{23}:0:C_{12}]$, the intersection point of the two components. Therefore, if we replace all instances of $x_{i}$ in the Weierstrass model with the corresponding expressions in \eref{eq:veronesemap}, we would essentially send all of the points in the $\{C_{13}=C_{12}x_{1}+C_{23}x_{3}=0\}$ component of a degenerate fiber to a single point given by $[U:V]=[0:1]$. In other words, we would blow down this component in the degenerate fibers.

\begin{figure}[t!]
\begin{center}
\begin{picture}(0,120)
\put(-130,0){\includegraphics[scale=0.7]{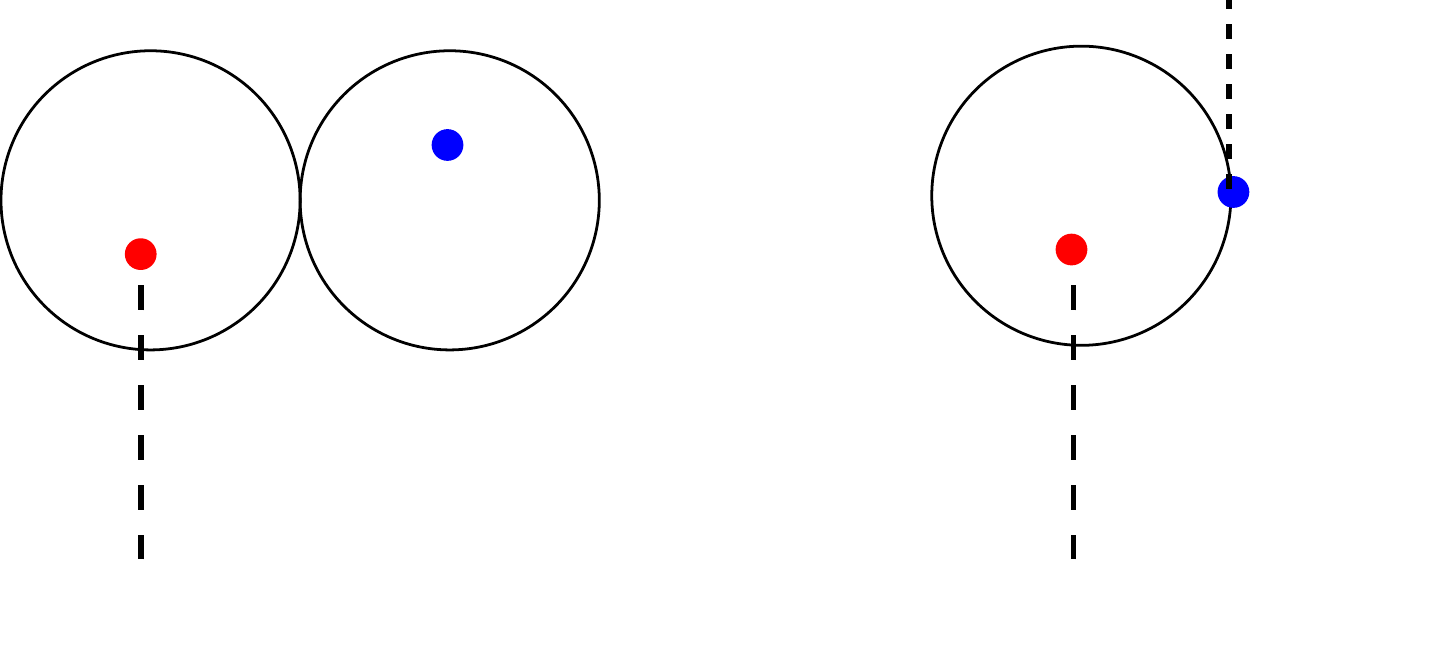}}
\put(-120,80){$\hat{s}_{13}$}
\put(-60,100){$\hat{s}_{12}$}
 
\put(0,90){$\xrightarrow{\text{blow-down}}$}
\put(65,80){$\hat{s}_{13}$}
\put(130,90){$\hat{s}_{12}$} 
\put(100,140){$(4,6,12)$}
\end{picture}
\caption{\label{fig:BlowdownConic}\emph{Illustration of the degenerate conic bundle base in codimension one. A blow-down produces a $(4,6,12)$ singular Weierstrass model in codimension two.}}
\end{center}
\end{figure}
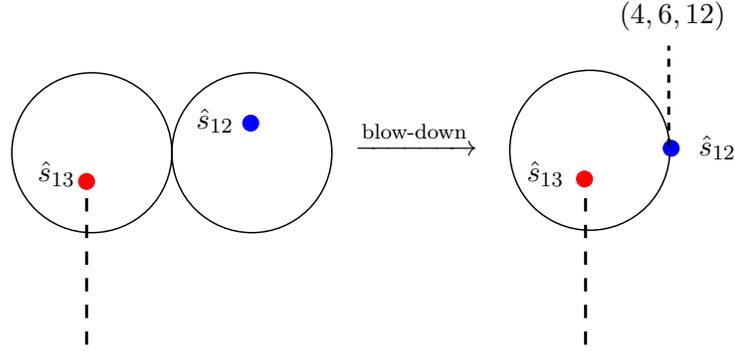

If we plug in the expressions in \eref{eq:veronesemap} into \eref{eq:fexconic}, we obtain the following expressions for $f$:
\begin{multline}
    f= \sum_{i=1}^{4}f^\prime_{i}(y) C_{13}^{i}C_{23}^{4-i}V^{4}U^{4-i}\left(V-U\right)^{i} + f_{0}(y)C_{23}^{4} U^4 V^4 \\
    + \sum_{j=1}^{4}f_{j}^{\prime\prime}(y)C_{12}^j C_{23}^{4-j}U^4 V^{4-j}(U-V)^{j}
\end{multline}
This new $f$ vanishes to order 4 at $\{C_{13}=U=0\}$. Similarly, $g$ takes the form
\begin{multline}
   g= \sum_{i=1}^{6}g^\prime_{i}(y) C_{13}^{i}C_{23}^{6-i}V^{6}U^{6-i}\left(V-U\right)^{i} + g_{0}(y)C_{23}^{6} U^6 V^6\\ + \sum_{j=1}^{6}g_{j}^{\prime\prime}(y)C_{12}^j C_{23}^{6-j}U^6 V^{6-j}(U-V)^{j},
\end{multline}
which vanishes to order 6 at $\{C_{13}=U=0\}$. Recall that $\{C_{13}=U=0\}$ is the locus to which all points on $\{C_{13}=C_{12}x_{1}+C_{23}x_{3}=0\}$ are sent, see Figure~\ref{fig:BlowdownConic}. In other words, blowing down an extra component of a degenerate fiber gives a codimension-two locus where $(f,g)$ vanish to orders $(4,6)$. The appearance of the $(4,6)$ locus indicates that, in the dual heterotic/M-theory picture, we have an M5 brane touching one of the $E_8$ fixed-planes. If the volume of the extra component represents an M5 brane position modulus, this behavior exactly matches our expectations.

\vspace{0.5cm}

As a final comment on the heterotic/F-theory dictionary in the context of conic bundles with sections, it should be noted that the example above illustrates that two of the essential characterizations of conic bundles used in the Sarkisov program mentioned in Section \ref{sec:conic_intro} -- namely the existence of a generic twist to the $\mathbb{P}^1$ fibration and the discriminant locus of the fibration -- have a clear role in a dual heterotic geometry. As in the case of standard heterotic/F-theory duality utilizing $\mathbb{P}^1$-bundles the generic twist (i.e. the choice of ${\cal O}(D)$ in ${\cal B}_3=\mathbb{P}(\pi: {\cal O} \oplus {\cal O}(D) \to B_2$) fixes a component of the second Chern class of a vector bundle over the CY threefold $\pi_h: X_2 \to B_2$. The novel feature in the present context is the discriminant locus of the $\mathbb{P}^1$ fibration which specifies a curve in the base ${ B}_2$ which as argued above is wrapped by 5-branes in the heterotic dual theory. Note that this is a direct 4-dimensional analog of the 6-dimensional base geometries leading to extra tensor multiplets (i.e. blow-ups in the base) mentioned in Section \ref{sec:6d_hetF}.

\subsubsection{Tuning gauge groups}
Now that we have an understanding of how to construct Weierstrass models over $\mathcal{B}_{3}$, let us discuss how to tune gauge algebras in the corresponding F-theory model. We are particularly interested in tuning gauge algebras supported on the sections $\hat{s}_{13}$ and $\hat{s}_{12}$, as it is often easier to understand heterotic/F-theory duality when such gauge algebras are present. For example, suppose we wanted to tune an $E_8\times E_8$ gauge group with the two $E_8$ factors supported on the sections. According to the rules from the Kodaira classification, $f$ should vanish to order 4 at both sections, and $g$ should vanish to order 5. We can easily accomplish this by letting $f$ be proportional to $x_{1}^4$ and g be proportional to $x_{1}^5$, giving us a Weierstrass equation of the form
\begin{equation}
    Y^2 = X^3 + f_{0}(y)x_{1}^4 X + x_{1}^5\left(g_{1}^\prime(y) x_{2} + g_{0}(y) x_{1} + g_{1}^{\prime\prime}(y)x_{3}\right).
\end{equation}
Of course, we must also include the defining relation \eref{eq:exampleconicdefrel} in order to fully describe the elliptic fibration. Alternatively, suppose we wanted to tune an $E_7\times E_8$ gauge group with the $E_7$ supported on $\hat{s}_{13}$ and the $E_{8}$ supported on $\hat{s}_{12}$. Then, $(f,g)$ would have to vanish to orders $(3,5)$ at $\{x_{1}=x_{3}=0\}$ but to order $(4,5)$ at $\{x_{1}=x_{2}=0\}$. The Weierstrass model would then take the form
\begin{equation}
    Y^2 = X^3 + x_{1}^3\left(f_{1}^\prime(y) x_{2} + f_{0}(y)x_{1} \right)X + x_{1}^5\left(g_{1}^\prime(y) x_{2} + g_{0}(y) x_{1} + g_{1}^{\prime\prime}(y)x_{3}\right).
\end{equation}

\subsection{Another example of a conic bundle without monodromy}

In this subsection, we will give one more example to illustrate an effect which sometimes arises in these compactifications. This phenomenon concerns the tuning of gauge groups in the Weierstrass model. As such we will be very brief regarding many details and only present those that are necessary to illustrate the point of interest.

We consider a 3-dimensional conic bundle $\mathcal{B}_3$ described by the following complete intersection configuration matrix.
\begin{equation}
\label{eq:conicCicy}
\left[
\begin{array}{c|cc}
\firstp & 1 & 1 \\
\secondp & 1 & 0 \\\hdashline[2pt/2pt]
\thirdp & 0 & 1 
\end{array}\right].
\end{equation}
We choose the $\mathbb{P}^n$ coordinates to be
\begin{align}
    \firstp:&\,[\firstxone:\firstxtwo:\firstxthree] & \secondp:&\,[\secondxone:\secondxtwo] & \thirdp:&\,[\thirdxone:\thirdxtwo:\thirdxthree].
\end{align}
The K\"ahler forms of these spaces descend to give $(1,1)$ forms on the conic bundle itself. In what follows we will denote these by $J_1,J_2$ and $J_3$ respectively. The ambient space factor $\mathbb{P}^2_{(3)}$ is the base of the conic bundle and the two equations encoded in the columns of (\ref{eq:conicCicy}), which we will denote by $P_{b_1}$ and $P_{b_2}$, describe the, generically $\mathbb{P}^1$, fiber inside $\mathbb{P}^2_{(1)}$ and $\mathbb{P}^1_{(2)}$.

\vspace{0.2cm}

A similar analysis of the sections of ${\cal B}_3$ to that performed for the previous example can be carried out. In the case at hand one finds that, in order to have two non-intersecting sections that satisfy all of the conditions outlined in Section \ref{bob}, they must both have class Poincare dual to $J_2$.
The sections essentially specify a particular point on $\secondp$ that does not vary with the position in the base $\thirdp$. For instance, one could choose the two sections to be $y_0=0$ and $y_1=0$.

\vspace{0.2cm}

The $\mathbb{P}^1$ fiber of the conic bundle ${\cal B}_3$ degenerates at codimension one in its base $B_2$. We write the defining relations of the conic bundle as follows,
\begin{align}
    P_{b_1}\equiv& l_{0}(\secondxone, \secondxtwo) \firstxone + l_{1}(\secondxone, \secondxtwo) \firstxtwo +l_{2}(\secondxone, \secondxtwo) \firstxthree =0\label{eq:P1conicnomono}\\
    P_{b_2}\equiv&m_0(\thirdxone, \thirdxtwo, \thirdxthree)\firstxone+m_1(\thirdxone, \thirdxtwo, \thirdxthree)\firstxtwo+m_2(\thirdxone, \thirdxtwo, \thirdxthree)\firstxthree = 0\label{eq:P2conicnomono},
\end{align}
where the $l_i$ and $m_i$ are linear expressions in the $y_i$ and $z_i$, respectively. Since $l_0(y)$, $l_{1}(y)$, and $l_{2}(y)$ are linear expressions in the $\secondp$ coordinates, they cannot be independent, and one of them must be a linear combination of the others. Let us assume that $l_{2}(y)$ can be written as
\begin{equation}
\alpha l_{0}(y) + \beta l_{1}(y) \label{eq:l2redefnomono}
\end{equation}
for some complex numbers $\alpha$ and $\beta$. Then, the locus in $B_2$ over which the $\mathbb{P}^1$ fiber degenerates is given by the following.
\begin{equation}
    \Delta_{b} \equiv m_{2}(z) - \alpha m_{0}(z) - \beta m_{1}(z) = 0.
\end{equation}
Over this locus the fiber splits into two $\mathbb{P}^1$'s. The first component, which we refer to as $\tilde{c}_a$, is specified by
\begin{align}
   [\firstxone:\firstxtwo:\firstxthree]=&[-\alpha:-\beta:1]  \label{eq:firstcompnomono}
\end{align}
with $[\secondxone:\secondxtwo]$ unrestricted.  The second component, which we refer to as $\tilde{c}_b$, is specified by
\begin{align}
    m_{0}(z)l_{1}(y)-m_{1}(z)l_{0}(y)=&0 &  m_{0}(z)\firstxone + m_{1}(z)\firstxtwo + m_{2}(z)\firstxthree =&0. \label{eq:secondcompnomono}
\end{align}
We refer to the divisor found by fibering $\tilde{c}_a$ over $\Delta_b$ as $\tilde{D}_a$, while we refer to the divisor found by fibering $\tilde{c}_b$ as $\tilde{D}_b$. 

The two sections mentioned above generically hit the $\tilde{c}_a$ component of the fiber over $\Delta_b$ at a single point and miss the component $\tilde{c}_b$ entirely. However, at a codimension one locus inside $\Delta_b$, the $[\secondxone:\secondxtwo]$ picked out by the definition of a section may satisfy
\begin{displaymath}
m_0(z)l_1(y)-m_1(z)-l_0(y) = 0.
\end{displaymath}
When this happens, the section wraps the $\tilde{c}_b$ component. 

\vspace{0.2cm}

We shall now describe the elliptic fibration over this conic bundle. We focus on elliptic fibrations described by Weierstrass models of the form
\begin{equation}
    Y^2  = X^3 + f X + g,
\end{equation}
where $X$ and $Y$ are coordinates for the elliptic fiber and $f$ and $g$ are respectively sections of the line bundles $\mathcal{O}(-4K_{\conicbundle})$ and $\mathcal{O}(-6K_{\conicbundle})$. Computing the first Chern class of ${\cal B}_3$ we find,
\begin{displaymath}
-K_{\conicbundle} = c_1(\conicbundle) = J_1 + J_2 + 2 J_3.
\end{displaymath}
Therefore, $f$ should be a polynomial of order $4$ in the $x_i$, order $4$ in the $y_i$, and order $8$ in the $z_i$. It can be written explicitly as
\begin{equation}
    f = \sum_{i_1=0}^4\sum_{i_2=0}^{4-i_1}\sum_{j=0}^4\sum_{k_1=0}^8\sum_{k_2=0}^{8-k_1} f_{i_1,i_2,j,k_1,k_2}\firstxone^{i_1}\firstxtwo^{i_2}\firstxthree^{4-i_1-i_2}\secondxone^{j}\secondxtwo^{4-j}\thirdxone^{k_1}\thirdxtwo^{k_2}\thirdxthree^{8-k_1-k_2} ,\label{eq:fnomono}
\end{equation}
where the $f_{i_1,i_2,j,k_1,k_2}$ are complex numbers. Similarly, $g$ should be a polynomial of order $6$ in the $x_i$, order $6$ in the $y_i$, and order $12$ in the $z_i$. It can be written as 
\begin{equation}
    g = \sum_{i_1=0}^6\sum_{i_2=0}^{6-i_1}\sum_{j=0}^6\sum_{k_1=1}^{12}\sum_{k_2=0}^{12-k_1} g_{i_1,i_2,j,k_1,k_2}\firstxone^{i_1}\firstxtwo^{i_2}\firstxthree^{6-i_1-i_2}\secondxone^{j}\secondxtwo^{6-j}\thirdxone^{k_1}\thirdxtwo^{k_2}\thirdxthree^{12-k_1-k_2}, \label{eq:gnomono}
\end{equation}
where the $g_{i_1,i_2,j,k_1,k_2}$ are also complex numbers. 
Of course, the full description of the elliptic fibration also includes the defining relations \eref{eq:P1conicnomono} and \eref{eq:P2conicnomono} for the base. 

\vspace{0.2cm}

We now turn our attention to the key issue we wish to address in this section: tuning gauge groups. To understand the heterotic dual of this F-theory model, we would like to tune gauge groups on the two chosen sections of the conic bundle. But tuning gauge groups in this conic bundle model can be somewhat subtle. For example, let us attempt to tune an $E_8\times E_8$ gauge symmetry with the $E_8$ factors living on the two chose conic bundle sections. We focus in particular on tuning $f$. The sections we choose are specified by $y_0=0$ and $y_1=0$ respectively as mentioned above. According to the Kodaira classification, $f$ must vanish to order 4 on each section, and one might imagine tuning $f$ to be proportional to $\secondxone^4 \secondxtwo^4$. However, $f$ is only order $4$ in the $y_i$, so this tuning is not possible. 

There is a way to circumvent this issue, at least at a naive level. For the first $E_8$ gauge factor, we tune $f$ to be proportional to $\secondxone^4$. For the second $E_8$ factor, note that, at points on $B_2$ away from the degeneration locus, the defining relations $P_{b_1}$ and $P_{b_2}$ associate each point on $\secondp$ with a unique point on $\firstp$. We therefore make $f$ proportional to $P_{b_1}$ evaluated at $y_0=1,y_1=0$, which is a linear expression in the $x_i$. In the end, we have an $f$ of the form
\begin{equation}
    f = \secondxone^4\left(P_{b_1}\Big|_{y_1=0}\right)^4\sum_{k_1=1}^{8}\sum_{k_2=1}^{8-k_1}f_{k_1,k_2}\thirdxone^{k_1}\thirdxtwo^{k_2}\thirdxthree^{8-k_1-k_2}.
\end{equation}
But the correspondence between points on $\secondp$ and points on $\firstp$ breaks down along the degeneration locus, and this trick should cause problems there. Indeed, $P_{b_1}$ vanishes for all points on the $\tilde{c}_a$ component specified by \eref{eq:firstcompnomono}. The tuned $f$ should therefore vanish to order 4 on $\tilde{D}_a$. This same effect would be seen if we chose to tune the $E_8$ on $\secondxone=0$ via $P_{b_1}$ rather than the $E_8$ on $y_1=0$.

We also want $g$ to vanish to order 5 on the two sections, but the expansion for $g$ is order $6$ in the $y_i$. If we attempt to tune $g$ using a trick similar to that above, $g$ is forced to vanish to order 4 on $\tilde{D}_a$. According to the Kodaira classification, codimension-one loci in the base where $f$ and $g$ vanish to order 4 support $E_7$ gauge groups, at least at the geometric level. In the end, tuning an $E_8\times E_8$ gauge symmetry on the sections has forced an $E_7$ gauge group along $\tilde{D}_a$. Similar arguments suggest that tuning other sufficiently large gauge groups on the sections would also force some gauge group on one of the components. 

This behavior seems to be an example of the general phenomenon discussed in \cite{Raghuram:2020vxm}. We will see a similar phenomenon for the conic bundle examined in Section \ref{sec:conicmonodromy}, suggesting that the forced gauge group occurs whenever both sections intersect the same degenerate component. This behavior may also be a 4-dimensional analogue of the ``$E_8$ rule'' \cite{Heckman:2013pva, Johnson:2016qar, Morrison:2016djb} proposed for 6-dimensional F-theory models. The $E_8$ rule states that if two curves carrying gauge algebras $G_1$ and $G_2$ intersect a $-1$ curve, then $G_{1}\times G_{2}$ should be contained in $E_8$. Typically, this rule has been stated in contexts where the $-1$ curve does not itself support any gauge algebra. In the conic bundle analyzed here, $\tilde{c}_a$ is the 4-dimensional analogue of the $-1$ curve. Since the sections intersect the same component, the $E_8$ rule would suggest something unusual should occur if the gauge algebras tuned on the sections are not contained in $E_8$. Clearly, $E_8\times E_8$ is not contained in $E_8$, so tuning $E_8\times E_8$ on the sections should lead to some issue. The forced $E_7$ gauge group therefore may be the model's response to an attempted violation of the $E_8$ rule. As an additional test for this proposition, the $E_8$ rule suggests that we should be able tune $\SO(8)\times \SO(8)$, which is contained in $E_8$, on the two sections. The Kodaira rules state that $f$ and $g$ must vanish to orders 2 and 3 on a divisor supporting an $I_0^*$ singularity, the singularity type associated with $\SO(8)$.\footnote{There are additional conditions, known as the split conditions, that distinguish between $G_2$, $\SO(7)$, $\SO(8)$ \cite{Bershadsky:1996nh}. However, these conditions do not significantly affect the argument here.} But since $f$ and $g$ are order $4$ and $6$ in the $y_i$'s, one can make $f$ and $g$ proportional to the appropriate powers of $\secondxone$ and $\secondxtwo$ without resorting to the trick used above. We therefore expect that it is possible to tune $\SO(8)\times\SO(8)$ without any additional gauge symmetry, in line with the expectations from the $E_8$ rule.


\section{Generalized conic bundles and monodromy}
\label{sec:conicmonodromy}
In the last section, we analyzed some simple initial examples of elliptic fourfolds whose bases were conic bundles with two sections and no monodromy, focusing on the corresponding F-theory model and its heterotic dual. We now wish to give a broader, more general discussion of conic bundles and elliptic fourfolds built from them. In this section, we focus on those aspects of conic bundles closely tied to the degenerations of the conic fibers and the associated physics. This discussion includes questions about monodromy effects and the presence of conic bundle sections. We also discuss the effects of non-flat fibers in a conic bundle. Finally, we describe how to build elliptic fibrations that use these conic bundles as bases.  Much of this survey is structured around a set of illuminating examples.

\subsection{Generalities about conic bundles}
\label{sec:sarkisov}
We begin by discussing some key mathematical results regarding conic bundles, many of which were developed by Sarkisov \cite{Sarkisov_1983}.\footnote{Many of these concepts are also summarized in  \cite{2018RuMaS..73..375P}.} There, a conic bundle over a non-singular variety $B_2$ is defined by the triple $({\cal B}_3,B_2,\pi)$, where $\pi:{\cal B}_3\to B_2$ is a rational map with an irreducible rational curve as the generic fiber. If $\pi$ is a flat morphism of non-singular varieties, the conic bundle is called \emph{regular}. If, for a regular conic bundle, $\pi^{-1}(d)$ is an irreducible divisor in V for any irreducible divisor $d\in B_2$, the conic bundle is \emph{standard}. The examples in Section \ref{sec:conicnomono} are regular conic bundles. But the degeneration divisor in the bases of those geometries uplifts to two divisors in the conic bundles found by fibering the two rational curve components over the degeneration divisors. These examples are therefore not standard conic bundles. However, if the conic bundles had exhibited monodromy effects that exchanged the two components, they would have no longer formed distinct divisors, and the conic bundles would have been standard.

The above definition does not necessarily involve the quadratic defining relation in $\mathbb{P}^2$ that we encountered in Section \ref{sec:conicnomono}. Sarkisov introduces the term \emph{embedded} conic bundle to describe such situations. Specifically, an embedded conic bundle is a conic bundle $({\cal B}_3,B_2,\tau|_{{\cal B}_3})$ where ${\cal B}_3$ is an irreducible reduced hypersurface in the projectivization $\mathbb{P}(\mathcal{E})$ of a rank-three locally free sheaf $\mathcal{E}$ on $B_2$. The map $\tau:\mathbb{P}(\mathcal{E})\to B_2$ is the standard projection. Then, at least in the neighborhood $U$ of a point on $B_2$, the conic bundle can be described the equation
\begin{equation}
    \sum_{i,j}C_{i,j}x_{i}x_{j},
\end{equation}
where the $x_{i}$ act as coordinates for the fibers of $\mathbb{P}(\mathcal{E})$ and $C_{ij}$ can vary over $U$. 
The first of the conic bundles discussed in Section \ref{sec:conicnomono} is an example of an embedded conic bundle. More broadly, a regular conic bundle can be written as an embedded conic bundle: according to point 1.5 in \cite{Sarkisov_1983}, the pushforward $\pi_*\mathcal{O}_{{\cal B}_3}(-K_{{\cal B}_3})$ of the anticanonical bundle $\mathcal{O}_{{\cal B}_3}(-K_{{\cal B}_3})$ on a regular conic ${\cal B}_3$ is a locally free sheaf of rank three, and we can find an embedding of ${\cal B}_3$ in the projectivization of  $\pi_*\mathcal{O}_{{\cal B}_3}(-K_{{\cal B}_3})$ where the fiber is a conic in $\mathbb{P}^2$. 

According to Proposition 1.8 of \cite{Sarkisov_1983}, many properties of an embedded conic bundle are encoded in the properties of $C_{ij}$. First, the rank of $C_{ij}$ is 3 at a particular point if and only if that point does not lie on the degeneration divisor. The fiber at such points should be an irreducible rational curve. Said another way, the degeneration divisor, consisting of points in the base where the conic fiber is degenerate, is essentially given by the determinant of $C_{ij}$, as the rank drops below three at points where this determinant vanishes. If a point in the base lies on the degeneration divisor but is not a singularity of the degeneration divisor, the rank of $C_{ij}$ at that point is 2. If the rank at a point is either 0 or 1, that point is a singularity of the degeneration divisor. We can also make more general statements about properties of the conic bundle on an open set $U\subset B_2$. For instance, the conic bundle is flat\footnote{Phrased more properly, the projection map is a flat morphism on $U$.} on $U$ if the rank of $C_{ij}$ is always greater than 0 on $U$. It is regular on $U$ if, in addition to the rank of $C_{ij}$ being greater than 0, the following conditions hold:
\begin{itemize}
    \item If the rank of $C_{ij}$ at point $p \in U$ is 2, the point is on the degeneration divisor but is not a singular locus of the degeneration divisor.
    \item The rank of $C_{ij}$ at a point $p \in U$ is 1 if and only if that point is a singularity of the degeneration divisor and
    \begin{equation}
        \rk\left(\frac{\partial^2}{\partial y_{i} \partial y_{j}}\det(C)\right) > 1 \label{eq:regconicrank1cond}
    \end{equation}
    at that point, where the $y_{i}$ are local parameters near the point. 
\end{itemize}

\subsection{Conic bundles with and without sections}
\label{sec:conicsections}
We start our discussion with a ``generic'' conic bundle represented as a degree two hypersurface in the ambient space $\mathbb{P}^2_{x}$ which itself is fibered over a base $B_2$. This conic bundle is described by a hypersurface of the form
\begin{align} 
\label{eq:conic}
p = \sum_{i,j=1}^{3}x_{i}C_{ij}x_{j}  = C_{11}x_1^2 + 2 C_{12}x_{1}x_{2} + 2 C_{13}x_{1}x_{3} + C_{22}x_2^2 + 2 C_{23}x_{2} x_{3}+ C_{33}x_3^2 \, . 
\end{align}
The $C_{ij}$ are some sections of line bundles on $B_2$ that are specified by the degree of the hypersurface equation and the twist of the ambient $\mathbb{P}^2$ over $B_2$. The discriminant can be computed as the determinant of the coefficient matrix $C_{ij}$ 
\begin{align}
\label{eq:discriminantconic}
\Delta(p) = C_{11} C_{22} C_{33}-C_{13}^2 C_{22} + 2 C_{12} C_{13} C_{23} - C_{11} C_{23}^2 - C_{12}^2 C_{33} \, .
\end{align}

Some of the most important properties of conic bundle bases in an F-theory context are the Hodge numbers of the spaces, particularly $h^{1,1}(\conicbundle)$ and $h^{2,1}(\conicbundle)$. The physical implications of $h^{1,1}(\conicbundle)$ are closely tied to degenerations of the conic fibers, as we will discussed in detail in the next subsection. Meanwhile, if $h^{2,1}(\conicbundle)$ is non-zero, the F-theory model built using $\conicbundle$ has $h^{2,1}(\conicbundle)$ additional U(1) gauge bosons, which arise from the Type IIB $C_{4}$ form \cite{Grimm:2010ks, Grimm:2012yq}.

The Hodge numbers for a conic bundle of the type we are considering here depend on $B_2$, but we can give a sense of some of their properties by looking at a class of examples. We choose $B_2 = \mathbb{P}^2$, taking the ambient $\mathbb{P}^2$ of the fiber to have a vanishing twist over this base, and considering a conic bundle which is a degree $[2,d]$ hypersurface in the full ambient space. In other words, \eref{eq:conic} is of degree 2 in the $[x_1:x_2:x_3]$ coordinates and of degree $d$ in the $B_2$ coordinates. We can compute the Hodge numbers using standard techniques. 
\begin{align}\label{eq:hodgegenconic}
h^{1,1}(\conicbundle)=& 2 & h^{2,1}(\conicbundle)&=\frac{9}{2}(d-1)d 
\end{align}
Since the Euler number is just the alternating sum of the Betti numbers, we have $\chi(\conicbundle)=\sum_i (-1)^i b_i = 2+2h^{1,1}-2h^{2,1}$ where we have used Hodge duality and the fact that $h^{0}({\cal B}_3,{\cal O})=1$ and $h^{1}({\cal B}_3,{\cal O})=0$. We thus obtain for the above geometry,
\begin{align}
\chi_{\conicbundle}=& 6 -9 ( d-1) d \, . 
\end{align}
The nontrivial dependence of the Euler and Hodge numbers on the degree $d$ is yet another feature of conic bundles that is not shared by simple $\mathbb{P}^1$ fibrations.The Euler number and the Hodge numbers for a degree $p=[1,d]$ surface in the same ambient space are independent of $d$ for example, which gives them a structure more similar to generalized Hirzebruch surfaces.

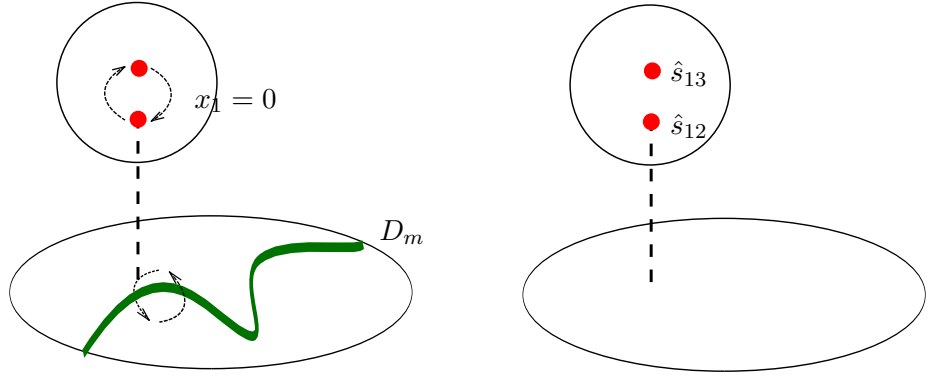
\begin{figure}[t!]
\begin{center}
\begin{picture}(0,120)
\put(-130,0){\includegraphics[scale=0.7]{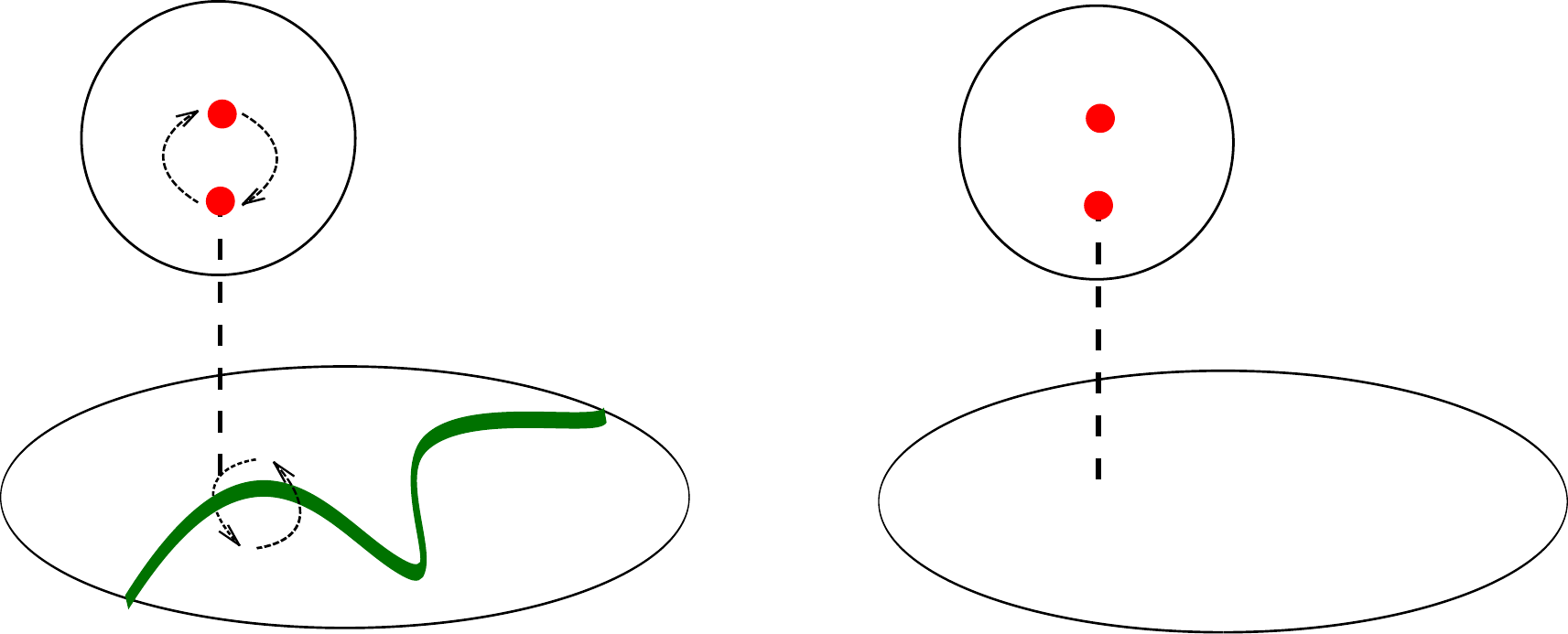}}
 
\put(-60,100){$x_{1}=0$}
  \put(10,50){$D_m$}

\put(120,110){$\hat{s}_{13}$}
\put(120,90){$\hat{s}_{12}$}

\end{picture}
\caption{\label{fig:ConicMono}\emph{Illustration of a conic bundle with two-sections of $x_1=0$ that are exchanged around the monodromy divisor $D_m$ on the left. If $D_m$ is trivial, the two-sections split globally into two sections $s_{12}, s_{13}$, as illustrated on the right.}}
\end{center}
\end{figure}

The threefolds described by \eref{eq:conic} are natural first examples of conic bundles, and one can construct well-defined F-theory models by building elliptic fibrations over them.  However, an important reason for studying conic bundles is that we want to analyze heterotic/F-theory duality in this new context. The duality is best understood when there are sections of the $\mathbb{P}^1$ fibration, but the generic conic bundle described above does not necessarily admit sections. Therefore, the generic conic bundles are not ideal examples for heterotic/F-theory duality. Here, we discuss two ways of specializing the generic conic bundle above to ensure the existence of sections. The first occurs when we choose divisor classes such that some of the $C_{ij}$ are globally constants, which can happen when the ambient $\mathbb{P}^2$ is non-trivially fibered over $B_2$. For example, suppose that $[C_{22}]$, $[C_{33}]$, and $[C_{23}]$ are trivial. In this case, we see that plugging $x_1=0$ into \eref{eq:conic} leads to
\begin{align}
p=C_{22} x_2^2 + 2 C_{23} x_2 x_3 + C_{33} x_3^2 \,.
\end{align}
If non-trivial the two solutions above are exchanged around the monodromy divisor $D_m: C_{23}^2 - 	C_{22} C_{33}=0$. Because $[C_{22}]$, $[C_{33}]$, and $[C_{23}]$ are trivial, we can factor this expression globally. As a result, the locus $x_1=0$, which is a two-section for general choices of the divisor classes, now consists of two independent sections as shown in Figure~\ref{fig:ConicMono}. We can in fact eliminate $C_{22}$ and $C_{33}$ by shifting the coordinates $x_2$ and $x_3$, giving us the equation
\begin{equation}
    p =  C_{11}x_{1}^2 + 2 C_{12}x_{1}x_{2} + 2 C_{13}x_{1}x_{3} + 2 C_{23}x_{2}x_{3} = 0\, ,.\label{eq:monodefrel}
\end{equation}
The two sections now take the form $x_{1}=x_{2}=0$ and $x_{1}=x_{3}=0$. These are not the only sections, as there are additional sections such as $C_{11}x_{1}+2C_{13}x_{3} = x_{2}=0$ and $C_{11}x_{1}+2 C_{12}x_{2} = x_{3}=0$. But the sections $x_{1}=x_{2}=0$ and $x_{1}=x_{3}=0$ have the attractive feature that they do not intersect and are therefore natural choices for the divisors that carry the heterotic $E_8$ factors. 
An example of such a conic bundle with a base $\mathbb{P}^2_y$ is given by the GLSM matrix
\begin{equation}
\label{eq:conicbase}
\begin{array}{cccccc|c}
x_1 &x_2 &x_3 & y_0 &y_1 & y_2 & p\\\hline
1   &   1  &   1  &   0   &   0   &   0   &  2 \\
0  &    d  &   d  &   1   &   1   &   1   &  2d \\
\end{array}
\end{equation}
The Hodge and Euler numbers for this conic bundle are
\begin{align}
\chi_{\conicbundle}=&6 + 6 d - 4 d^2 & h^{1,1}(\conicbundle) =& 3 & h^{2,1}(\conicbundle)=&  (d-1) (2 d - 1)) \, .\label{eq:conicbasehodge}
\end{align}
Note that $h^{1,1}(\conicbundle)$ is greater than $1+h^{1,1}(B_2)$; this will be important in the next subsection. 

One can also find a conic bundle with sections by tuning the complex structure. We can globally set one or more $C_{ij}$ to zero to obtain a (potentially singular) conic bundle with sections. For instance, we can set $C_{33}=0$ in \eref{eq:conic}, resulting in sections at $x_1=x_2=0$ and $x_1 = C_{22}x_2+2C_{23}x_3=0$. These sections in fact intersect at $x_{1}=x_{2}=C_{23}=0$, a somewhat undesirable property from a heterotic/F-theory perspective: one assumes that the sections do not intersect in more typical heterotic/F-theory setups. Of course, we can still construct valid F-theory models with this conic bundle, even if it is more difficult to understand the dual heterotic physics. This conic bundle is also singular at $x_{1}=x_{2}=C_{13}=C_{23}=0$, as can be seen from the conifold structure of the conic bundle defining relation:
 \begin{equation}
  x_{1}\left(C_{11}x_1+2C_{12}x_2+2C_{13}x_3\right) + x_{2}\left(C_{22}x_{2}+2C_{23}x_{3}\right). \label{eq:conicbundlesecalt}
 \end{equation}
Note that the rank of $C_{ij}$ is 2 at $\{C_{13}=C_{23}=0\}$ even though the degeneration locus
\begin{equation}
    \det(C) = -C_{22} C_{13}^2+2 C_{12} C_{23} C_{13}-C_{11} C_{23}^2 = 0
\end{equation}
is singular at $C_{13}=C_{23}=0$. By the conditions for regular conics described in Section \ref{sec:sarkisov}, this conic bundle is not regular. The resolution of these singularities and a discussion of the resulting Hodge numbers, as part of a more general discussion of the degenerate conic fibers over the descriminant (\ref{eq:discriminantconic}), is the subject of the next subsection.

\subsection{Degenerate conic fibers}
\label{sec:degconics}

In general, conic bundles, including the three conic bundles introduced in the previous subsection, may have fibers that degenerate into a union of two rational curves along a discriminant locus in the base. We have already discussed the physical interpretation of these degenerations in F-theory models in 6-dimensions, where the conic bundle is (complex) 2-dimensional. The degenerations, which contribute to $h^{1,1}$ of the conic bundle, signal that the dual heterotic/M-theory model has M5 branes in the bulk of the $S^1/\mathbb{Z}_2$ interval. The modulus that controls the volume of the extra rational curve corresponds to the position modulus of this M5 brane along the $S^1/\mathbb{Z}_2$ interval. One would expect many aspects of this interpretation to carry over to 4-dimensional F-theory models, where the conic bundle is a complex threefold. Specifically, the dual heterotic model would involve an elliptic fibration over the conic bundle base $B_2$. In the M-theory dual of this heterotic model, we would expect to have an M5 brane in the bulk of the $S^1/\mathbb{Z}_2$ interval lying along the discriminant locus. But in 4-dimensions there is the additional complication that degenerations at codimension-one do not contribute to $h^{1,1}(\conicbundle)$ when there is monodromy. If one tracks the two rational curves along a closed path in the discriminant locus, one many find that they are exchanged after returning to the starting point. 
This effect identifies the two components of the degenerate fibers, and one cannot shrink one of the rational curve components without shrinking the entire fiber. The volume of the extra rational curve is then not a true modulus, and one should not expect additional contributions to $h^{1,1}$.  

As an example, consider a conic bundle described by the defining relation
\begin{equation}
p = C_{11}x_1^2 + C_{22}x_2^2 + C_{33}x_3^2.
\end{equation}
As before, $[x_1:x_2:x_3]$ are coordinates of the $\mathbb{P}^2$ ambient space in which the conic fiber is embedded, and the $C_{ij}$ are sections of line bundles over $B_2$. The conic fiber degenerates into two rational curves along the codimension-one loci $C_{11}=0$, $C_{22}=0$, and $C_{33}=0$.
To explicitly see the degeneration along, for instance, $C_{11}=0$, one can plug $C_{11}=0$ into the defining relation to obtain
\begin{equation}
p\Big|_{C_{11}=0}= C_{22} x_{2}^2 + C_{33}x_{3}^2.
\end{equation}
If we restrict attention to a single point in the base along $C_{11}=0$, the coefficients $C_{22}$ and $C_{33}$ are essentially numbers, and the quadratic $C_{22} x_{2}^2 + C_{33}x_{3}^2$ seems to factor into two components:
\begin{equation}
p\Big|_{C_{11}=0}= \left(\sqrt{C_{22}} x_{2} + i \sqrt{C_{33}} x_{3}\right)\left(\sqrt{C_{22}} x_{2} - i \sqrt{C_{33}} x_{3}\right).
\end{equation}
This suggests that, at a single point in the base along $C_{11}=0$, the fiber has degenerated into two components. But if $C_{22}$ or $C_{33}$ are sections of nontrivial line bundles, the quadratic does not truly factorize in a manner which is polynomial in the coordinates of the two dimensional base $B_2$, and the two components should be somehow identified. Indeed, if we move on a path along $C_{11}=0$ that encircles $C_{11}=C_{33}=0$ (traversing a path such that $C_{33}$ becomes $e^{2\pi i}C_{33}$), the two components are exchanged. Because the two components are identified, one cannot independently adjust their volumes and the degeneration locus should not contribute to $h^{1,1}$. 
 \begin{figure}[t!]
\begin{center}
\begin{picture}(0,120)
\put(-130,0){\includegraphics[scale=0.7]{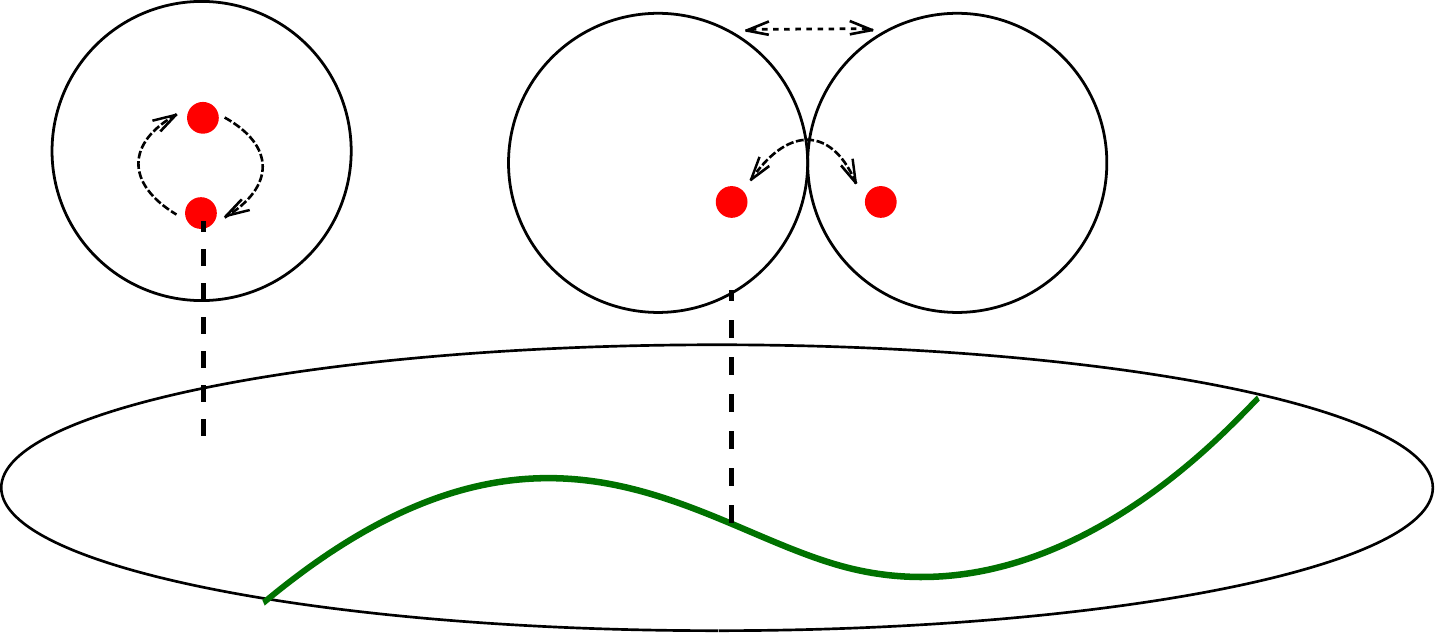}}
 \put(120,25){$C_{ii}=0$} 
\end{picture}
\caption{\label{fig:ConicMonoDegen}\emph{Illustration of a degenerate conic bundle with no section. The two-section monodromy extends to the two fibral curves of the degenerate locus.}}
\end{center}
\end{figure}
The generic conic bundle in \eref{eq:conic} exhibits a similar monodromy effect. The algebra is more involved for the general conic, so we do not go through the detailed procedure of demonstrating this monodromy here. However, the Hodge numbers for the generic conic bundles over a $\mathbb{P}^2$ base, which are listed in \eref{eq:hodgegenconic}, reflect this monodromy. For these examples, $h^{1,1}(\mathbb{P}^2)$ is 1, and $h^{1,1}(\conicbundle)$ is $h^{1,1}(\mathbb{P}^2)+1 = 2$. Therefore, the degenerations do not give additional contribution to $h^{1,1}(\conicbundle)$, as expected for situations with monodromy. An example depiction is given in Figure~\ref{fig:ConicMonoDegen}

 However, the examples most amenable to a dual heterotic interpretation admit sections, and the two examples above with monodromy do not. In some sense, the more pressing question is whether the degenerations contribute to $h^{1,1}(\conicbundle)$ in a conic bundle with sections. Intuitively, one might expect that monodromy does not occur when there are sections. At least at most points along the degeneration locus, the section would hit only one of the components. This would distinguish the two components, suggesting that monodromy could not identify the two rational curves. In turn, one would expect that the codimension-one degeneration loci should contribute to $h^{1,1}$ when there are sections. Of course, sections can behave in more complicated ways at codimension two; for instance, they can wrap rational curves. One might therefore worry that this behavior could spoil this naive argument. Nevertheless, for the two conic bundles with sections described in Section \ref{sec:conicsections}, the codimension-one loci with degenerations contribute to $h^{1,1}(\conicbundle)$, and monodromy does not seem to be present.

Let us first focus on the conic bundle described by \eref{eq:monodefrel}, where $[C_{22}]$, $[C_{33}]$, and $[C_{23}]$ are trivial. For simplicity, we set $C_{23}$ to 1. Over points along the discriminant 
 \begin{align}
 \Delta=-C_{11} + 2 C_{12} C_{13} \, .
 \end{align}
the conic reduces to two components: 
 \begin{align}
 p=2 (C_{13} x_1 + x_2) (C_{12} x_1 + x_3) \, .
 \end{align}
Each component is hit by only one of the sections $x_1=x_2=0$ and $x_1=x_3=0$. The two components are distinguished by the behavior of the sections, and we would expect that this conic bundle does not exhibit monodromy. As a result, $h^{1,1}(\conicbundle)$ should be $1+h^{1,1}(B_2) + 1$, with the additional contribution due to the degenerations over $\Delta=0$. This expectation is borne out in the construction of \eref{eq:conicbase}, where the conic bundle base $B_2$ is a $\mathbb{P}^2$ space. Here, $h^{1,1}(\mathbb{P}^2)$ is 1, so we would expect that $h^{1,1}(\conicbundle)$ is 3. This number is in exact agreement the calculated $h^{1,1}(\conicbundle)$ in \eref{eq:conicbasehodge}. For this conic bundle with sections, we see that the degenerate fibers at codimension-one lead to additional contributions to $h^{1,1}(\conicbundle)$. As mentioned previously, we expect that, in this case, the degenerations signal the presence of a bulk M5 brane in the dual heterotic M-theory model. The extra contribution to $h^{1,1}$, which represents the ability to change the volume of the extra fiber component along the discriminant locus in the F-theory model, should correspond to the M5 brane position modulus in the dual model.

\begin{figure}[t!]
\begin{center}
\begin{picture}(0,120)
\put(-180,30){\includegraphics[scale=0.7]{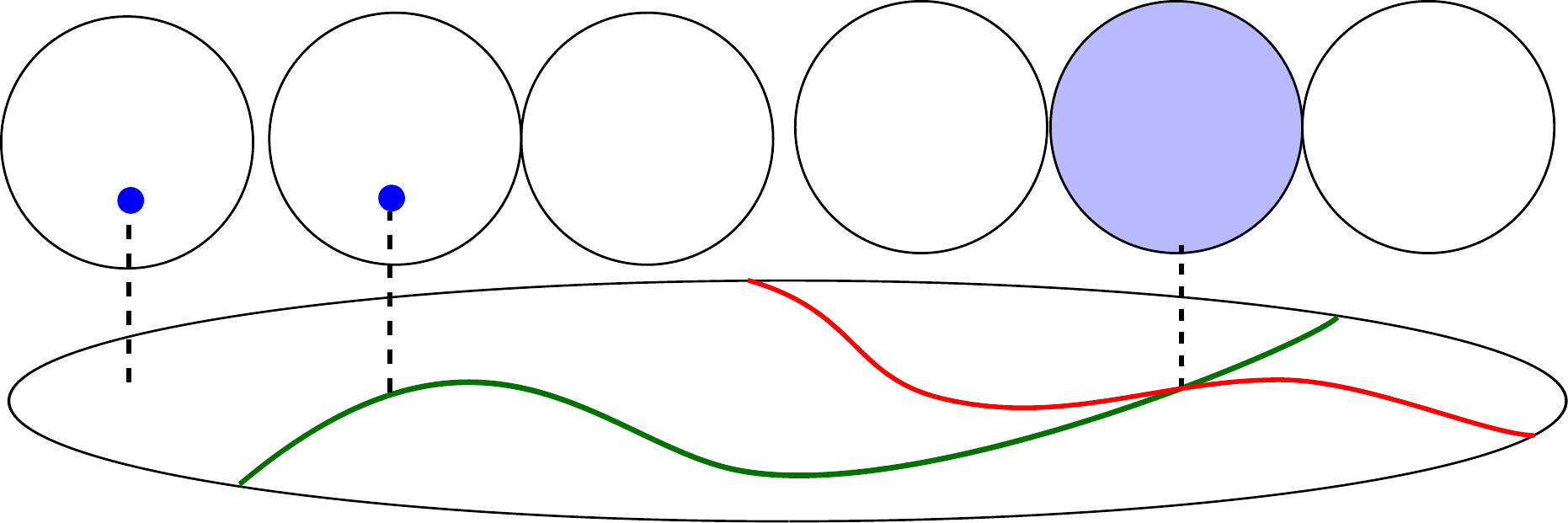}}
 \put(-105,45){$\Delta=0$} 
  \put(135,45){$C_{13}=0$} 
\end{picture}
\caption{\label{fig:BlowUpBundle}\emph{Illustration of the degeneration of the blow-up bundle. Over the discriminant locus it degenerates into two components and in codimension two into three components, one is wrapped by the section.}}
\end{center}
\end{figure}

For the other example with sections, described by \eref{eq:conicbundlesecalt}, we must resolve the singularities at $x_{1}=x_{2}=C_{13}=C_{23}=0$. We enhance the conic ambient space by introducing a new hypersurface $e_1$ and letting $x_1\to e_1\hat{x}_1$ and $x_2\to e_1\hat{x}_2$. In the process, we add $\{\hat{x}_1\hat{x}_2, e_1 x_3\}$ to the Stanley-Reisner ideal. The conic bundle defining relation then becomes
\begin{equation}
 \hat{p} = C_{11} e_{1}\hat{x}_1^2 + 2 C_{12}  e_1 \hat{x}_1\hat{x}_2+ 2 C_{13}\hat{x}_1 x_3 + C_{22}  e_1  \hat{x}_2^2 +2C_{23}\hat{x}_2 x_3 \label{eq:conicbundlesecaltres}
\end{equation}
after a proper transform. The two sections, which were previously $x_{1}=x_{2}=0$ and $x_{1}=C_{22}x_2+2C_{23}x_3=0$, are now given by
\begin{equation}
[e_1: \hat{x}_1:\hat{x}_2:x_3] = [0:-C_{23}:C_{13}:1] \label{eq:sec1conicb}
\end{equation}
and
\begin{equation}
[e_1: \hat{x}_1:\hat{x}_2:x_3] = [-2C_{23}:0:1:C_{22}].\label{eq:sec2conicb}
\end{equation}
These sections still intersect\footnote{There are now two scalings associated to the fiber which ensure that this is true. This is apparent if one considers the GLSM configuration matrix of an example such as that in (\ref{eq:tunedconicexample}).} at $C_{23}=0$. At generic points in the base along the discriminant locus
\begin{equation}
C_{11} C_{23}^2 -2 C_{12} C_{13} C_{23} + C_{22} C_{13}^2=0,
\end{equation}
the fiber still consists of two rational curves. For example, away from $C_{13}=0$, we can write these components as
\begin{equation}
2 C_{13}^2  x_3 + e_1\left(C_{11} C_{13} \hat{x}_1 + 2 C_{12} C_{13}\hat{x}_2 - C_{11}C_{23}\hat{x}_2\right) = 0
\end{equation}
and 
\begin{equation}
C_{13}\hat{x}_1 + C_{23} \hat{x}_2 = 0.
\end{equation}
The two different components are hit by the two different sections. At $C_{23}=C_{13}=0$, the codimension-two locus in the base where the singularity occurred prior to resolution, the defining relation takes the form
\begin{equation}
p|_{C_{23}=C_{13}=0}=e_1\left(C_{11} \hat{x}_{1}^2 + 2 C_{12} \hat{x}_{1} \hat{x}_{2} + C_{22} \hat{x}_2^2\right).
\end{equation}
After resolution, the fiber here is now the union of three rational curves: $e_1=0$ and the two rational curves found by factoring $C_{11} \hat{x}_{1}^2 + 2 C_{12} \hat{x}_{1} \hat{x}_{2} + C_{22} \hat{x}_2^2=0$. The extra curve is that associated to the resolution of the conifold singularities. The $\hat{x}_1=0$ section described by   \eref{eq:sec2conicb} hits the $e_1=0$ component at a point. The expressions for the $e_1=0$ section in  \eref{eq:sec1conicb}, however, is ill-defined at $C_{23}=C_{13}=0$, thanks to the Stanley-Reisner ideal. This section in fact wraps the $e_1=0$ component at  $C_{23}=C_{13}=0$ as shown in Figure~\ref{fig:BlowUpBundle}.

As with the previous example, the two rational curves at codimension one in the base are not exchanged under monodromy. The codimension-one degenerations should therefore contribute to $h^{1,1}$. We can verify this by considering a situation described by the GLSM configuration matrix
  \begin{align}
\begin{array}{cccc|ccc|c} 
x_1 & x_2 &x_3 & e_1 & y_1 & y_2 & y_3 & \hat{p} \\ \hline
1 & 1&1 &0&   0 & 0 & 0& 2 \\ 
0 & 0 & 1 &1 & 0 & 0 & 0 & 1 \\ 
0 & 0 & 0 &0 & 1 & 1 & 1 & 2 \\ 
\end{array} \label{eq:tunedconicexample}
\end{align} 
Here, the base $B_2$ is a $\mathbb{P}^2$ with coordinates $[y_0:y_1:y_2]$, which has an $h^{1,1}$ of 1. The Hodge numbers for this conic bundle are
\begin{align}
h^{1,1}(\conicbundle) =& 3 & h^{2,1}(\conicbundle) =& 6.
\end{align}
We see that $h^{1,1}(\conicbundle)$ is one greater than $h^{1,1}(B_2)+1$, confirming that the degenerations contribute to $h^{1,1}$. As before, the codimension-one degenerations suggest the presence of a bulk M5 brane in the dual heterotic/M-theory picture. 

To summarize, it appears that when a conic bundle admits sections, the degenerations at codimension-one are not affected by monodromy. The 6-dimensional interpretation of degenerations in terms of dual M5 branes in the bulk therefore seems to carry over to these 4-dimensional models without issue. The codimension-one degenerations in these models also contribute to $h^{1,1}$ of the conic bundle. It is important to note, however, that even in models without sections, there can be codimension-one degenerations unaffected by monodromy that do contribute to $h^{1,1}$. Whenever the conic bundle lacks sections, much of the conventional intuition for heterotic/F-theory duality may no longer be applicable. Since those conic bundles exhibiting monodromy also seem to lack sections, understanding the heterotic interpretation of degenerations affected by monodromy would require a broader understanding of if/how heterotic/F-theory duality holds when there are no sections (see \cite{Heckman:2013sfa} for related ideas). Thus, while we may have a heterotic interpretation of the degenerations unaffected by monodromy, degenerations affected by monodromy seem more difficult to understand from a heterotic perspective. Perhaps conifold transitions that introduce sections, such as the one described above, could shed some light on possible heterotic duals of these models.

 \subsection{Non-flat conic fibers}
It is also possible to engineer a conic bundle with non-flat fibers at codimension two in the base. Consider the following GLSM charge matrix.
  \begin{align}
\begin{array}{ccc|cc|ccc|c }
x_1  & x_2 &x_3 & e_1& e_2 & y_0 & y_1 & y_2 & \hat{p} \\ \hline
1 & 1&1 &0&   0 & 0 & 0&0 & 2 \\ 
-1 & 0 & 0 &1 & 0 & -1 & 0 & 0 & -1 \\ 
0 & -1 & 0 &0 & 1 & -1 & 0 & 0 & -1 \\  
0 & 0 & 0 &0 &0 & 1 & 1 & 1 & 2 \\ 
\end{array}\, 
\end{align}
The ambient variety admits two smooth triangulations. We chose one of them, leading to a Stanley-Reisner ideal of
\begin{align}
\{ x_1 y_0, x_2 e_1, x_2y_0, x_1x_2x_3, e_1y_1y_2, y_0y_1y_2, y_1y_2e_2, x_1x_3e_2, x_3e_1e_2 \}\, .
\end{align}
The hypersurface is specified by the equation
\begin{align}
\hat{p}=C_{11} x_1^2 y_0 e_1^2 + C_{22} x_2^2 y_0 e_2^2 + C_{33} y_0 x_3^2 + 2C_{12} x_1 x_2 + 
 2C_{13} e_1 y_0 x_1 x_3 + 2C_{23} x_2 x_3  y_0 e_2 \, .
\end{align}

This conic bundle has degenerate fibers along the codimension-one locus $y_0=0$ in the base. At generic points along this locus, the conic fiber consists of two rational curves. However, when both $y_0=0$ and $C_{12}$=0, we find $\hat{p}$ to be trivially fulfilled and the coordinates $\{ x_3, e_1, e_2\}$ to be unconstrained. Hence, the conic fiber becomes a surface, namely a whole $\mathbb{P}^2$. The number of points in the base where this happens is given by the intersection number $n_{nf}=[y_0]\cdot [C_{12}]$.  

This structure has an important impact on the hodge numbers. For the conic bundle at hand,
\begin{align}
h^{1,1} &= 5 &\textnormal{and} && h^{2,1} = 3.
\end{align} 
There should be a contribution to $h^{1,1}$ of $1+h^{1,1}(\mathbb{P}^2) =2$ which is analogous to that seen in $\mathbb{P}^1$ bundles, and the codimension-one degeneration locus at $y_0=0$ should also contribute 1 to $h^{1,1}$. This leaves us with an excess contribution to $h^{1,1}$ of 2 due to the non-flat fibers. Each non-flat fiber is a new divisor in the conic bundle, so the presence of non-flat fibers should contribute $n_{nf}$ to $h^{1,1}$. In this case, $C_{12}$ is of degree 2 in the $y_i$ coordinates, suggesting that $n_{nf}=2$. Thus, the non-flat fibers exactly account for the additional contributions to $h^{1,1}$. Similarly, if we change  $[\hat{p}]$ to $[2,-1,-1,d]$, the Hodge numbers for various values of $d$ are
\begin{align}
\begin{array}{|c|c|c|}\hline
d & h^{11}& h^{2,1}  \\ \hline
2 & 5 & 3 \\ \hline
3 & 6 & 17 \\ \hline
4 & 7 & 40 \\ \hline
\end{array}
\end{align}
In general, $C_{12}$ should be of degree $d$ in the $y_i$ coordinates, and $n_{nf}$ should be $d$. If we increase $d$ by 1, we see that $h^{1,1}$ also increases by 1, in line with the statement that each non-flat fiber should contribute $1$ to $h^{1,1}$. 

The non-flat nature of this conic bundle agrees with the conditions outlined in Section \ref{sec:sarkisov}. If we consider the blown-down version of this conic bundle with $e_1$ and $e_2$ set to 1, we have
\begin{equation}
    C_{ij} = \begin{pmatrix}C_{11} y_0 & C_{12} & C_{13} y_{0}\\C_{12} & C_{22} y_0 & C_{23} y_{0} \\ C_{13} y_{0} & C_{23} y_{0} & C_{13} y_{0}\end{pmatrix}.
\end{equation}
Clearly, the rank of $C_{ij}$ is 0 at $\{C_{12}=y_{0}=0\}$, so conic bundle cannot be flat according to the conditions in Section \ref{sec:sarkisov}.

\subsection{Elliptic fourfold completions} 
We now discuss how to build elliptic fibrations using conic bundles bases. Since most of the examples considered in this section result from specializations of the generic conic bundle in \eref{eq:conic}, let us first consider building elliptic fibrations over this conic bundle. As in Section \ref{sec:conicnomono}, we assume that the elliptic fibration is in the Weierstrass form
\begin{equation}
    Y^2 = X^3 + f X Z^4 + g Z^6,
\end{equation}
where $f$ and $g$ are sections of $\mathcal{O}(-4K_{\conicbundle})$ and $\mathcal{O}(-6K_{\conicbundle})$, respectively.  Before writing out expressions for $f$ and $g$, we need to calculate $-K_{\conicbundle}$ or, alternatively, $c_{1}(\conicbundle)$. The conic bundle is embedded in a $\mathbb{P}^2$ bundle over $B_2$, with $[x_1:x_2:x_3]$ serving as the coordinates of the $\mathbb{P}^2$. We take $x_1$ to be a section of the line bundle $\mathcal{L}_1$ on the $\mathbb{P}^2$ bundle. Meanwhile, $x_2$ and $x_3$ are respectively sections of $\mathcal{L}_1\otimes\mathcal{O}([C_{13}]-[C_{23}])$ and $\mathcal{L}_1\otimes\mathcal{O}([C_{12}]-[C_{23}])$, where the $[C_{ij}]$ are divisor classes on $B_2$. From the adjuction formula, the total Chern class for $\conicbundle$ is
\begin{equation}
    c(\conicbundle) = c(B_2) \frac{\left(1+c_{1}(\mathcal{L}_1)\right)\left(1+c_{1}(\mathcal{L}_1)+[C_{13}]-[C_{23}]\right)\left(1+c_{1}(\mathcal{L}_1)+[C_{12}]-[C_{23}]\right)}{1+2c_{1}(\mathcal{L}_1) + [C_{12}] +[C_{13}] - [C_{23}]},
\end{equation}
and as a result
\begin{equation}
    c_1(\conicbundle) = c_{1}(B_2) + c_{1}(\mathcal{L}_{1}) - [C_{23}].
\end{equation} 
The expression for $c_1(\conicbundle)$ suggests that we can expand $f$ and $g$ as series of orders $4$ and $6$ in the $x_i$:
\begin{align}
    f =& \sum_{i=0}^{4}\sum_{j=0}^{4-i}f_{i,j} x_{1}^ix_{2}^j x_{3}^{4-i-j} & g=& \sum_{i=0}^{6}\sum_{j=0}^{6-i}g_{i,j} x_{1}^i x_{2}^j x_{3}^{6-i-j}.
\end{align}
Of course, we must also include \eref{eq:monodefrel} to properly describe the elliptic fibration. 

As an example, consider the conic bundle described by  \eref{eq:conicbase} over a base $\mathbb{P}^2$. Recall that this example is essentially a special case of the generic conic bundle where $[C_{22}]$, $[C_{23}]$, and $[C_{33}]$ are trivial, leading to sections and a degeneration locus not subject to monodromy. We can build an elliptic fibration using the configuration matrix 
\begin{align}
\begin{array}{ccc|ccc|ccc| c c }
X&Y&Z& x_1 & x_2 & x_3 & y_0 & y_1 & y_2&    P_w & P_b    \\ \hline
2 & 3  & 1 & 0 & 0 & 0 &0 & 0 & 0  &6& 0  \\ 
0 & 0 & -1 & 1 & 1 & 1 & 0& 0 & 0&    0 & 2   \\ 
0 & 0 & -3 & 0 & d &d & 1 & 1 & 1 &   0& 2d  \\ 
\end{array}\label{eq:c23trivconicellfib}
\end{align}
The Hodge and Euler numbers of the full elliptic fibration for $d=1,2$ are given below:
\begin{equation}
\begin{array}{|c|c|c|c||c|c|}  \hline
\multicolumn{1}{|c|}{d}& h^{1,1}& h^{2,1} & h^{3,1}& h^{2,2} & \chi \\ \hline
1 & 4 & 0 & 2316& 9324 & 	13968 \\ \hline
2& 4 & 3 & 1599 & 6450 & 9648 \\ \hline 
\end{array}
\end{equation}
As written, $P_w$ has  $X Y Z$, $Y Z^3$ and $X^2 Z^2$ terms, and the elliptic fibration is therefore in Tate form. However, $P_w$ can easily be converted to Weierstrass form by shifts of $X$ and $Y$. For this conic bundle, the Weierstrass polynomials are of degrees $deg[f_{i,j}] =12-d(4-i)$ and $deg(g_{i,j})= 18-d(6-i)$ in the coordinates of the $\mathbb{P}^2$ base for all $j$. 
 
We can also use the toric description to further investigate how contributions to the cohomology groups of the fourfold depend on the conic bundle base.  In particular, we want to examine how the cohomology groups change when we tune in a section to the conic bundle. Let us focus on a specific example. We start with a fourfold $Y_4$ over a generic conic bundle whose base is $\mathbb{F}_0$. The conic bundle, described by an equation of the form \eref{eq:conic},  does not admit a section. However, we can tune sections in the conic bundle and resolve the resulting singularities, leading to a conic bundle defining relation of the form \eref{eq:conicbundlesecaltres}. In the process, the conic fiber ambient space changes from $\mathbb{P}^2$ to $\text{dP}_1$. The GLSMs for the full elliptic fibration are given below; the entries highlighted in red are those introduced by the resolution procedure. 
\begin{align}
\label{eq:tateConF0}
\begin{array}{ccc|ccc|cc|cc|c|c c }
X&Y&Z& x_1 & x_2 & x_3 & s_0 & s_1 & t_0&  t_1 & \textcolor{red}{e_1} & P_w & P_{b_1}    \\ \hline
2 & 3  & 1 & 0 & 0 & 0 &0 & 0 & 0 & 0 & \textcolor{red}{0} &6& 0  \\ 
0 & 0 & -1 & 1 & 1 & 1 & 0& 0 & 0&0 &  \textcolor{red}{0}& 0 & 2   \\ 
0 & 0 & -1 & 0 & 0 &0 & 1 & 1 & 0 & 0 & \textcolor{red}{0} & 0& 1  \\  
0 & 0 & -1 & 0 & 0 &0& 0 & 0 & 1 & 1 &  \textcolor{red}{0}  &  0 & 1 \\   
 \textcolor{red}{0} &  \textcolor{red}{0} &  \textcolor{red}{0}&  \textcolor{red}{-1} &  \textcolor{red}{-1} & \textcolor{red}{0}& \textcolor{red}{0} &  \textcolor{red}{0} &  \textcolor{red}{0} &  \textcolor{red}{0} &  \textcolor{red}{1}  &   \textcolor{red}{0} &  \textcolor{red}{-1} \\ 
\end{array}
\end{align}
Again, while the elliptic fibration as written is in Tate form, it can easily be converted to Weierstrass form by shifts in $X$ and $Y$. The Hodge and Euler numbers for the fourfold before the tuning and resolution are
\begin{align}
  h^{1,1}&= 4 & h^{2,1}&=3 & h^{3,1}&= 1119 & h^{2,2}=& 4530 & \chi =& 6768. 
\end{align}
Afterwards, they are
\begin{align}
  h^{1,1}&= 5 & h^{2,1}&=2 & h^{3,1}&= 1117 & h^{2,2}=& 4528 & \chi =& 6768.
\end{align}
The change in $h^{1,1}$ reflects the fact that, after tuning the sections, there is an extra codimension-one locus present due to the absence of monodromy effects in the conic bundle base. Notably, the transition changes the Hodge numbers in a way that preserves the Euler number.  This behavior agrees with the observations in  \cite{Greiner:2017ery,Oehlmann:2021man}.

\subsection{Tuning gauge groups}
As elliptic fibrations built over conic bundles give rise to well-defined F-theory models, we should be able to tune in gauge groups in the usual fashion. In fact, tuning nonabelian gauge groups on sections of conic bundles is particularly important for understanding heterotic duals. The tuning process is not significantly different for these conic bundle models to more standard cases in the literature. However,  the tunings may appear somewhat unfamiliar compared to the tunings seen for $\mathbb{P}^1$ bundles, primarily because $f$ and $g$ are of orders $4$ and $6$ in the $x_i$ coordinates.

For example, let us consider an elliptic fibration over the conic bundle described by \eref{eq:monodefrel} with trivial $[C_{23}]$. Recall that this conic bundle admits two sections: $x_{1}=x_{2}=0$ and $x_{1}=x_{3}$. An example of such an elliptic fibration is described by the GLSM matrix in \eref{eq:c23trivconicellfib}, and we can perform variable shifts to ensure that the elliptic fibration is in Weierstrass form. As before, $f$ and $g$ can be expanded as
\begin{align}
    f =& \sum_{i=0}^{4}\sum_{j=0}^{4-i}f_{i,j} x_{1}^ix_{2}^j x_{3}^{4-i-j} & g=& \sum_{i=0}^{6}\sum_{j=0}^{6-i}g_{i,j} x_{1}^i x_{2}^j x_{3}^{6-i-j}.
\end{align}
Suppose we want to tune an $E_8\times E_7$ gauge symmetry, with the $E_8$ group supported on $x_1=x_2=0$ and the $E_7$ group supported on $x_1=x_3=0$. From the Kodaira classification, $f$ should vanish to order 4 on $x_1=x_2=0$ and to order 3 on $x_1=x_3=0$. This can be accomplished by tuning $f$ to have only $x_1^4$ and $x_1^3 x_2$ terms . Similarly, $g$ should vanish to order 5 on both $x_1=x_2=0$ and $x_1=x_3=0$, which can be accomplished tuning $g$ to be proportional to $x_1^5$. In the end, $f$ and $g$ take the form
\begin{align}
f=& f_{3,1} x_1^3 x_2+ f_{4,0}x_1^4 & g=& g_{5,1} x_1^5 x_2+g_{6,0} x_1^6 +  g_{5,0} x_1^5 x_3 .
\end{align}

As a second instructive example, we consider tuning a gauge group over a two-section of the generic conic bundle and then tuning the conic bundle to convert the two-section to two genuine sections. We start with the Tate model as specified in \eref{eq:tateConF0} in the phase where we have a conic; in other words, this is the phase with $h^{1,1}=4$. We consider the Tate model
\begin{align}
P_w = Y^2 + X^3 + XYZ a_{[1,1,1]} + X^2 Z^2 a_{[2,2,2]} + Y Z^3 a_{[3,3,3]} + X Z^4 a_{[4,4,4]} + Z^6 a_{[6,6,6]} \, .  
\end{align}
with the $a_{[i,i,i]}$ being of triple degree $i$ in the various base classes.  We tune an $\SU(5)$ on the two-section $x_1=0$ by making the  $a_{[i,i,i]}$ coefficients proportional to appropriate powers of $x_1$:
\begin{align}
 a_{[2,2,2]} \rightarrow& x_1 b_{[1,2,2]} & a_{[3,3,3]} \rightarrow& x_1^2 b_{[1,3,3]} & a_4 \rightarrow& x_1^3 b_{[1,4,4]}& a_6 \rightarrow& x_1^5 a_{[1,6,6]} \, .
\end{align}
We then tune in a section of the conic by setting $C_{33}$ to 0. As discussed previously,  the conic bundle has additional $\mathbb{Z}_2$ singularities at $x_{1}=x_{2}=C_{13}=C_{23}=0$, which can be resolved by letting $x_1 \rightarrow x_1 e_1, x_2 \rightarrow x_2 e_1$. The tuning forces the previously irreducible $\SU(5)$ divisor $x_1=0$ to split into two components given by 
\begin{equation} \{ x_1= C_{22} e_1 x_2+ 2 C_{23} x_3=0 \} 
\end{equation}
and
\begin{equation}
\{ e_1= C_{13} x_1 + C_{23} x_2=0 \}  .
\end{equation}
The gauge algebra has therefore enhanced from $\SU(5)$ to $\SU(5)\times\SU(5)$. These components still intersect over the codimension two locus $x_1 = e_1 = C_{23}=0$, in line with the previous observations that the two sections intersect.  The orders of vanishing of $(f,g,\Delta)$ enhance to $(0,0,10)$ at this locus, suggesting that the intersection locus supports bi-fundamental matter.

\vspace{0.5cm}

In general, the effects noted above raise interesting questions about the possible heterotic duals of these theories. At the locus in moduli space where the two sections exist, the gauge symmetry on each section indicates an $SU(5)$ symmetry arising from heterotic bundles on each heterotic fixed plane (of the $S^1/\mathbb{Z}_2$ interval of heterotic M-theory). We can then ask about what the heterotic effect would be of the CY 4-fold deformation that recombines these two sections into a multi-section supporting a single $SU(5)$ symmetry? This is clearly outside the usual weakly coupled heterotic/F-theory correspondence. However, at an intuitive level, this merging of sections seems to indicate some identification/merging of heterotic fixed planes/bundles. This is similar in spirit to effects seen in dimensional reductions of the CHL string \cite{Chaudhuri:1995fk,Chaudhuri:1995bf,Park:1996it}. It is beyond the scope of the present work to fully study log semi-stable degenerations and heterotic/F-theory duality in this context. However it would be interesting to consider heterotic stable degeneration limits of these geometries in more detail in the future.

\section{Conclusions and future directions}
In this work we have explored $\mathbb{P}^1$-fibered manifolds as bases for CY elliptic fibrations within F-theory. Our goals were to clearly outline the properties of such base manifolds and their features of interest for F-theory (and heterotic/F-theory duality). The geometry of the simple $\mathbb{P}^1$-fibrations considered in this work (both $\mathbb{P}^1$-bundles and conic bundles) is very well understood in the mathematics literature, but many of aspects of this literature have remained unused within F-theory compactifications. In this work we have summarized key results in a context accessible to physicists and have demonstrated that F-theory compactifications on these spaces lead to a number of novel new physical features in both $6$ and $4$-dimensions.

The majority of work in heterotic/F-theory duality has restricted consideration to $n$-fold bases ${\cal B}_n$ to the CY elliptic fibration that are of the form ${\cal B}_n=\mathbb{P}(\pi: {\cal O} \oplus {\cal O}(D) \to B_{n-1})$. In the previous sections we have demonstrated that even in the context of bases ${\cal B}_n$ which are $\mathbb{P}^1$-bundles, more general choices of vector bundle over ${B}_{n-1}$ are possible than that described above, in 4-dimensional compactifications of F-theory. Moreover, they can come equipped with rational rather than holomorphic sections in 4-dimensions. Even in the simple context of $6$-dimensional compactifications of F-theory we have uncovered previously unexplored physics from simple geometries -- the Hirzebruch surfaces, $\mathbb{F}_n$, and the transitions between them. In particular we find new F-theory transitions (facilitated by superconformal loci) that allow the F-theory base geometries to change. These transitions were not visible in the simple toric descriptions utilized previously. 

More generally, the new $3$-dimensional base geometries for CY 4-folds we describe here in Sections \ref{Sec:4dftheory}-\ref{sec:conicmonodromy} are of interest because of their ubiquitous nature in the landscape of possible elliptic CY base manifolds (see \cite{Di_Cerbo_2021,birkar2020boundedness} characterizations of CY 4-fold base geometries in the context of the minimal model program). By studying their properties we enhance the known datasets of elliptic CY manifolds studied in F-theory and hence the possible effective fields that can be obtained. Such progress is crucial, especially in $4$-dimensional compactifications of F-theory, since the effective physics that can be obtained from such theories has been proven to sweep out vast swathes of the string landscape and hence has played a key role in recent investigations into the string Swampland.  

In addition, the novel geometry we have built in this paper demonstrates that there are many new ``weakly coupled" limits of F-theory that should -- in principle -- be related to heterotic string theory. In particular, the results of Section \ref{Sec:4dftheory} indicate that rational sections to the $\mathbb{P}^1$-fibration seem to change the nature of heterotic/F-theory duality by forming an elliptic CY threefold heterotic dual that is defined over the zero locus of the rational section rather than the base to the $K3$ fibration. It would be interesting to study heterotic/F-theory duality (and the detailed mapping of moduli spaces) in more detail for these examples. We hope these results can shed some light on how this familiar duality may be extended.

In Sections \ref{sec:conicnomono} and \ref{sec:conicmonodromy} we discussed the role that can be played by degenerations of the $\mathbb{P}^1$ fibration of the base of the elliptic fibration in compactifications to 4-dimensions. We saw that, in the absence of monodromy, the discriminant locus in the base $B_2$ is associated to the locus wrapped by 5-branes in the dual heterotic theory as expected. In the presence of monodromy associated to the $\mathbb{P}^1$ components of the degenerate fibers however, such an interpretation does not seem to hold.

\vspace{0.5cm}

There are a number of important open questions, one of which is how the effective physics of the jumping transitions of Section \ref{sec:6DFtheory} can be understood more directly, including the nature of the superconformal sectors that generally appear to facilitate the transitions between $\mathbb{F}_1 \to \mathbb{F}_3$ (and between $\mathbb{F}_0 \to \mathbb{F}_2$ when certain gauge groups are tuned in the Weierstrass model). Focusing on the $\mathbb{F}_1 \to \mathbb{F}_3$ transition for a moment, it should be noted that the jumping of the effective cone that occurs in this transition must change the spectrum of BPS strings. If one wraps a $D3$ brane along the curve of self-intersection $-3$ in the base, one obtains a half-BPS string\footnote{This string will of course be massive until the $-3$ curve is shrunk to zero size (a limit which is available while keeping the volume of $\mathbb{F}_3$ finite).}  in the $6$-dimensional $(1,0)$ supergravity theory whose charge vector has self-intersection $-3$. This BPS string\footnote{For brevity we refer to half-BPS strings as simply BPS strings. We also refer to the string found by wrapping a $D3$ brane on a $-n$ curve as a ``$(-n)$-string".} is seemingly no longer available once the complex structure has been deformed and the $-3$ curve ceases to be effective. Clearly the states associated to this changing $(-3)$-string must experience a decay/recombination in the spectrum. Such a transition is likely an example of ``wall-crossing" phenomena and deserves to be studied in detail. Note that in the limit that the $-3$ curve supports only an $SU(3)$ gauge group (i.e. the usual ``non-higgsable cluster" over $\mathbb{F}_3$) no transition is possible. The transition is only possible when the theory is tuned to support an $SO(8)$ gauge symmetry, intersecting another component of the discriminant (with $(f,g,\Delta)$ vanishing to orders $(4,6,12)$). The intersection point is indicative of ``superconformal matter". The case of $SO(8) \times SO(8)$ tuning mentioned in Section \ref{f1f3sec} is suggestive in that in this context it has previously been observed that superconformal matter charged under $SO(8) \times SO(8)$ can be Higgsed to force the two $SO(8)$ factors to recombine \cite{Del_Zotto_2015}. In the context explored here this Higgsing process facilitates the transition $\mathbb{F}_3 \to \mathbb{F}_1$ (and leaves a single $SO(8)$ symmetry supported on the $-1$ curve inside $\mathbb{F}_1$). It would be interesting to explore the effective physics of these transitions both from the point of view of the SCFT physics as well as the BPS strings described above. We hope to return to such questions in the future.

\section*{Acknowledgments}

The authors would like to thank Tony Pantev, Seung-Joo Lee and Washington Taylor for useful discussions. The work of M.K. was supported by IBS under the project code, IBS-R018-D1. The work of P.K.O.  is supported by a grant of the Carl Trygger Foundation for Scientific Research. The work of L.A. and J.G. is supported in part by NSF grant PHY-2014086. 

\appendix
\section{Some topology of $\mathbb{P}^1$ bundles} \label{sometop}

In the main text we will require a number of straightforward topological results concerning the structure of projective bundles and their sections. We derive these in this appendix.

\vspace{0.1cm}

Let us begin by computing some results concerning the topology of projective bundles that are a $\mathbb{P}^1$ bundle over a 2-dimensional base. Consider the usual decomposition of the tangent bundle of a fibration \cite{HARTSHORNE}, applied to the case of a 3-dimensional projective bundle. 
\begin{eqnarray} \label{tang}
0 \to T_{{\cal B}_3|{ B}_2} \to T{\cal B}_3 \to \pi^*(T{ B}_2) \to 0
\end{eqnarray}
From (\ref{tang}), we see that
\begin{eqnarray} \label{tang2}
c_1(T{\cal B}_3 )=c_1(T_{{\cal B}_3|{ B}_2} ) +c_1(\pi^*(T{ B}_2)) \;.
\end{eqnarray}
We can then use an Euler sequence describing the relative tangent bundle $T_{{\cal B}_3|{ B}_2}$ \cite{HARTSHORNE}.
\begin{eqnarray} \label{euly}
0 \to {\cal O}_{{\cal B}_3} \to \pi^* (V_2^{\vee}) \otimes {\cal O}(S) \to T_{{\cal B}_3|{ B}_2} \to 0
\end{eqnarray}
This enables us to write,
\begin{eqnarray} \label{euler}
c_1(T_{{\cal B}_3|{ B}_2} ) = 2 c_1({\cal O}(S)) - c_1(\pi^*V_2) \;.
\end{eqnarray}
Combining (\ref{tang2}) and (\ref{euler}) and using that $c_1(\pi^*V_2)=\pi^*c_1(V_2)$ and $c_1({\cal O}(S))=S$ we find the following.
\begin{eqnarray} \label{c1b3}
c_1(T{\cal B}_3) = 2S - \pi^*c_1(V_2)+\pi^*c_1(T{ B}_2)
\end{eqnarray}
A similar analysis, using the same sequences (\ref{tang}) and (\ref{euly}) given above, results in the following two expressions for the other Chern characters.
\begin{eqnarray} \label{cotherb3}
\textnormal{ch}_2(T{\cal B}_3)&=&\pi^*\textnormal{ch}(V_2)-\pi^*c_1(V_2)\cdot S +S\cdot S+\pi^*\textnormal{ch}_2({ B}_2) \\ 
\textnormal{ch}_3(T{\cal B}_3)&=&-\frac{1}{2} \pi^*c_1(V_2)\cdot S\cdot S + \pi^*\textnormal{ch}_2(V_2) \cdot S+\frac{1}{3} S\cdot S\cdot S
\end{eqnarray}
These results for the Chern characters of ${\cal B}_3$ can then be used to trivially compute the Chern classes given in (\ref{mrthechern}). 

\vspace{0.2cm}

Next we will derive an expression for the self intersection of $S$. From adjunction we have the following.
\begin{eqnarray}
0\to TS \to T{\cal B}_3|_S \to {\cal O}_S(S)  \to 0
\end{eqnarray}
This then implies that we have the following relationship between Chern classes.
\begin{eqnarray}
c_1(T{\cal B}_3)\cdot S = c_1(TS) + S\cdot S
\end{eqnarray}
Using the result (\ref{c1b3}), we find the following equation.
\begin{eqnarray} \nonumber
&&S\cdot( 2S - \pi^*c_1(V_2)+\pi^*c_1(T{ B}_2))=c_1(TS)+S\cdot S \\ \label{tang30}
\Rightarrow &&S\cdot S  =S\cdot \pi^*c_1(V_2)-S\cdot \pi^*c_1(T{ B}_2) + c_1(TS)
\end{eqnarray}
If we have a holomorphic section, so that $S$ is diffeomorphic to ${ B}_2$, we have that $S\cdot \pi^*c_1(T{ B}_2) =c_1(TS)$, so that
\begin{eqnarray} \label{inter}
\Rightarrow && S.(S-\pi^*c_1(V_2)) =0 \;.
\end{eqnarray}

\vspace{0.1cm}

There is also a standard result relating the topology of $S$ and $V_2$ for any choice of such divisor and bundle which says \cite{Szurek},
\begin{eqnarray} \label{gen}
S\cdot S-\pi^* c_1(V_2) \cdot S + \pi^* c_2(V_2) =0\;.
\end{eqnarray}
Comparing (\ref{inter}) and (\ref{gen}) we see that $c_2(V_2)=0$ if $V_2$ is the bundle related to a holomorphic section. More generally, comparing (\ref{tang30}) and (\ref{gen}) we see that the condition $c_2(V_2)=0$ implies that $S\cdot \pi^*c_1(T{\cal B}_2) =c_1(TS)$.

\section{Holomorphic and rational sections to $\mathbb{P}^1$ bundles}\label{Section_appendix}
In this appendix we will consider $\mathbb{P}^1$-bundles ${\cal B}_n$ built as the projectiviation of a rank 2 bundle ${\cal B}_n=\mathbb{P}(V_2)$ and address the question of how many sections exist to the rational fibration. As described in \eref{serre_bundle2} in Section \ref{sec:p1bun}, any rank 2 bundle $V_2$ over a complex surface can be defined via extension as
\begin{eqnarray} \label{theform}
0 \to L_1 \to V_2 \to L_2 \otimes {\mathcal I}_z \to 0
\end{eqnarray}
Here, $L_1$ and $L_2$ are line bundles and ${\mathcal I}_z$ is an ideal sheaf associated to a codimension 2 locus in the base. The projectivization of a bundle is defined only up to twisting by a line bundle and comes equipped with a set of divisors that intersect a generic fibers $S_{\cal L}$ in a hyperplane, that is a point \cite{HARTSHORNE}. These are defined such that
\beq\label{sdef3}
\pi_*(\mathcal{O}(S_{\cal L}))=V_{\cal L} \;,
\eeq
where $V_{\cal L}=V_2 \otimes {\cal L}$. The divisor classes $S_{\cal L}$ are related as $S_{\cal L}=S_{\cal O} + \pi^*c_1({\cal L})$.

If an element in a divisor class ${\cal S}_{\cal L}$ is to define the image of a section of the fibration then it must be effective. To see when this is the case this let us recall a standard result \cite{HARTSHORNE} relating sheaf valued cohomologies on the total space of a fibration $\pi: {\cal B}_n \to { B}_{n-1}$ to those on the base.
\begin{eqnarray} \label{cohrel}
H^m({\cal B}_n,{\cal F})= \bigoplus_{p+q=m} H^p({ B}_{n-1},R^q\pi_* {\cal F})
\end{eqnarray}
In particular then, given (\ref{sdef3}), we have that,
\begin{eqnarray}
h^0({\cal B}_n,{\cal O}(S_{\cal L})) = h^0({ B}_{n-1},V_{\cal L})\;.
\end{eqnarray}
Thus $S_{\cal L}$ is an effective class if and only if the bundle $V_{\cal L}$ has global holomorphic sections. If $S_{\cal L}$ is effective then clearly it does indeed induce a section of the projective bundle (i.e. it provides a map $\sigma: { B}_{n-1} \to {\cal B}_n$ such that $\pi \circ \sigma=\id_{{ B}_{n-1}}$). 

From the above discussion it is easy to see that there will always be at least one effective $S_{\cal L}$. Twisting the sequence (\ref{theform}) by $L_1^{\vee}$ we arrive at the following.
\begin{eqnarray} \label{alwayseff}
0 \to {\cal O} \to V_{L_1^{\vee}} \to L_2 \otimes {\mathcal I}_z \otimes L_1^{\vee} \to 0
\end{eqnarray}
Examining the long exact sequence in cohomology associated to (\ref{alwayseff}) we find that $h^0({ B}_2,V_{L_1^{\vee}}) \geq 1$ and thus $S_{L_1^{\vee}}$ is effective.

\vspace{0.2cm}

Having determined the existence of sections to $\pi: {\cal B}_n \to { B}_{n-1}$ we should next ask about their nature. Specifically, we would like to know when these sections intersect {\it every} fiber, rather than simply the generic fiber, at a single point. We shall refer to these sections, whose zero-loci are strictly diffeomorphic to the base ${ B}_{n-1}$ (instead of merely birational to it) as holomorphic and sections not of this form as rational\footnote{We will assume in what follows that the sections we are considering are smooth.}. 

In understanding whether a section is holomorphic, the following result is useful \cite{HARTSHORNE}:
\begin{theorem}\label{secthhm1}
Let $\mathbb{P}(V_2)$ be defined as above and let $g: {\cal S} \rightarrow { B}_{n-1}$ be any morphism. Then defining a morphism $f: {\cal S} \rightarrow \mathbb{P}(V_2)$ is equivalent to specifying a surjective morphism of sheaves on ${\cal S}$, $g^* V_2 \rightarrow \mathcal{L} \rightarrow 0$ where $\mathcal{L}$ is an invertible sheaf on ${\cal S}$.
\end{theorem}
Consider applying this theorem to the case where ${\cal S}$ is a \emph{holomorphic} section. In such a setting $g$ is a diffeomorphism -- the section is isomorphic to ${ B}_{n-1}$. Then, there is a correspondence between such sections and surjective morphisms over the base of the form $V_2 \to L \to 0$ with $L$ a line bundle on ${ B}_{n-1}$. Such a surjection can always be completed to a short exact sequence by adding in a kernel,
\begin{eqnarray} \label{simpext}
0 \to {\cal K} \to V_2 \to L \to 0\;,
\end{eqnarray}
and as such, if a holomorphic section exists then $V_2$ must be describable as an extension of two line bundles (that is it must be possible to take the ideal sheaf ${\mathcal I}_z$ in (\ref{theform}) to be trivial).

Their exists a standard result which tells us where a given element of a divisor class $S_{\cal L}$ will wrap an entire fiber of the projective bundle, rather than intersecting the fiber at a point \cite{Eisenbud_Harris}. An element of the class $S_{\cal L}$ is associated to a global section of the bundle $V_{\cal L}$. The loci over the base where the section wraps the fiber is given by those points where the section of $V_{\cal L}$ vanishes. The section is holomorphic iff there are no such points. It will be useful to note that every ${\cal B}_n$ that can be written as the projectivization of an extension of the form (\ref{simpext}) admits such a holomorphic section. A simple twist takes this bundle to
\begin{eqnarray} \label{twist2}
0 \to {\cal O} \to V_{\cal K}^{\vee}  \to L \otimes {\cal K}^{\vee} \to 0 \;.
\end{eqnarray}
If we are working in the case where $n =3$, for example, a section $S_{\cal L}$ is holomorphic iff $c_2(V_{\cal L})=0$, which can easily be shown to be the case for (\ref{twist2}). To see this we note that the first Chern classes of the base and section match iff $c_2(V_{\cal L})=0$, as we show in Appendix \ref{sometop}.  Given that $S_{\cal L}$ is birational to the base, it will be diffeomorphic to ${ B}_{2}$ iff their first Chern classes match.

\vspace{0.2cm}

We would also like to isolate the necessary and sufficient conditions for their to be two holomorphic sections that do not intersect one another. This is the case that is studied in standard discussions of heterotic/F-theory duality. To that end, consider the case where we have a bundle $V_2$ which can be defined by the sequence,
\begin{eqnarray} \label{first}
0 \to {\cal L}_1 \to V_2 \to {\cal L}_2 \to 0 
\end{eqnarray}
for which there is an associated holomorphic section $S$. For there to be another holomorphic section there must be another sequence,
\begin{eqnarray} \label{second}
0 \to {\cal L}_1 \otimes {\cal L}_2^{\vee} \otimes {\cal L}_3 \to V_{{\cal L}_2^{\vee} \otimes {\cal L}_3} \to {\cal L}_3 \to 0 \;,
\end{eqnarray}
with associated second section $S' = S+ \pi^* \left(c_1({\cal L}_3) - c_1({\cal L}_2)\right)$. Requiring that $S \cdot S'=0$ leads us, via a short computation using the results given in Section \ref{sec:p1bun}, to conclude that ${\cal L}_3={\cal L}_1^{\vee}$ and $c_1({\cal L}_1)\cdot c_2({\cal L}_2) =0$. The first of these two conditions, together with the fact that $V_2^{\vee} = V_2 \det (V_2)^{\vee}$, leads us to see that (\ref{second}) can be written as follows,
\begin{eqnarray}\label{third}
0 \to {\cal L}_2^{\vee} \to V_2^{\vee} \to {\cal L}_1^{\vee} \to 0,
\end{eqnarray}
and $S'=S- \pi^*(c_1({\cal L}_1)+c_1({\cal L}_2))$. The $c_1({\cal L}_1)\cdot c_2({\cal L}_2) =0$ condition is actually automatically satisfied by the assumption that both of our two sections were holomorphic so that $c_2(V_2)=0$. Thus we see that, in any case where we have two disjoint holomorphic sections, the twists of $V_2$ that those sections correspond to will be dual to one another.

At this stage, a further mathematical result is extremely useful \cite{eisenbudharris}. If ${\cal L} \subset V_2^{\vee}$ is a line sub-bundle then $\mathbb{P}({\cal L}) \subset \mathbb{P}(V_2^{\vee}) = \mathbb{P}(V_2)$ is the image of a section ${ B}_{n-1} \to \mathbb{P}(V_2)$, and indeed every section has this form\footnote{Note that for our case of the projectivization of a rank 2 bundle $\mathbb{P}(V_2)=\mathbb{P}(V_2^{\vee})$. This is because $V_2= V_2^{\vee} \otimes \det V_2$ for any rank 2 bundle and projectivizations of bundles related by twists yield the same object as described above.}. For such a section, the normal bundle to the section inside $\mathbb{P}(V_2)$ is given by ${\cal O}(S) \otimes \pi^* (V_2^{\vee}/{\cal L})$. For our sections $S$ and $S- \pi^*(c_1({\cal L}_1)+c_1({\cal L}_2))$, this means that $V_2^{\vee}/{\cal L} = {\cal O}$ and ${\cal L}_1^{\vee} \otimes {\cal L}_2^{\vee}$ respectively. Thus we can write the following two sequences for $V_2^{\vee}$,
\begin{eqnarray}
&&0 \to {\cal L}_1^{\vee} \otimes {\cal L}_2^{\vee} \to V_2^{\vee} \to {\cal O} \to 0 \\ \nonumber
&&0 \to {\cal O} \to V_2^{\vee} \to {\cal L}_1^{\vee} \otimes {\cal L}_2^{\vee} \to 0 \;,
\end{eqnarray}
and thus $V_2 = {\cal O} \oplus {\cal L}_1\otimes {\cal L}_2$.

In conclusion then, if ${\cal B}_n$ admits two disjoint holomorphic sections, then it admits a description where it splits as a sum of two line bundles. Conversely, any bundle that is written as a sum of line bundles can be twisted to be of the form $V_2''={\cal O} \oplus {\cal L}$ for some ${\cal L}$. That bundle then admits two disjoint holomorphic sections associated to the (trivial) extensions
\begin{eqnarray}
0\to {\cal O} \to V_2'' \to {\cal L} \to 0
\end{eqnarray}
and it's dual, both of which are effective because the extensions are split.

\vspace{0.2cm}

The final possibility is that all sections that appear are rational. In such a case, the morphism  $V_2 \to L \to 0$ that is implied by Theorem \ref{secthhm1} is over ${\cal S}$ which is not diffeomorphic to ${ B}_{n-1}$. In such an instance, by process of elimination from the above discussion, $V_2$ must be described by an extension of the form (\ref{theform}), including a non-trivial ideal sheaf ${\mathcal I}_z$. 

To conclude this discussion, let us consider this last case of rational sections in a little more detail.  From Theorem \ref{secthhm1} there must exist a surjection $g^*V_2 \rightarrow \mathcal{L}_2$ on $S$ which could be ``completed" into an extension sequence on $S$, 
\begin{eqnarray}
0\rightarrow \mathcal{L}_1 \rightarrow g^*V_2 \rightarrow \mathcal{L}_2 \rightarrow 0. \label{pulledback_def}
\end{eqnarray}
By applying the pushforward functor $Rg_*$ on the short exact sequence above, we cannot obtain a simple extension of two line bundles on $B_2$ because this corresponds to the case that ${\cal B}_3$ admits a holomorphic section. The only possibility that remains of course is that under push-forward we obtain \eref{theform}. Therefore without loss generality we can compare \eref{pulledback_def} and \eref{theform} and conclude that, 
\begin{eqnarray}\label{defratext}
&&\mathcal{L}_1 \simeq g^*L_1, \\ 
&&\mathcal{L}_2 = (g^*L_2)\otimes \mathcal{O}_S(e), \label{pullback_v2}
\end{eqnarray}
where $e= \sum_{i}^n e_i$ for some exceptional (i.e. $(-1)$-curves) in $S$ where $n$ is the number of points in the set $\{z\}$ (i.e. defining the ideal sheaf). 

Running the logic above in reverse, it is clear that if ${\cal B}_3 = \mathbb{P}(V)$ where $V$ is given by \eref{theform} with $\{z\}\ne 0$, then ${\cal B}_3$ admits only a rational section. The zero locus of such a rational section $S$ (which we shall denote $\tilde{B}_2$ in other sections) is only birational to the base and corresponds to the blow up of $B_2$ over the $n$ points of $\{z\}$.

\newpage

\bibliographystyle{utphys}
\bibliography{Refs2}

\providecommand{\href}[2]{#2}\begingroup\raggedright\begin{thebibliography}{10}

\bibitem{Vafa:1996xn}
C.~Vafa, ``{Evidence for F theory},''
  \href{http://dx.doi.org/10.1016/0550-3213(96)00172-1}{{\em Nucl.\ Phys.\ B}
  {\bfseries 469} (1996) 403--418},
  \href{http://arxiv.org/abs/hep-th/9602022}{{\ttfamily arXiv:hep-th/9602022}}.

\bibitem{Morrison:1996na}
D.~R. Morrison and C.~Vafa, ``{Compactifications of F theory on Calabi-Yau
  threefolds. 1},'' \href{http://dx.doi.org/10.1016/0550-3213(96)00242-8}{{\em
  Nucl.\ Phys.\ B} {\bfseries 473} (1996) 74--92},
  \href{http://arxiv.org/abs/hep-th/9602114}{{\ttfamily arXiv:hep-th/9602114}}.

\bibitem{Morrison:1996pp}
D.~R. Morrison and C.~Vafa, ``{Compactifications of F theory on Calabi-Yau
  threefolds. 2.},'' \href{http://dx.doi.org/10.1016/0550-3213(96)00369-0}{{\em
  Nucl. Phys. B} {\bfseries 476} (1996) 437--469},
  \href{http://arxiv.org/abs/hep-th/9603161}{{\ttfamily arXiv:hep-th/9603161}}.

\bibitem{Bershadsky:1996nh}
M.~Bershadsky, K.~A. Intriligator, S.~Kachru, D.~R. Morrison, V.~Sadov, and
  C.~Vafa, ``{Geometric singularities and enhanced gauge symmetries},''
  \href{http://dx.doi.org/10.1016/S0550-3213(96)90131-5}{{\em Nucl.\ Phys.\ B}
  {\bfseries 481} (1996) 215--252},
  \href{http://arxiv.org/abs/hep-th/9605200}{{\ttfamily arXiv:hep-th/9605200}}.

\bibitem{Friedman:1997ih}
R.~Friedman, J.~W. Morgan, and E.~Witten, ``{Vector bundles over elliptic
  fibrations},'' \href{http://arxiv.org/abs/alg-geom/9709029}{{\ttfamily
  arXiv:alg-geom/9709029}}.

\bibitem{Friedman:1997yq}
R.~Friedman, J.~Morgan, and E.~Witten, ``{Vector bundles and F theory},''
  \href{http://dx.doi.org/10.1007/s002200050154}{{\em Commun.\ Math.\ Phys.}
  {\bfseries 187} (1997) 679--743},
  \href{http://arxiv.org/abs/hep-th/9701162}{{\ttfamily arXiv:hep-th/9701162}}.

\bibitem{Gross:1993fd}
M.~Gross, ``{A Finiteness theorem for elliptic Calabi-Yau threefolds},''
  \href{http://arxiv.org/abs/alg-geom/9305002}{{\ttfamily
  arXiv:alg-geom/9305002}}.

\bibitem{grassi}
A.~Grassi, ``{On minimal models of elliptic threefolds},'' {\em Math. Ann.}
  {\bfseries 290} (1991) 287.

\bibitem{Di_Cerbo_2021}
G.~Di~Cerbo and R.~Svaldi, ``Birational boundedness of low-dimensional elliptic
  calabi–yau varieties with a section,''
  \href{http://dx.doi.org/10.1112/s0010437x2100717x}{{\em Compositio
  Mathematica} {\bfseries 157} no.~8, (Jul, 2021) 1766–1806}.
  \url{http://dx.doi.org/10.1112/S0010437X2100717X}.

\bibitem{birkar2020boundedness}
C.~Birkar, G.~D. Cerbo, and R.~Svaldi, ``Boundedness of elliptic calabi-yau
  varieties with a rational section,'' 2020.

\bibitem{Anderson:2017aux}
L.~B. Anderson, X.~Gao, J.~Gray, and S.-J. Lee, ``{Fibrations in CICY
  Threefolds},'' \href{http://dx.doi.org/10.1007/JHEP10(2017)077}{{\em JHEP}
  {\bfseries 10} (2017) 077}, \href{http://arxiv.org/abs/1708.07907}{{\ttfamily
  arXiv:1708.07907 [hep-th]}}.

\bibitem{Gray:2014fla}
J.~Gray, A.~S. Haupt, and A.~Lukas, ``{Topological Invariants and Fibration
  Structure of Complete Intersection Calabi-Yau Four-Folds},''
  \href{http://dx.doi.org/10.1007/JHEP09(2014)093}{{\em JHEP} {\bfseries 09}
  (2014) 093}, \href{http://arxiv.org/abs/1405.2073}{{\ttfamily arXiv:1405.2073
  [hep-th]}}.

\bibitem{Anderson:2014gla}
L.~B. Anderson and W.~Taylor, ``{Geometric constraints in dual F-theory and
  heterotic string compactifications},''
  \href{http://dx.doi.org/10.1007/JHEP08(2014)025}{{\em JHEP} {\bfseries 08}
  (2014) 025}, \href{http://arxiv.org/abs/1405.2074}{{\ttfamily arXiv:1405.2074
  [hep-th]}}.

\bibitem{Halverson:2015jua}
J.~Halverson and W.~Taylor, ``{$ {\mathrm{\mathbb{P}}}^1 $-bundle bases and the
  prevalence of non-Higgsable structure in 4D F-theory models},''
  \href{http://dx.doi.org/10.1007/JHEP09(2015)086}{{\em JHEP} {\bfseries 09}
  (2015) 086}, \href{http://arxiv.org/abs/1506.03204}{{\ttfamily
  arXiv:1506.03204 [hep-th]}}.

\bibitem{Friedman:1998}
R.~Friedman, {\em {Algebraic Surfaces and Holomorphic Vector Bundles}}.
\newblock Springer, 1998.

\bibitem{Sarkisov_1983}
V.~G. Sarkisov, ``On conic bundle structures,''
  \href{http://dx.doi.org/10.1070/im1983v020n02abeh001354}{{\em Mathematics of
  the {USSR}-Izvestiya} {\bfseries 20} no.~2, (Apr, 1983) 355--390}.

\bibitem{Donagi:2012ts}
R.~Donagi, S.~Katz, and M.~Wijnholt, ``{Weak Coupling, Degeneration and Log
  Calabi-Yau Spaces},'' \href{http://arxiv.org/abs/1212.0553}{{\ttfamily
  arXiv:1212.0553 [hep-th]}}.

\bibitem{Heckman:2013sfa}
J.~J. Heckman, H.~Lin, and S.-T. Yau, ``{Building Blocks for Generalized
  Heterotic/F-theory Duality},''
  \href{http://dx.doi.org/10.4310/ATMP.2014.v18.n6.a7}{{\em Adv. Theor. Math.
  Phys.} {\bfseries 18} no.~6, (2014) 1463--1503},
  \href{http://arxiv.org/abs/1311.6477}{{\ttfamily arXiv:1311.6477 [hep-th]}}.

\bibitem{brauer}
J.~Giraud, {\em {Dix exposes sur la cohomologie des schemas}}.
\newblock Amsterdam, North-Holland Publishing Company, 1968.

\bibitem{HARTSHORNE}
R.~Hartshorne, {\em Algebraic geometry}.
\newblock Graduate texts in mathematics. Springer, New York, 1977.

\bibitem{Projectivebundlesonacomplextorus}
G.~Elencwajg and M.~Narasimhan, ``Projective bundles on a complex torus.,''
  {\em Journal fur die reine und angewandte Mathematik} {\bfseries 1983}
  no.~340, (1983) 1--5.

\bibitem{Aspinwall_1998}
P.~S. Aspinwall, ``Aspects of the hypermultiplet moduli space in string
  duality,'' \href{http://dx.doi.org/10.1088/1126-6708/1998/04/019}{{\em
  Journal of High Energy Physics} {\bfseries 1998} no.~04, (Apr, 1998)
  019–019}. \url{http://dx.doi.org/10.1088/1126-6708/1998/04/019}.

\bibitem{Eisenbud_Harris}
D.~Eisenbud and J.~Harris, {\em {The Geometry of Schemes}}.
\newblock Springer, 2000.

\bibitem{2018RuMaS..73..375P}
Y.~G. {Prokhorov}, ``{The rationality problem for conic bundles},''
  \href{http://dx.doi.org/10.1070/RM9811}{{\em Russian Mathematical Surveys}
  {\bfseries 73} no.~3, (June, 2018) 375},
  \href{http://arxiv.org/abs/1712.05564}{{\ttfamily arXiv:1712.05564
  [math.AG]}}.

\bibitem{Braun:2016sks}
A.~P. Braun and T.~Watari, ``{Heterotic-Type IIA Duality and Degenerations of
  K3 Surfaces},'' \href{http://dx.doi.org/10.1007/JHEP08(2016)034}{{\em JHEP}
  {\bfseries 08} (2016) 034}, \href{http://arxiv.org/abs/1604.06437}{{\ttfamily
  arXiv:1604.06437 [hep-th]}}.

\bibitem{Candelas:1996ht}
P.~Candelas, E.~Perevalov, and G.~Rajesh, ``{F theory duals of nonperturbative
  heterotic E(8) x E(8) vacua in six-dimensions},''
  \href{http://dx.doi.org/10.1016/S0550-3213(97)00375-1}{{\em Nucl.\ Phys.\ B}
  {\bfseries 502} (1997) 613--628},
  \href{http://arxiv.org/abs/hep-th/9606133}{{\ttfamily arXiv:hep-th/9606133}}.

\bibitem{Morrison:2012np}
D.~R. Morrison and W.~Taylor, ``{Classifying bases for 6D F-theory models},''
  \href{http://dx.doi.org/10.2478/s11534-012-0065-4}{{\em Central Eur. J.
  Phys.} {\bfseries 10} (2012) 1072--1088},
  \href{http://arxiv.org/abs/1201.1943}{{\ttfamily arXiv:1201.1943 [hep-th]}}.

\bibitem{Donagi:1998vw}
R.~Donagi, ``{Heterotic / F theory duality: ICMP lecture},'' in {\em
  {Mathematical physics. Proceedings, 12th International Congress, ICMP'97,
  Brisbane, Australia, July 13-19, 1997}}, pp.~206--213.
\newblock 2, 1998.
\newblock \href{http://arxiv.org/abs/hep-th/9802093}{{\ttfamily
  arXiv:hep-th/9802093}}.

\bibitem{Aspinwall:1998bw}
P.~S. Aspinwall, ``{Aspects of the hypermultiplet moduli space in string
  duality},'' \href{http://dx.doi.org/10.1088/1126-6708/1998/04/019}{{\em JHEP}
  {\bfseries 04} (1998) 019},
  \href{http://arxiv.org/abs/hep-th/9802194}{{\ttfamily arXiv:hep-th/9802194}}.

\bibitem{Duff:1996rs}
M.~Duff, R.~Minasian, and E.~Witten, ``{Evidence for heterotic / heterotic
  duality},'' \href{http://dx.doi.org/10.1016/0550-3213(96)00059-4}{{\em Nucl.\
  Phys.\ B} {\bfseries 465} (1996) 413--438},
  \href{http://arxiv.org/abs/hep-th/9601036}{{\ttfamily arXiv:hep-th/9601036}}.

\bibitem{Anderson:2016cdu}
L.~B. Anderson, X.~Gao, J.~Gray, and S.-J. Lee, ``{Multiple Fibrations in
  Calabi-Yau Geometry and String Dualities},''
  \href{http://dx.doi.org/10.1007/JHEP10(2016)105}{{\em JHEP} {\bfseries 10}
  (2016) 105}, \href{http://arxiv.org/abs/1608.07555}{{\ttfamily
  arXiv:1608.07555 [hep-th]}}.

\bibitem{Hubsch:1992nu}
T.~Hubsch, {\em {Calabi-Yau manifolds: A Bestiary for physicists}}.
\newblock World Scientific, Singapore, 1994.

\bibitem{Berglund:2016yqo}
P.~Berglund and T.~H\"ubsch, ``{On Calabi\textendash{}Yau generalized complete
  intersections from Hirzebruch varieties and novel $K3$-fibrations},''
  \href{http://dx.doi.org/10.4310/ATMP.2018.v22.n2.a1}{{\em Adv. Theor. Math.
  Phys.} {\bfseries 22} (2018) 261--303},
  \href{http://arxiv.org/abs/1606.07420}{{\ttfamily arXiv:1606.07420
  [hep-th]}}.

\bibitem{Berglund:2016nvh}
P.~Berglund and T.~Hubsch, ``{A Generalized Construction of Calabi-Yau Models
  and Mirror Symmetry},''
  \href{http://dx.doi.org/10.21468/SciPostPhys.4.2.009}{{\em SciPost Phys.}
  {\bfseries 4} no.~2, (2018) 009},
  \href{http://arxiv.org/abs/1611.10300}{{\ttfamily arXiv:1611.10300
  [hep-th]}}.

\bibitem{Braun:2018ovc}
A.~P. Braun, C.~R. Brodie, A.~Lukas, and F.~Ruehle, ``{NS5-Branes and Line
  Bundles in Heterotic/F-Theory Duality},''
  \href{http://dx.doi.org/10.1103/PhysRevD.98.126004}{{\em Phys. Rev. D}
  {\bfseries 98} no.~12, (2018) 126004},
  \href{http://arxiv.org/abs/1803.06190}{{\ttfamily arXiv:1803.06190
  [hep-th]}}.

\bibitem{okonek}
C.~Okonek, M.~Schneider, and H.~Spindler, {\em Vector bundles on complex
  projective spaces}.
\newblock Modern Birkhauser Classics. Birkhauser, 1980.

\bibitem{donaldson1990geometry}
S.~Donaldson, S.~Donaldson, and P.~Kronheimer, {\em The Geometry of
  Four-manifolds}.
\newblock Oxford mathematical monographs. Clarendon Press, 1990.
\newblock \url{https://books.google.com/books?id=LbHmMtrebi4C}.

\bibitem{loop_group}
A.~Pressley and G.~Segal, {\em {Loop Groups}}.
\newblock Oxford Mathematical Monographs, 1988.

\bibitem{Morrison:2012js}
D.~R. Morrison and W.~Taylor, ``{Toric bases for 6D F-theory models},''
  \href{http://dx.doi.org/10.1002/prop.201200086}{{\em Fortsch. Phys.}
  {\bfseries 60} (2012) 1187--1216},
  \href{http://arxiv.org/abs/1204.0283}{{\ttfamily arXiv:1204.0283 [hep-th]}}.

\bibitem{Anderson:2015iia}
L.~B. Anderson, F.~Apruzzi, X.~Gao, J.~Gray, and S.-J. Lee, ``{A new
  construction of Calabi\textendash{}Yau manifolds: Generalized CICYs},''
  \href{http://dx.doi.org/10.1016/j.nuclphysb.2016.03.016}{{\em Nucl. Phys. B}
  {\bfseries 906} (2016) 441--496},
  \href{http://arxiv.org/abs/1507.03235}{{\ttfamily arXiv:1507.03235
  [hep-th]}}.

\bibitem{CICYpackage}
L.~Anderson, J.~Gray, Y.-H. He, S.-J. Lee, and A.~Lukas, ``{CICY package, based
  on methods described in arXiv:0911.1569, arXiv:0911.0865, arXiv:0805.2875,
  hep-th/0703249, hep-th/0702210},''.

\bibitem{Katz:2011qp}
S.~Katz, D.~R. Morrison, S.~Schafer-Nameki, and J.~Sully, ``{Tate's algorithm
  and F-theory},'' \href{http://dx.doi.org/10.1007/JHEP08(2011)094}{{\em JHEP}
  {\bfseries 08} (2011) 094}, \href{http://arxiv.org/abs/1106.3854}{{\ttfamily
  arXiv:1106.3854 [hep-th]}}.

\bibitem{Morrison:2014lca}
D.~R. Morrison and W.~Taylor, ``{Non-Higgsable clusters for 4D F-theory
  models},'' \href{http://dx.doi.org/10.1007/JHEP05(2015)080}{{\em JHEP}
  {\bfseries 05} (2015) 080}, \href{http://arxiv.org/abs/1412.6112}{{\ttfamily
  arXiv:1412.6112 [hep-th]}}.

\bibitem{Grassi:2011hq}
A.~Grassi and D.~R. Morrison, ``{Anomalies and the Euler characteristic of
  elliptic Calabi-Yau threefolds},''
  \href{http://dx.doi.org/10.4310/CNTP.2012.v6.n1.a2}{{\em Commun. Num. Theor.
  Phys.} {\bfseries 6} (2012) 51--127},
  \href{http://arxiv.org/abs/1109.0042}{{\ttfamily arXiv:1109.0042 [hep-th]}}.

\bibitem{Anderson:2015cqy}
L.~B. Anderson, J.~Gray, N.~Raghuram, and W.~Taylor, ``{Matter in
  transition},'' \href{http://dx.doi.org/10.1007/JHEP04(2016)080}{{\em JHEP}
  {\bfseries 04} (2016) 080}, \href{http://arxiv.org/abs/1512.05791}{{\ttfamily
  arXiv:1512.05791 [hep-th]}}.

\bibitem{Kulikov}
V.~Kulikov, ``Degenerations of k3 surfaces and enriques surfaces,'' {\em
  Mathematics of the USSR-Izvestiya} {\bfseries 11} (1977) 957--989.

\bibitem{Persson_Pinkham}
U.~Persson and H.~Pinkham, ``Degeneration of surfaces with trivial canonical
  bundle,'' {\em Annals of Mathematics} {\bfseries 113} (1981) 45 -- 66.

\bibitem{Donagi:2008ca}
R.~Donagi and M.~Wijnholt, ``{Model Building with F-Theory},''
  \href{http://dx.doi.org/10.4310/ATMP.2011.v15.n5.a2}{{\em Adv. Theor. Math.
  Phys.} {\bfseries 15} no.~5, (2011) 1237--1317},
  \href{http://arxiv.org/abs/0802.2969}{{\ttfamily arXiv:0802.2969 [hep-th]}}.

\bibitem{Cvetic:2015uwu}
M.~Cvetic, A.~Grassi, D.~Klevers, M.~Poretschkin, and P.~Song, ``{Origin of
  Abelian Gauge Symmetries in Heterotic/F-theory Duality},''
  \href{http://dx.doi.org/10.1007/JHEP04(2016)041}{{\em JHEP} {\bfseries 04}
  (2016) 041}, \href{http://arxiv.org/abs/1511.08208}{{\ttfamily
  arXiv:1511.08208 [hep-th]}}.

\bibitem{Blumenhagen:2010pv}
R.~Blumenhagen, B.~Jurke, T.~Rahn, and H.~Roschy, ``{Cohomology of Line
  Bundles: A Computational Algorithm},''
  \href{http://dx.doi.org/10.1063/1.3501132}{{\em J. Math. Phys.} {\bfseries
  51} (2010) 103525}, \href{http://arxiv.org/abs/1003.5217}{{\ttfamily
  arXiv:1003.5217 [hep-th]}}.

\bibitem{cohomCalg:Implementation}
``{cohomCalg package}.'' Download link, 2010.
\newblock \url{https://github.com/BenjaminJurke/cohomCalg}. High-performance
  line bundle cohomology computation based on \cite{Blumenhagen:2010pv}.

\bibitem{Grimm:2012yq}
T.~W. Grimm and W.~Taylor, ``{Structure in 6D and 4D N=1 supergravity theories
  from F-theory},'' \href{http://dx.doi.org/10.1007/JHEP10(2012)105}{{\em JHEP}
  {\bfseries 10} (2012) 105}, \href{http://arxiv.org/abs/1204.3092}{{\ttfamily
  arXiv:1204.3092 [hep-th]}}.

\bibitem{Grimm:2010ks}
T.~W. Grimm, ``{The N=1 effective action of F-theory compactifications},''
  \href{http://dx.doi.org/10.1016/j.nuclphysb.2010.11.018}{{\em Nucl. Phys. B}
  {\bfseries 845} (2011) 48--92},
  \href{http://arxiv.org/abs/1008.4133}{{\ttfamily arXiv:1008.4133 [hep-th]}}.

\bibitem{Morrison:2012ei}
D.~R. Morrison and D.~S. Park, ``{F-Theory and the Mordell-Weil Group of
  Elliptically-Fibered Calabi-Yau Threefolds},''
  \href{http://dx.doi.org/10.1007/JHEP10(2012)128}{{\em JHEP} {\bfseries 10}
  (2012) 128}, \href{http://arxiv.org/abs/1208.2695}{{\ttfamily arXiv:1208.2695
  [hep-th]}}.

\bibitem{Raghuram:2020vxm}
N.~Raghuram, W.~Taylor, and A.~P. Turner, ``{Automatic enhancement in 6D
  supergravity and F-theory models},''
  \href{http://dx.doi.org/10.1007/JHEP07(2021)048}{{\em JHEP} {\bfseries 07}
  (2021) 048}, \href{http://arxiv.org/abs/2012.01437}{{\ttfamily
  arXiv:2012.01437 [hep-th]}}.

\bibitem{Heckman:2013pva}
J.~J. Heckman, D.~R. Morrison, and C.~Vafa, ``{On the Classification of 6D
  SCFTs and Generalized ADE Orbifolds},''
  \href{http://dx.doi.org/10.1007/JHEP05(2014)028}{{\em JHEP} {\bfseries 05}
  (2014) 028}, \href{http://arxiv.org/abs/1312.5746}{{\ttfamily arXiv:1312.5746
  [hep-th]}}. [Erratum: JHEP 06, 017 (2015)].

\bibitem{Johnson:2016qar}
S.~B. Johnson and W.~Taylor, ``{Enhanced gauge symmetry in 6D F-theory models
  and tuned elliptic Calabi-Yau threefolds},''
  \href{http://dx.doi.org/10.1002/prop.201600074}{{\em Fortsch. Phys.}
  {\bfseries 64} (2016) 581--644},
  \href{http://arxiv.org/abs/1605.08052}{{\ttfamily arXiv:1605.08052
  [hep-th]}}.

\bibitem{Morrison:2016djb}
D.~R. Morrison and T.~Rudelius, ``{F-theory and Unpaired Tensors in 6D SCFTs
  and LSTs},'' \href{http://dx.doi.org/10.1002/prop.201600069}{{\em Fortsch.
  Phys.} {\bfseries 64} (2016) 645--656},
  \href{http://arxiv.org/abs/1605.08045}{{\ttfamily arXiv:1605.08045
  [hep-th]}}.

\bibitem{Greiner:2017ery}
S.~Greiner and T.~W. Grimm, ``{Three-form periods on Calabi-Yau fourfolds:
  Toric hypersurfaces and F-theory applications},''
  \href{http://dx.doi.org/10.1007/JHEP05(2017)151}{{\em JHEP} {\bfseries 05}
  (2017) 151}, \href{http://arxiv.org/abs/1702.03217}{{\ttfamily
  arXiv:1702.03217 [hep-th]}}.

\bibitem{Oehlmann:2021man}
P.-K. Oehlmann, ``{Non-flat elliptic four-folds, three-form cohomology and
  strongly coupled theories in four dimensions},''
  \href{http://arxiv.org/abs/2102.10722}{{\ttfamily arXiv:2102.10722
  [hep-th]}}.

\bibitem{Chaudhuri:1995fk}
S.~Chaudhuri, G.~Hockney, and J.~D. Lykken, ``{Maximally supersymmetric string
  theories in D \ensuremath{<} 10},''
  \href{http://dx.doi.org/10.1103/PhysRevLett.75.2264}{{\em Phys. Rev. Lett.}
  {\bfseries 75} (1995) 2264--2267},
  \href{http://arxiv.org/abs/hep-th/9505054}{{\ttfamily arXiv:hep-th/9505054}}.

\bibitem{Chaudhuri:1995bf}
S.~Chaudhuri and J.~Polchinski, ``{Moduli space of CHL strings},''
  \href{http://dx.doi.org/10.1103/PhysRevD.52.7168}{{\em Phys. Rev. D}
  {\bfseries 52} (1995) 7168--7173},
  \href{http://arxiv.org/abs/hep-th/9506048}{{\ttfamily arXiv:hep-th/9506048}}.

\bibitem{Park:1996it}
J.~Park, ``{Orientifold and F theory duals of CHL strings},''
  \href{http://dx.doi.org/10.1016/S0370-2693(97)01492-5}{{\em Phys. Lett. B}
  {\bfseries 418} (1998) 91--97},
  \href{http://arxiv.org/abs/hep-th/9611119}{{\ttfamily arXiv:hep-th/9611119}}.

\bibitem{Del_Zotto_2015}
M.~Del~Zotto, J.~J. Heckman, A.~Tomasiello, and C.~Vafa, ``{6d Conformal
  Matter},'' \href{http://dx.doi.org/10.1007/JHEP02(2015)054}{{\em JHEP}
  {\bfseries 02} (2015) 054}, \href{http://arxiv.org/abs/1407.6359}{{\ttfamily
  arXiv:1407.6359 [hep-th]}}.

\bibitem{Szurek}
M.~Szurek and J.~Wisniewski, ``{Fano bundles over $P^3$ and $Q_3$},''
  \href{http://dx.doi.org/10.2140/pjm.1990.141.197}{{\em Pacific Journal of
  Mathematics} {\bfseries 141} (1990) 197–208}.

\bibitem{eisenbudharris}
D.~Eisenbud and J.~Harris,
  \href{http://dx.doi.org/10.1017/CBO9781139062046}{{\em 3264 and All That: A
  Second Course in Algebraic Geometry}}.
\newblock Cambridge University Press, {New York}, {2016}.

\end{thebibliography}\endgroup
\end{document}